\documentclass[a4paper,12pt]{article}

\usepackage{jheppub,xcolor,url}
\usepackage{natbib}
\usepackage[T1]{fontenc} 
\usepackage{caption,subcaption,float, overpic}
\usepackage{tabularx}
\usepackage{comment}
\title{Condensate phases of nuclear matter from AdS Hardwall models}

\author{Akash Singh, K. P. Yogendran}
\affiliation{Department of Physical Sciences,\\
IISER Mohali\\Sector 81, Knowledge City\\Punjab, 140306, India}

\emailAdd{akashsingh@iisermohali.ac.in, yogendran@iisermohali.ac.in}

\abstract{This work develops our previous study of confined phases at finite densities in AdS/QCD by systematically exploring the possibility of baryonic condensates. Using phenomenologically motivated boundary conditions in an AdS hardwall model, we show that both baryonic and quark type condensates dominate the phase diagram at low temperatures. We also undertake a careful scan of the parameter space to extract robust conclusions.
}
\gdef\@fpheader{}

\begin{document} 
\maketitle
\flushbottom

\section{Introduction}
Compact stars are remnants of old massive stars that have gone through a supernova explosion, providing a unique opportunity to study matter at high density and low temperature. Inside these stars, matter ranges from nuclei at low densities in the crust through highly neutron-rich uniform matter in the outer core leading to a variety of possible condensates, such as baryon superfluids (and/or superconductors), meson condensates, and even quark pairs in the inner core (a fairly modern summary can be found in the excellent book \cite{Haensel:2007yy}). All of these can have a considerable impact on the characteristics of the compact star \cite{Pethick:2015jma,Alford:2007xm}. QCD matter at high densities and low temperatures thus presents an exciting opportunity to explore a novel condensed matter system with multiple mass scales. However, quantitative analysis of the QCD phase diagram and/or equation of state (EOS) in this regime is a hard problem with multiple unanswered questions \cite{Brambilla:2014jmp}. Perturbative QCD analysis is only reliable at asymptotically high densities and temperatures where the quarks are free degrees of freedom. Lattice QCD and various effective theory methods, such as in-medium chiral perturbation theory and mean field theory approximations \cite{Schmitt:2010pn} are employed at low densities and temperatures. However, they are not under control at mid-to-high density. Therefore, investigating the QCD phase diagram as a whole and finding the EOS is essential to understanding the structure of compact stars.    

Since the formulation of Gauge/Gravity duality (or holography) in the seminal work by Maldacena \cite{Maldacena:1997re}, the AdS/CFT correspondence has been used to study the non-perturbative strong coupling effects in the field theory using the gravitational description. Significant effort has been devoted to understanding QCD-like boundary dual theory through modified bulk gravity theories going far beyond the original conformal duality. This area of research, referred to as AdS/QCD, has been extensively explored using both phenomenologically motivated bottom-up approach that captures the essential features of the boundary theory, such as hardwall \cite{Erlich:2005qh}, softwall \cite{Karch:2006pv}, and V-QCD models \cite{Jarvinen:2011qe}, or a more rigorous top-down approach based on the first principle string theory (see \cite{Sakai:2004cn,Sakai:2005yt,Karch:2002sh,Karch:2007br, Kruczenski:2003be, Constable:1999ch, Singh:2022obu}). 
Holography has been shown to be an insightful tool in a variety of contexts, including the small shear-viscosity-to-entropy ratio \cite{Kovtun:2004de}, the Regge behavior of hadrons \cite{Afonin:2021cwo}, the Chiral symmetry breaking \cite{Erlich:2005qh}, and holographic superconductors and superfluids \cite{Gubser:2008px, Hartnoll:2008kx, Hartnoll:2008vx, Sonner:2010yx}, among others. A significant amount of work has been done in obtaining the EOS of dense QCD matter inside the star using holography \cite{BallonBayona:2007vp, Megias:2010ku, DeWolfe:2010he, Kim:2014pva, Hoyos:2016zke, Jokela:2018ers, Annala:2019puf, BitaghsirFadafan:2019ofb, Mamani:2020pks, BitaghsirFadafan:2020otb, Kovensky:2021kzl, Hoyos:2021uff, Ghoroku:2021fos, Demircik:2021zll, Hippert:2023bel, Bartolini:2023wis, Braga:2024nnj}.

From the phenomenology of QCD at high densities, we can expect three kinds of condensates to occur in such systems. Firstly, mesons such as pions, kaons, etc., can undergo Bose condensation restricted by charge conservation and beta-equilibrium. Secondly, neutron, proton and quark condensates associated with superfluidity and superconductivity. At higher densities, quark pairs forming CFL type phases have also been suggested. Finally, we have the all-pervasive chiral condensate associated with the QCD vacuum and its vanishing at deconfinement. Many authors have studied the structure of a neutron star with phenomenologically motivated condensates (for instance, see \cite{2017PhRvL.119p1104A} and citations thereof). 
The various effective field theory studies indicate a strong dependence on the parameter values and the operators included. Additionally, these do not necessarily respect causality and stability constraints.

On the other hand, the holographic approach is rather easier and is surprisingly consistent with QCD phenomenology in the regions (of temperature, density, and couplings) where it has been compared. The equations of state obtained from AdS duals respect causality and stability criteria at least as long as the bulk matter obeys standard energy conditions. The hardwall approach is particularly interesting in this context since the hardwall {\it geometry} is likely to be common to all such approaches to QCD. Such studies will present us with a range of possibilities which will help us consolidate our expectations about QCD proper. 

We therefore attempt to construct a simple holographic model which gives a complete phase diagram {\em including baryonic condensates}. This study should be viewed as an intermediate step in building a {\em complete} holographic model for realistic nuclear matter at high baryonic and isospin densities. A key step to studying QCD condensates was the identification of a dual to a confined phase at low temperatures.  This was explored at length in \cite{Singh:2024amm}. This work develops the previous exploration by adding bulk degrees of freedom dual to condensates in the boundary theory. 

This document is organized as under. 
The first section explains the new features arising from the addition of a complex scalar dual to the condensing operator. In particular, we discuss the choices of the scaling dimension and baryonic charge. Following a short summary of the finite density phases without condensates, we discuss the phase diagrams obtained with the simplest boundary condition to understand the effect of the various parameters separately. Using this knowledge, we construct phase diagrams using phenomenologically motivated boundary conditions and show that condensed phases dominate at low temperatures. These include a baryonic condensate at low densities, giving way to quark condensates at higher densities. 
We end with a discussion motivating further studies. 

\section{AdS Hardwall models}\label{hardwallmodel}
The hardwall model is based on imposing IR cutoff(s) in the bulk radial coordinate. This breaks conformal symmetry and enables the possibility of a confinement deconfinement transition \cite{Herzog:2006ra}.

In a recent study \cite{Singh:2024amm}, we further explored these models at finite baryon density by analyzing a new class of solutions called charged AdS (CAdS). Incorporating phenomenological boundary conditions at the IR boundary, we were able to obtain a satisfactory phase diagram in the $\m_B-T$ plane. In this work, we expand on the previous work by including a complex scalar field that allows for the possibility of breaking the $U(1)_B$ symmetry spontaneously. The work of \cite{Hartnoll:2008kx} showed that in the black hole background, at sufficiently high density, this indeed happens. The question, then, is whether this spontaneous breakdown occurs in the confined phases identified in our previous work.

Thus, we write a minimally coupled Einstein-Maxwell-Scalar action in 5-dimensions,
\be\label{action1} 
S=\frac{1}{2\k^2}\int d^5x \sqrt{g}\left(R-2\L\right) - \frac{1}{g_5^2}\int d^5x \sqrt{g} \frac{F^2}{4 } -\l_s\int d^5x \sqrt{g} \left(|D\psi|^2 + m^2|\psi|^2\right),
\ee
which has been shown \cite{Hartnoll:2008kx} to lead to symmetry breaking and condensates. It is remarkable that the scalar field condenses without the need for a potential. Following the reasoning in \cite{Singh:2024amm}, we set the normalization as,
\be\label{kg} 
\k^2=\frac{4\p^2 L^3}{N_c^2}; \qquad g_5^2=\frac{24 \p^2 L}{\q N_f N_c}; 
\ee 
In these units, $[F]=[l]^{-2}$, $[\y]=[l]^{-1}$, $[\l_s]=[l]^{-1}$ and $[m]=[l]^{-1}$ and all coordinates have length dimensions. 
Note that if the scalar field action is also multiplied by $N_f$, then we may interpret it as dual to the condensate of all flavors of quark rather than a single flavor. In our work, we shall regard the scalar field as representing the condensates of only the lightest baryons (or quarks). However, the action for the scalar field will not involve a factor of $N_f.$ as we use the scaling of $\y$ to set $\l_s=1.$
The cosmological constant in the Einstein-Hilbert part of the action is set to be  $2\L=-\frac{12}{L^2},$ so that the asymptotic solution is $AdS_5$ with a length scale $L$.

The standard AdS/CFT dictionary states that the mass $m$ of the scalar field controls the scaling dimension $\D$ of the scalar operator in the boundary (or scaling dimension of the condensate operator) \cite{Klebanov:1999tb,McGreevy:2009xe} via the equation

\be
m^2L^2=\D(\D-d) \implies \D_\pm =\frac{d}{2}\pm \sqrt{\frac{d^2}{4}+m^2L^2}.
\ee
The gauge transformations for scalar and vector fields are,
\bea   
\psi\xrightarrow{} e^{i q \a(x)}\psi\\
A_\m \xrightarrow{} A_\m +\del_\m \a
\eea 
This fixes $D_\m=\del_\m-i q A_\m$.

The charge $q$ of the scalar field controls the interpretation of the scalar field as a quark or baryon condensate in the following sense. In a string theory view, $B_{NS}$ counts the number of fundamental strings/quarks which combines with the $U(1)$ gauge field $2\pi\a' F$ to form a gauge invariant combination (on the worldvolume of branes). Thus, the world volume gauge field of the D7-brane counts the number of quarks \cite{Kobayashi:2006sb}.

Thus, if the {\em scalar} field $\psi$ has charge $2$ under gauge transformations, we can say that it is a quark condensate. 
The color neutrality of the condensate is captured by an $N_c$ prefactor in the action of the scalar field expressing that all colors are equally condensed. On the other hand, if it has a charge $2 N_c$ under the gauge field, we can say that it represents a baryon condensate.

If we regard $\psi$ as being related to $N_c$ fundamental strings, then its (bulk) mass can be expected to be $m\sim \frac{N_c}{l_s}$ if the strings are weakly bound to each other. Whereas, if they are strongly coupled, then the field can be interpreted as an effective degree of freedom that captures only the binding energy among the baryons and $m\sim\frac{1}{L}.$ This assignment of the mass need not be related to the charge assignment earlier: for weakly interacting strings (quarks), the charge and mass (scaling dimension) will both be proportional to $N_c$, but when they are strongly interacting, the mass can be much smaller with significant binding energy.

\subsection{Equations of motion}\label{eom}

We assume a diagonal metric ansatz,
 \be 
 ds^2=-\frac{g(z)}{h(z)}L^2 dt^2+\frac{L^2}{z^2}d\Vec{x}^2+\frac{L^2}{g(z)}dz^2
 \ee
 which implies that the (conformal) boundary is at $z=0$. 
The gauge field ansatz, $A_0=\phi(z)$, $A_{i}=0$ where $i=1,2,3$ with gauge condition $A_z=0$. The scalar field ansatz is real $\psi(z)$. The equations of motion with these ans\"atze,
\bea
g'(z)-g(z) \left(\frac{4}{z}+\frac{h'(z)}{2 h(z)}\right)+4 z-2\z z \left(\frac{ h(z) \phi '(z)^2}{4 L^2 }+\frac{m^2 L^2  \psi (z)^2}{2}\l_s g_5^2  \right)=0\label{EOMg} \\
\frac{h'(z)}{h(z)}-\frac{4}{z}-2\z  \l_s g_5^2 z\left(\psi '(z)^2+\frac{q^2  \psi (z)^2 \phi (z)^2}{g(z)^2}h(z)\right)=0\label{EOMh} \\
\phi ''(z)-\phi '(z) \left(\frac{3}{z}-\frac{h'(z)}{2 h(z)}\right)-\l_s g_5^2\frac{2 q^2 \psi(z)^2 L^2}{g(z)}\phi (z)=0\label{EOMph} \\
\psi ''(z)+\psi '(z) \left(-\frac{3}{z}+\frac{g'(z)}{g(z)}-\frac{h'(z)}{2 h(z)}\right)- m_{eff}^2 L^2 \psi (z)=0\label{EOMy} 
\eea
where, $\z=\frac{2 \k^2}{3 g_5^2}=\frac{\q N_f L^2}{9N_c}$, and $m_{eff}^2=\frac{m^2}{g(z)}-\frac{q^2 h(z) \phi (z)^2}{ g(z)^2 L^2}$.
The parameter $\z$ controls the effect of the gauge field on the background and large $N_c$ corresponds to the probe approximation. These equations are seen to be invariant under the following two  transformations:
\bea
(\Vec{x},t,z)\to \a(\Vec{x},t, z),\; \f\to\frac{\f}{\a},\; g\to \a^2 g,\; h\to \a^4 h,\\ 
h\to \frac{h}{\b^2} \qquad \f \to \f \b.
\eea
The former corresponds to a scale symmetry, while the latter is expected from time rescaling invariance. These can be fully eliminated by a hardwall cutoff $z_0$ and setting $h(\e)=1$, respectively.

\subsection{Boundary conditions and Parameters}\label{bc}

As the differential equations show, we need two conditions on $\phi,\psi$ (\eqref{EOMph},\eqref{EOMy}) and one each on $h,g$ (\eqref{EOMh},\eqref{EOMg}) to obtain a unique solution.

The asymptotic analysis of the equation of motion near the UV boundary ($z\to \e$) gives: 
\begin{itemize}
\item $\phi\sim \mu-Qz^2+\dots$
\item $\y \sim \y_{+}z^{\Delta_+}+\y_- z^{\Delta_-}+\dots\qquad$  
\item $g\sim z^2(1+c z^4+\dots)$
\item $h\sim z^4(1+ \Tilde{h}\y_-^2  z^{2\Delta_-}+\dots)$
\end{itemize}

Two out of these six boundary conditions are determined by physical constraints. We require the solutions to have the same chemical potential $\m=\m_Q=\frac{\m_B}{N_c}$ at the UV boundary. To model the spontaneous breaking of symmetry, the scalar field requires that the non-normalizable mode (source) vanishes for all solutions. This leads to conditions $\f(\e) = \m$ and $\y_{-} = 0$. 

Requiring $h(\e)=1$ after fixing the hardwall cutoff $z_0$ gives us one more boundary condition. Consequently, we have to fix the remaining three boundary conditions on $g,\f,\y$ at the IR boundary.

\section{Confined phases at finite density: A review}\label{Confined}

To begin with, we list the possibilities at finite chemical potential but without a condensate \cite{Singh:2024amm}. For a trivial scalar field $\psi=0$, the equations of motion \eqref{EOMg}, \eqref{EOMh}, \eqref{EOMph} can be solved analytically. At finite temperature and chemical potential, three possible solutions are, 
\begin{itemize}
    \item Thermal AdS (thAdS)
        \be 
        g(z)=z^2 \: ;\qquad h(z)=z^4 \: ;\qquad
        \phi(z)=\m
        \ee 
with zero charge density
    \item Charged black hole (CBH)
        \be  \label{CBH} 
        g(z)=z^2\left(1-c z^4 + \frac{\z \mu ^2}{z_H^4 L^2}z^6\right)\: ;\qquad h(z)=z^4 \: ;\qquad \phi(z)=\m\left(1-\frac{z^2}{z_H^2}\right) 
        \ee
        where $z_H$ is the black hole horizon. The parameter $c$ is fixed by the condition that $z_H$ is the smallest root of $g(z_H)$ for each $\mu$ (quark chemical potential).

\item Charged AdS (CAdS)
\be \label{CAdS} 
  g(z)=z^2\left(1-c z^4 + \frac{\z Q}{L^2}z^6\right)\: ;\qquad h(z)=z^4 \;  ; \quad \phi(z)=\m-Q z^2;
\ee
with $\frac{\z^2 Q ^{4}}{L^4c ^3}>\frac{4}{27}$ implying that the function $g(z)$ never has a root. 
\end{itemize}
Here, $Q$ and $c$ (or equivalently, $g(z_0)$) are free parameters that have to be fixed by applying boundary conditions to completely determine the solutions.

A simple choice of boundary condition on the gauge field is to set $\f(z_0)=0$ which determines $Q(\mu)=\frac{\mu}{z_0 ^2}$ with dimensions of number density.

On the other hand, if there are sources for the gauge field that are present behind the cutoff surface, then it is natural to set $Q(\mu)$ such that the expression for pressure correctly reproduces the expected contribution from these sources. In this view, we imagine the IR region in the bulk $z>z_{0}$ which involves a distribution of baryon sources such as wrapped branes or fundamental strings. These have been shown to be ``pulled" to the far IR regions \cite{Imamura:1998hf, Brandhuber:1998xy,Gorsky:2015pra} in AdS geometries which justifies the expectation (the last reference importantly points out that the chiral condensate prevents the baryons from ``falling all the way''). The distribution of these branes in the IR regions can be expected to capture the pressure and energy density of this collection of baryons \cite{Kovensky:2021kzl} which feeds into the region $z<z_0$ via the boundary conditions. 

In conventional literature, finite baryon density is studied by using phenomenological models such as the Skyrme model, NJL or chiral perturbation theory (depending on the density). In our present work, we will use the thermodynamic quantities computed from these EFT approaches to determine the IR boundary conditions as detailed in our earlier study \cite{Singh:2024amm}. 

\be
\frac{N_c ^2}{8\pi^2} \left(\frac{g(z_0)}{z_0 ^4\sqrt{h(z_0)}}+\frac{\z Q^2}{L^2z_0 ^2}\sqrt{h(z_0)}\right)=p_B+\frac{N_c ^2}{ 8\pi ^2 }\frac{g_0}{z_0^6}.
\ee 
The total pressure can then be equated to the pressure of nuclear matter $p_B(T,\mu)$ (modeled phenomenologically) together with other contributions $z_0$ and $g_0$ from glueballs and even mesons. In this work, we will simply assume that the latter is a constant $g_0=6$ and is fixed by phenomenology albeit at $T=\m=0$ \cite{Singh:2024amm}.  This equation determines the IR boundary condition $g(z_0).$ 
Additionally, the boundary condition for the scalar potential is also specified by phenomenology:
\be\label{fprime}
\f'(z_0)=\frac{g_5 ^2 z_0 ^3}{2\sqrt{h(z_0)}}\rho_Q
\ee
where $\rho_Q$ is the quark number density determined by the phenomenological models that determine the pressure $p_B.$ This equation is nothing but Gauss' law which attributes the electric flux at the cutoff surface on the RHS to the baryonic number density located behind the cutoff surface (the normalization is determined by the usual AdS/CFT). Note that this also fixes the UV chemical potential:
\be
\f(\e)=\mu_Q = \frac{\mu_B}{N_c}
\ee
in which case the cutoff value of the scalar potential is determined by the numerical solution via 
\be
\f(\e)-\f(z_0)=\int dz\  \f'(z).
\ee
Thus, we may use a shooting method starting from the IR with these boundary conditions. 

\subsection{Phase diagram without condensates}
For a self-contained presentation, we will summarize the results of our earlier study \cite{Singh:2024amm} which did not consider the possibility of condensation.

At baryon densities up to $3\rho_0$ (here $\rho_0=0.157/fm^3$ is the nuclear saturation density), calculations using in-medium chiral perturbation theory \cite{Fiorilla} suggest a van der Waals gas description with a gas-liquid transition as the density increases. This model is sensible only at densities below $3.3\rho_0$ and temperature $T\lesssim50$ MeV while the liquid gas transition occurs below $\rho_0$ and $T\sim20$ MeV (these numbers are fit to the data mentioned in \cite{Fiorilla}).

At higher densities, the baryons overlap strongly and one could expect that a quark description such as the Nambu-Jona-Lasinio (NJL) model is the better description (however, see \cite{McLerran:2021zvt}). If this is to be true, we expect that the latter description has higher pressure indicating a change in the effective degrees of freedom. The NJL model is variously augmented by including a variety of condensates and possibly the Polyakov Loop to include effects of confinement. In our work, we will use the two flavor NJL without the Polyakov loop as a description of the IR physics. The idea being that using the AdS geometry together with the $g_0$ parameter will capture some aspects of confinement dynamics represented by the Polyakov loop. We note that the NJL model includes a UV cutoff $\L$ and can be expected to be sensible for chemical potentials $\m_Q<\L$. The reason for using the {\em two} flavor model is that the in-medium chiral perturbation theory which results in the vdW gas description that does not consider the strange quark baryons.

\begin{figure}[h]
    \centering
    \subfloat[Pressure at T=1]{\includegraphics[width=0.45\linewidth]{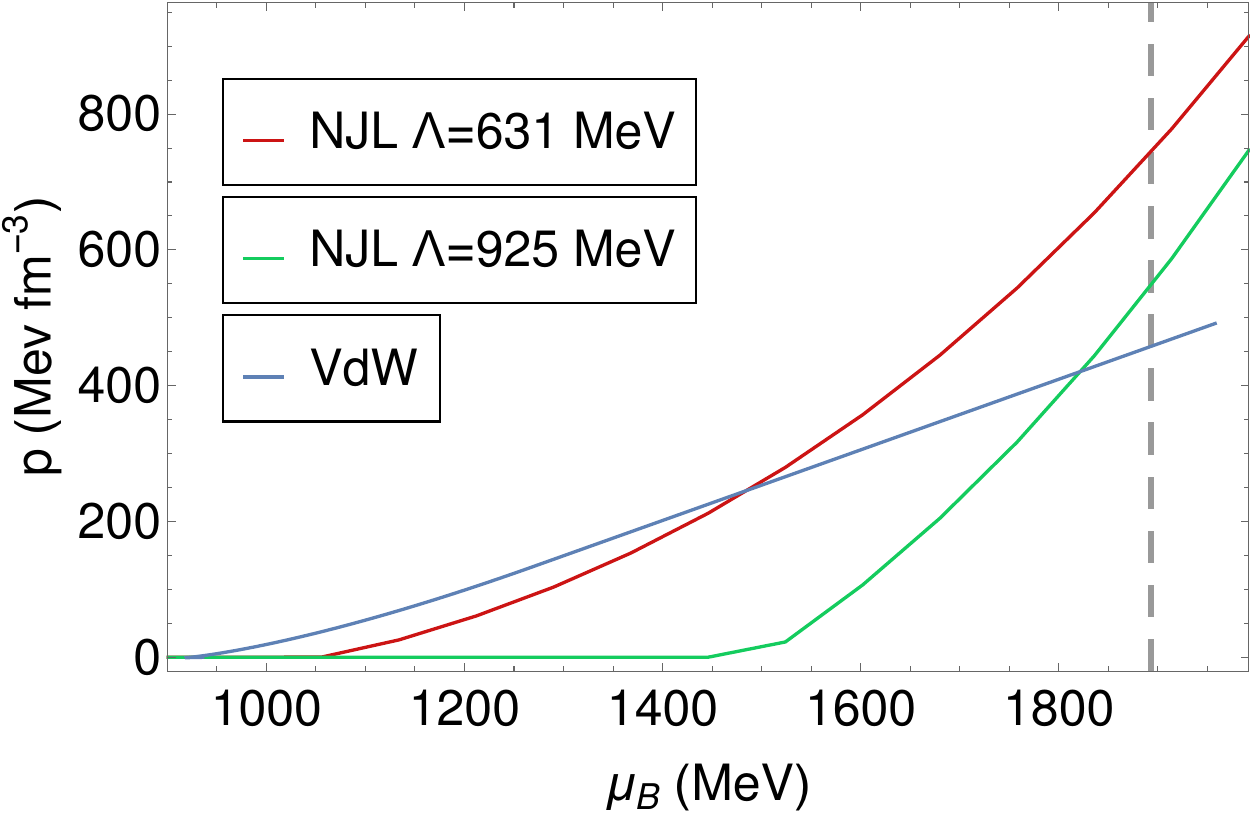}}\hfill
    \subfloat[Pressure at T=30]{\includegraphics[width=0.45\linewidth]{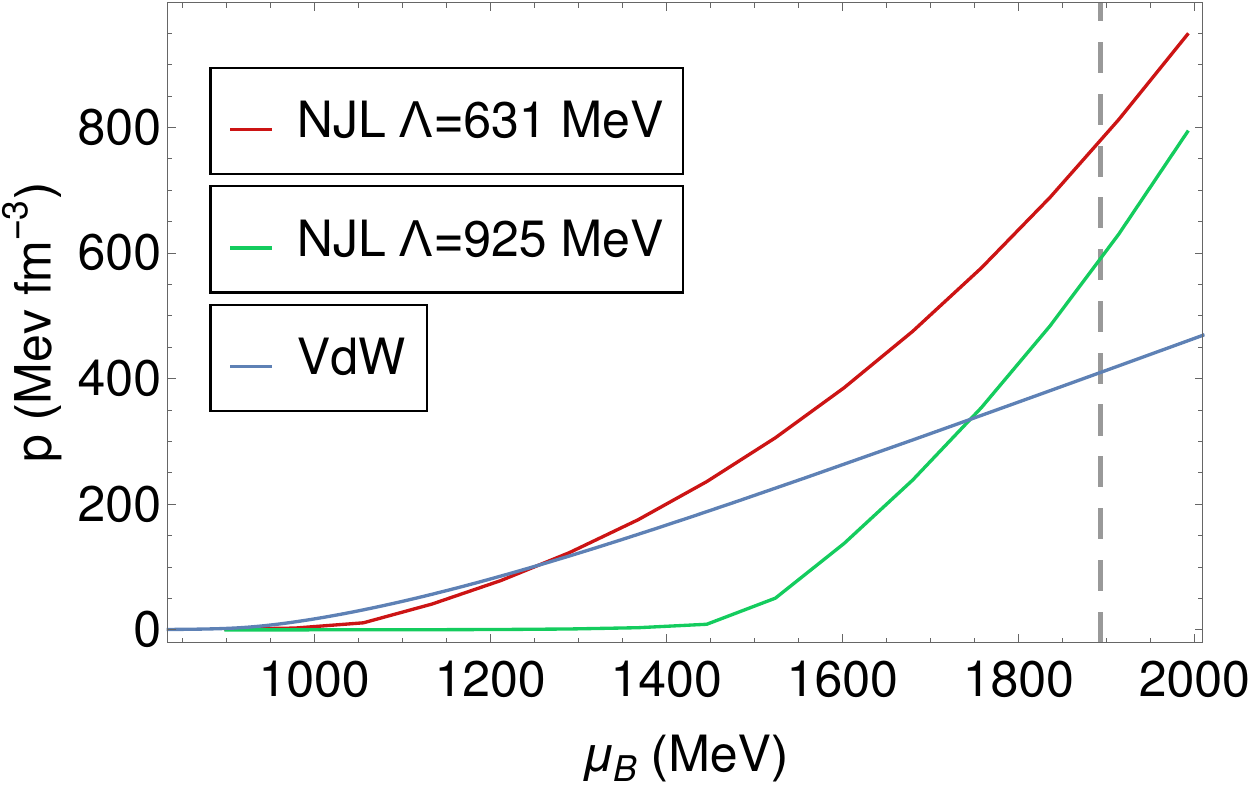}}
    \caption{Phase transition between NJL and vdW.}
    \label{fig:NJLVDWtransition}
\end{figure}

Using $p_B$ and $\rho_Q$ as computed from these models to fix the boundary conditions,
we showed that a first-order phase transition occurs from the NJL phase to the vdW-type phase with a decrease in density. It appears reasonable that the NJL description is better at higher densities. The phase transition point, shown in figure \ref{fig:NJLVDWtransition} for two different values of temperature, however, depends strongly on the choice of the parameters in the NJL model \cite{Asakawa:1989bq}. The green and red curves represent NJL with $\Lambda = 925$ MeV and $\Lambda = 631$ MeV, respectively. The blue curve represents the vdW phase, whereas the grey dashed line depicts the limit of validity for $\Lambda = 631$ MeV. 
For either choice of the NJL cutoff, we see that the quark description leads to higher pressures at large densities. However, for $\L=631$, the transition from the vdW description occurs at around $\m_B=1500$ MeV, which is quite close to the $3\L$ limit of validity. But, it is noteworthy that the transition chemical potential is significantly altered by temperature and already at $T=30$ MeV, even the NJL model with $\L=631$ MeV cutoff predicts a change in description at $\m_B\sim 1300$ MeV. For a value of $\Lambda=925$ MeV, the transition at $T=0$ MeV occurs at $\mu_B=1800$ MeV, which is quite smaller than $3\Lambda\simeq2800$ MeV. 
In a complete holographic description, we may expect that the bulk solitons (those that model the baryons) also overlap strongly in the deep IR contributing to this change in description.

At even higher densities, a deconfined phase of quarks may be expected and is usually modeled, holographically, by a charged black hole. Thus, we can search for a transition to a higher density phase modeled by a charged black hole in AdS.

\begin{figure}[h]
    \centering
    \includegraphics[width=0.5\linewidth]{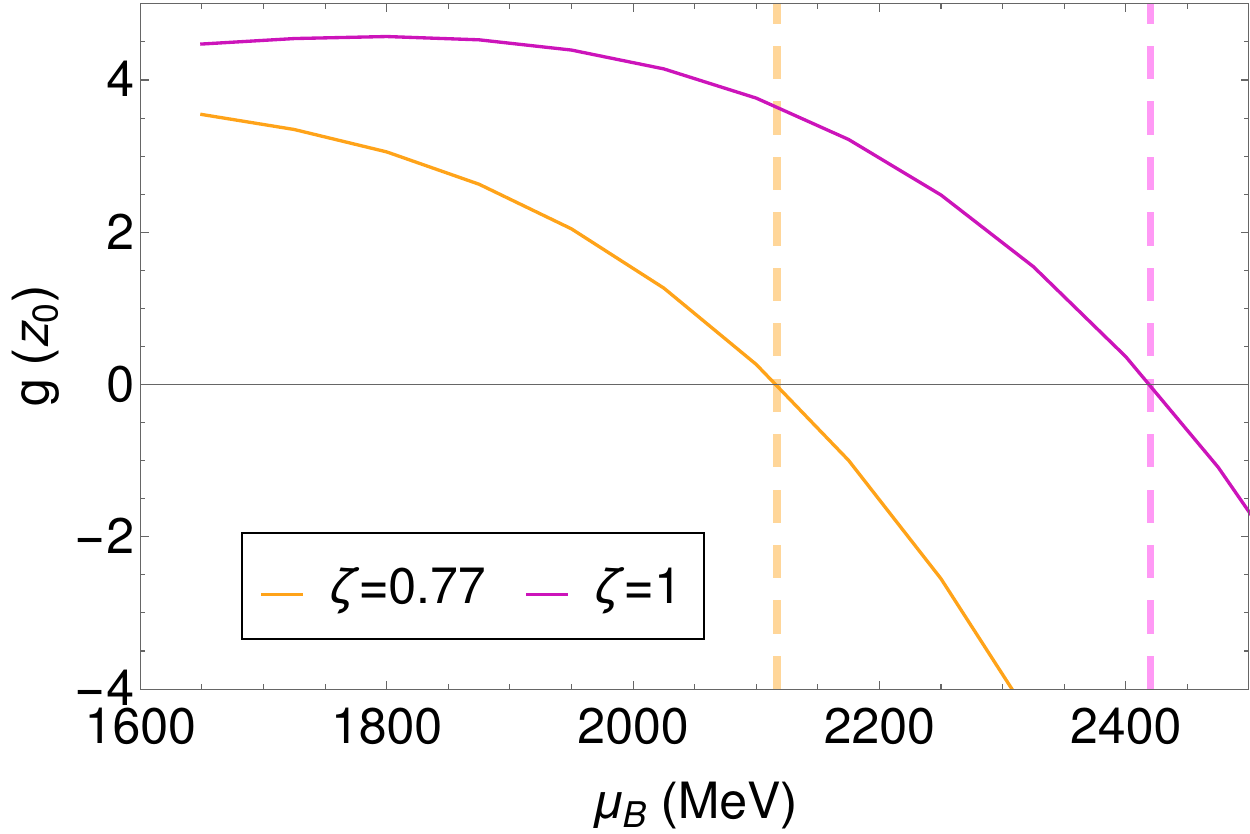}
    \caption{$g(z_0)$ as function of $\m_B$ for  $\Lambda = 925$ MeV}
    \label{fig:NJL_g0}
\end{figure}
However, as illustrated in figure \ref{fig:NJL_g0}, the metric function $g(z)$ develops a horizon as we increase the baryon chemical potential $\m_B$ within the NJL phase for $\L=925$ MeV. That is to say, the blackening factor, $g(z)$, vanishes at the IR cutoff $z_0.$ However, it should be noted that the $g(z)$ never develops horizon for $\L=631$ MeV.The value of $\m_B$ required for the horizon to form is significantly dependent on the parameter $\z$, with larger values of $\z$ requiring a greater $\m_B$. The vertical orange and magenta dashed lines represent the value of $\m_B$ at which the horizon appears for $\z=0.77$ and $1$, respectively. Additionally, the temperature of the black hole thus formed does not agree with that of the NJL model. Thus, this observation motivates two conclusions: at higher densities, the NJL model is not a good description. And, our assumption that $g_0$ is independent of the chemical potential likely needs to be modified. 

\begin{figure}[h]
    \centering
    \subfloat[Pressure for $\z=0.77$]{\includegraphics[width=0.45\linewidth]{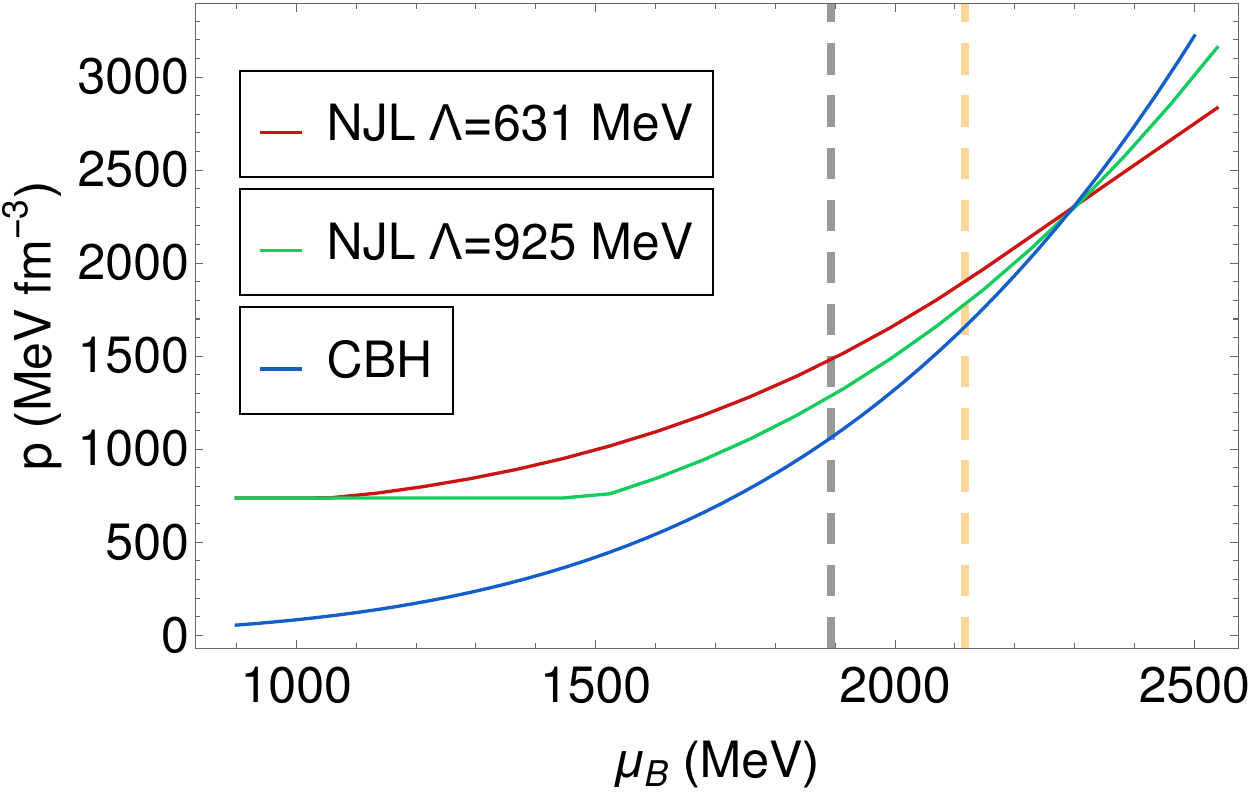}}\hfill
    \subfloat[Pressure for $\z=1$]{\includegraphics[width=0.45\linewidth]{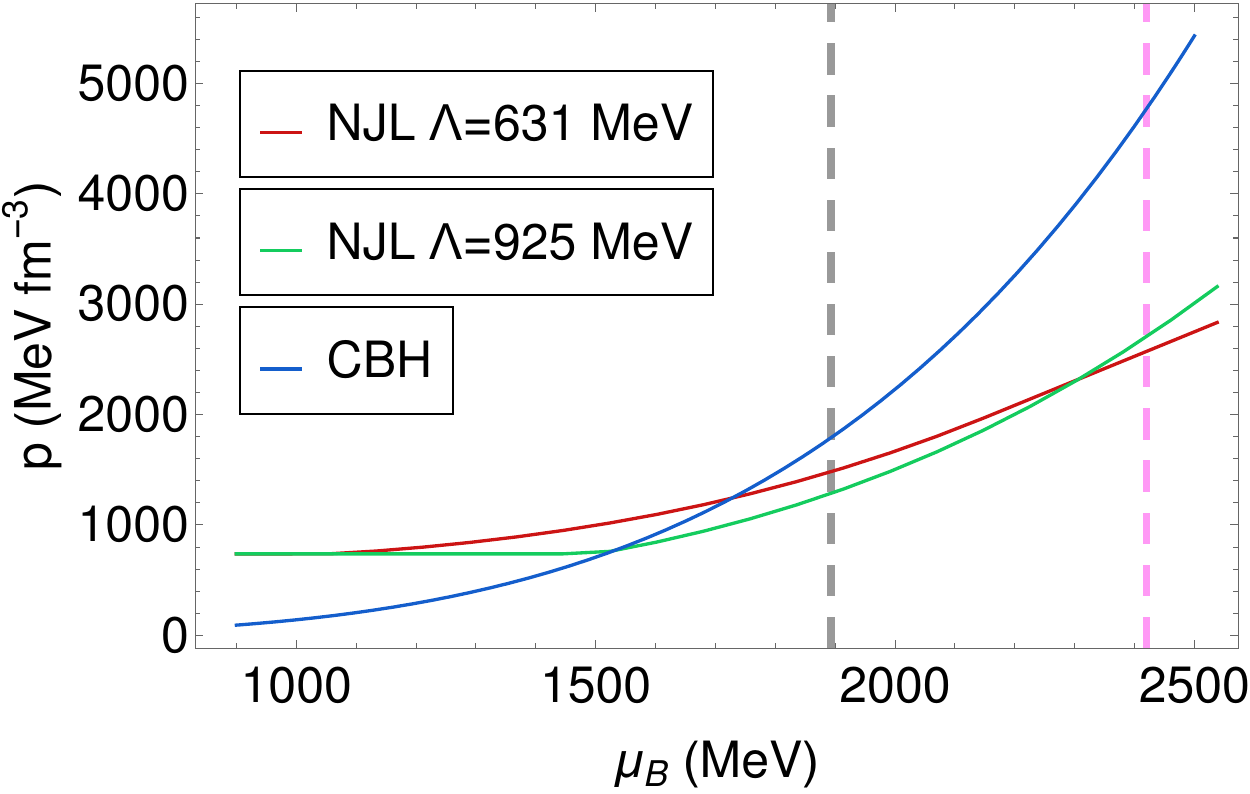}}
    \caption{Phase transition between charged black hole and NJL for $T=1$ MeV. }
    \label{fig:NJLBH1}
\end{figure}

The phase transition between charged black hole and NJL for $\z=0.77$ and $1$ is shown in figure \ref{fig:NJLBH1}. The green and red curves represent NJL with $\Lambda = 925$ MeV and $\Lambda = 631$ MeV, respectively. The blue curve represents the CBH phase, whereas the grey dashed line denotes the region of validity for $\Lambda = 631$ MeV. The vertical dashed lines in orange and magenta have the same meaning as discussed above.

\begin{figure}[h]
    \centering
    \subfloat[Pressure for $\z=0.77$]{\includegraphics[width=0.45\linewidth]{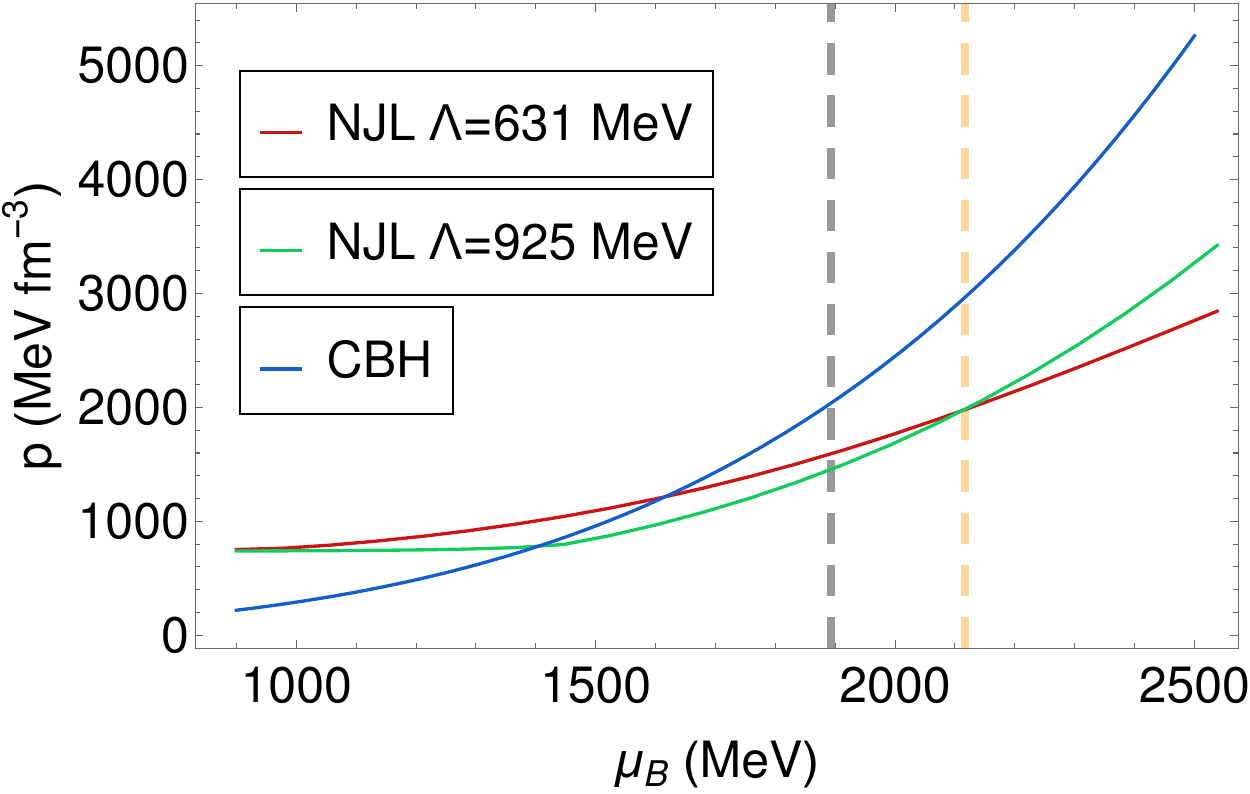}}\hfill
    \subfloat[Pressure for $\z=1$]{\includegraphics[width=0.45\linewidth]{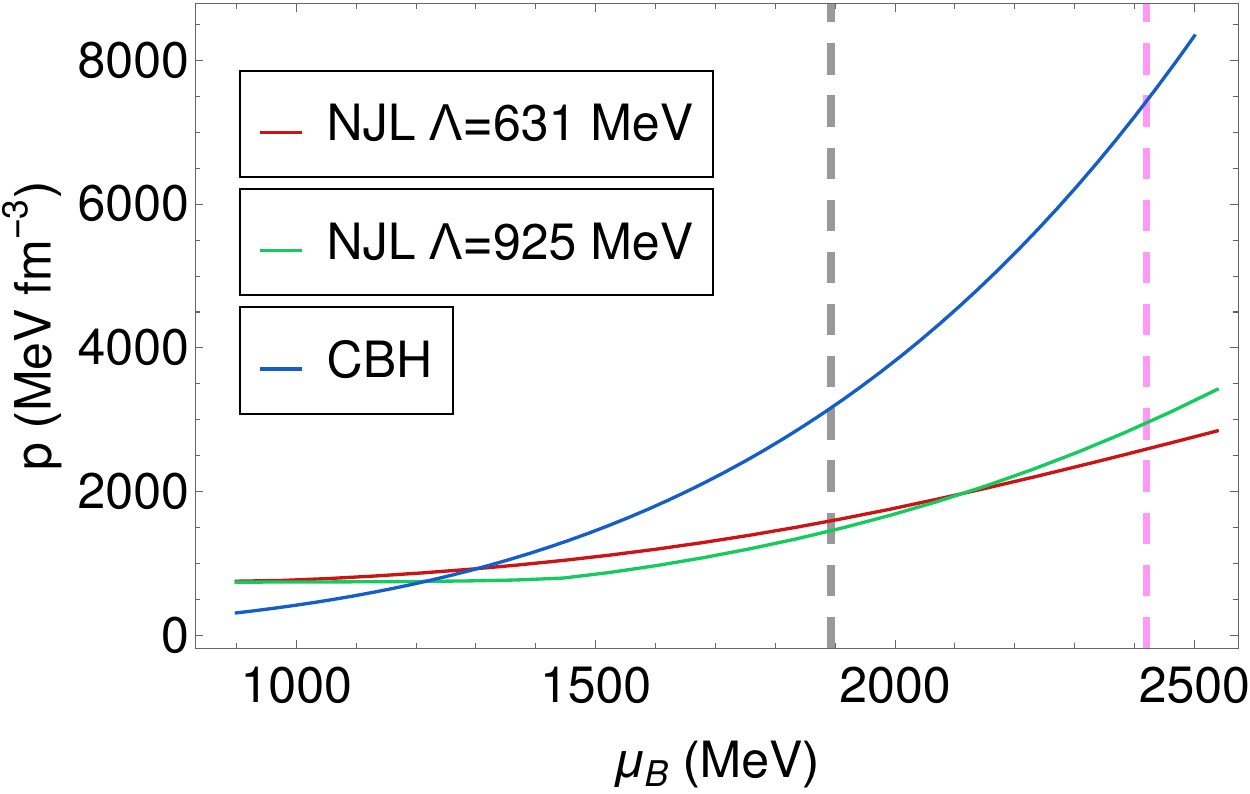}}
    \caption{Phase transition between charged black hole and NJL for $T=60$ MeV.}
    \label{fig:NJLBH60}
\end{figure}

We see that the pressure of the black hole geometry is larger at high chemical potential. In this case, while the location of the first-order phase transition is strongly dependent on $\z$, this comparison suggests that the high density phase is always better modeled by the black hole. 
For $\z=0.77$, the black hole transition occurs well beyond the region of validity of the NJL model with $\L=631$ MeV. With the NJL cutoff set to $925$ MeV, the black hole transition occurs after the horizon appears in the bulk metric. 
For $\z=1$, the transition between the CBH and NJL phase with both choices of $\L$ lies within the validity range.  
Additionally, as shown in figure \ref{fig:NJLBH60}, at slightly higher temperature values, the phase transition shifts to within the region of validity for both values of $\L$. 
Therefore, the occurrence of the confinement/deconfinement transition may be treated as a robust prediction from these approaches.  In both cases, the deconfinement transition seems to be correlated with the vanishing of the chiral condensate (shown in figure 9 of \cite{Singh:2024amm}).

\section{Condensate geometries}

Baryonic matter at high density can, in general, be expected to exhibit various levels of clustering due to various attractive interactions. At intermediate densities, this can lead to {\it Baryon} pairs and could entail the formation of condensates involving dibaryons. These superfluid/superconducting phases have been argued to occur in the same range of densities as above ($20<\mu<80$ MeV \cite{Sedrakian:2018ydt} above nuclear mass $M\sim 930$ MeV) albeit for a much smaller range of temperatures ($0<T<10$ MeV). At somewhat higher densities, {\it quark} pairing can also occur in various forms, including the diquarks \cite{Abuki:2010jq} and Color-Flavor Locked phases \cite{Alford:1998mk}. Since we are working with $N_f=2$, the 2SC phase is more relevant. For a detailed study of these phases, see \cite{Rajagopal:2000wf}. It is thus conceivable that spontaneous breaking of baryon number symmetry occurs as evinced by the corresponding condensates.

In the AdS bulk, such symmetry breaking leading to condensates has been extensively studied by introducing a charged scalar, which takes over the phase diagram at high densities. However, nearly all of these studies have been conducted in the black hole background, which, as summarized in the preceding section, can be expected to appear at high densities. 

An exception is the work of Basu et al., \cite{Basu:2016mol}, where the authors study nontrivial scalar field backreacted within {\em global} AdS (with spherical boundary topology), called Boson Stars. 
In our case, we require the boundary to be flat and since we allow for the possibility of condensates in the confined phase as well, we must study bulk geometries without a black hole but involving a nontrivial profile for a complex scalar field without a non-normalizable mode. We turn to a systematic exploration of these phenomena.

\subsection{Charged black hole}
At finite temperature and chemical potential in {\em four} bulk dimensions, the black hole solution with scalar hair \cite{Gubser:2008px} was discussed by \cite{Hartnoll:2008kx} as corresponding to a superfluid phase of the boundary theory. We will revisit these in {\em five} bulk dimensions ( \cite{DAlmeida:2018ldi}) but in the presence of an IR cutoff, which can be expected to contribute a few new ingredients. 

\begin{figure}[t]
    \centering
     \subfloat[Reduced Condensate]{\includegraphics[width=0.45\linewidth]{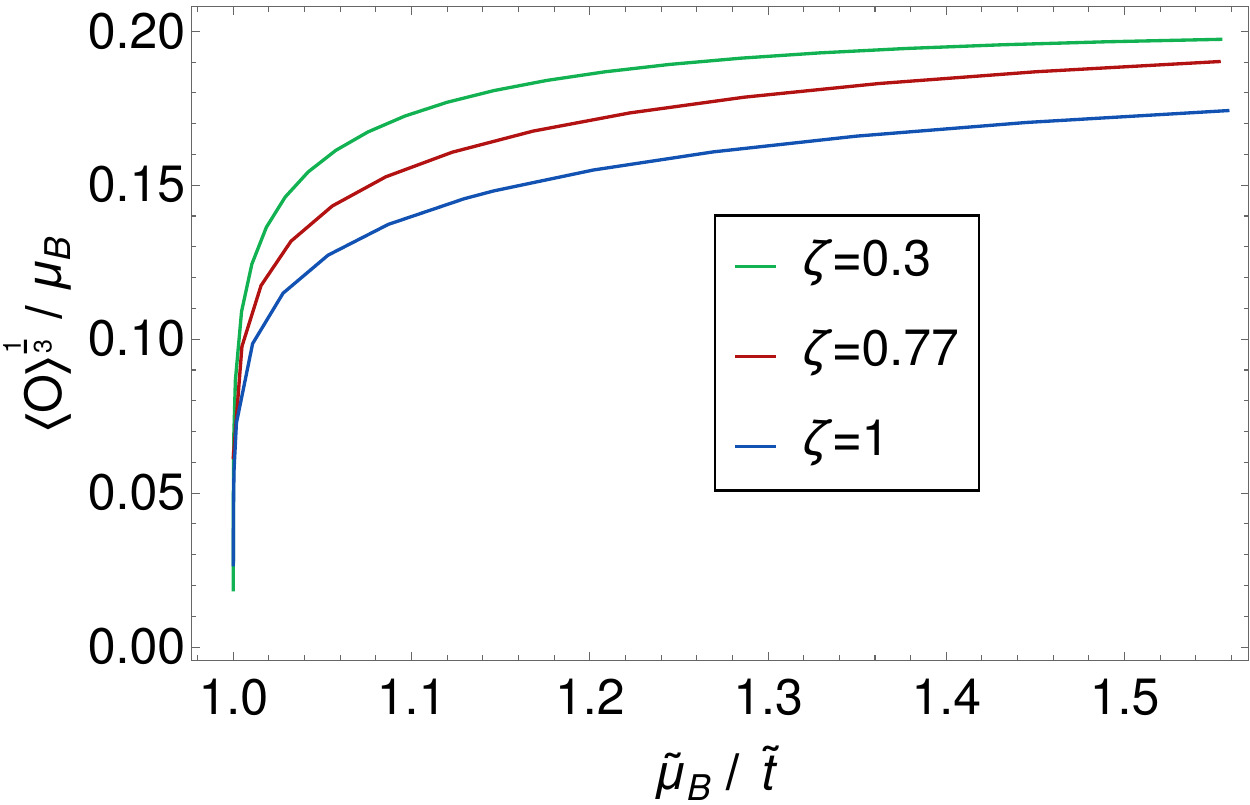}}\hfill
     \subfloat[Relative condensate number density]{\includegraphics[width=0.45\linewidth]{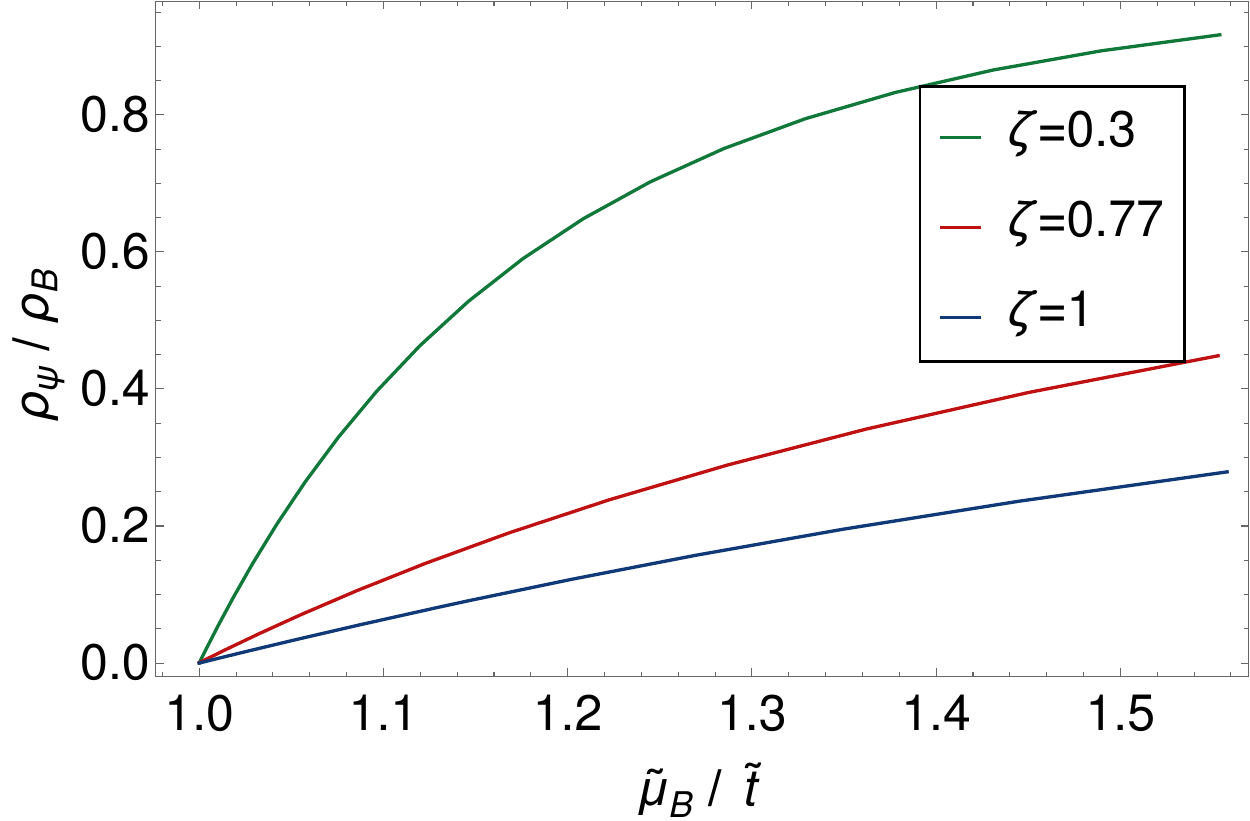}}
   \caption{Thermodynamics quantities for different $\z$. The reduced temperature and chemical potential are defined as $\Tilde{t}=\frac{T}{T_c}$ and $\Tilde{\m}_B=\frac{\m_B}{\m_B^c}$}
   \label{fig:CondCBH}   
\end{figure}

If the bulk geometry contains a black hole, the scalar field satisfies a regularity condition $g'\y' -m^2 L^2\y=0$ on the horizon $z_H$. The latter can also be viewed as a finite action condition. At high chemical potentials, we want the condensate operator in the boundary to be made up of a pair of quarks. Thus, following discussion in section \ref{hardwallmodel}, we set $\Delta=3$ and $q=2$.  We numerically solve the equations for different values of $\z$ and plot the condensate and relative condensate number density as a function of ratio $\Tilde{\m}_B /\Tilde{t}$ in figure \ref{fig:CondCBH}.  Here, the  {\em condensate} number density can be computed as the charge stored in the scalar field 
\be\label{rhoCond}
\rho_\psi=i q g_5^2\int_\e ^{z_0} \sqrt{g} g^{00} q(\psi^* D_0\psi-\psi D_0\psi^*)dz = 2g_5^2 q^2 L^2\int_\e ^{z_0} \frac{\sqrt{h} \phi}{z^3 g}\psi^2 dz 
\ee

The reduced temperature and the chemical potential are defined as $\Tilde{t}=\frac{T}{T_c}$ and $\Tilde{\m}_B=\frac{\m_B}{\m_B^c}$, respectively. As the chemical potential increases, the fraction of baryon density in the condensate also rises, as shown in the second panel. This occurs because, at higher densities, even the normal component experiences increased pressure. To counterbalance this, the condensate component must also increase.

It is important to mention that the gauge coupling $\z$ does not qualitatively change the nature of the solution or the behavior of the condensate. On the other hand, the amount of matter inside the condensate increases with smaller values of $\z$. In other words, the fraction saturates to $1$ faster for a smaller value of $\z$.  The $\z=0.3$ solutions saturates for $\Tilde{\m}_B / \Tilde{t}\sim 2.$

\begin{figure}[h]
    \centering
    \subfloat[Reduced pressure]{\includegraphics[width=0.3\linewidth]{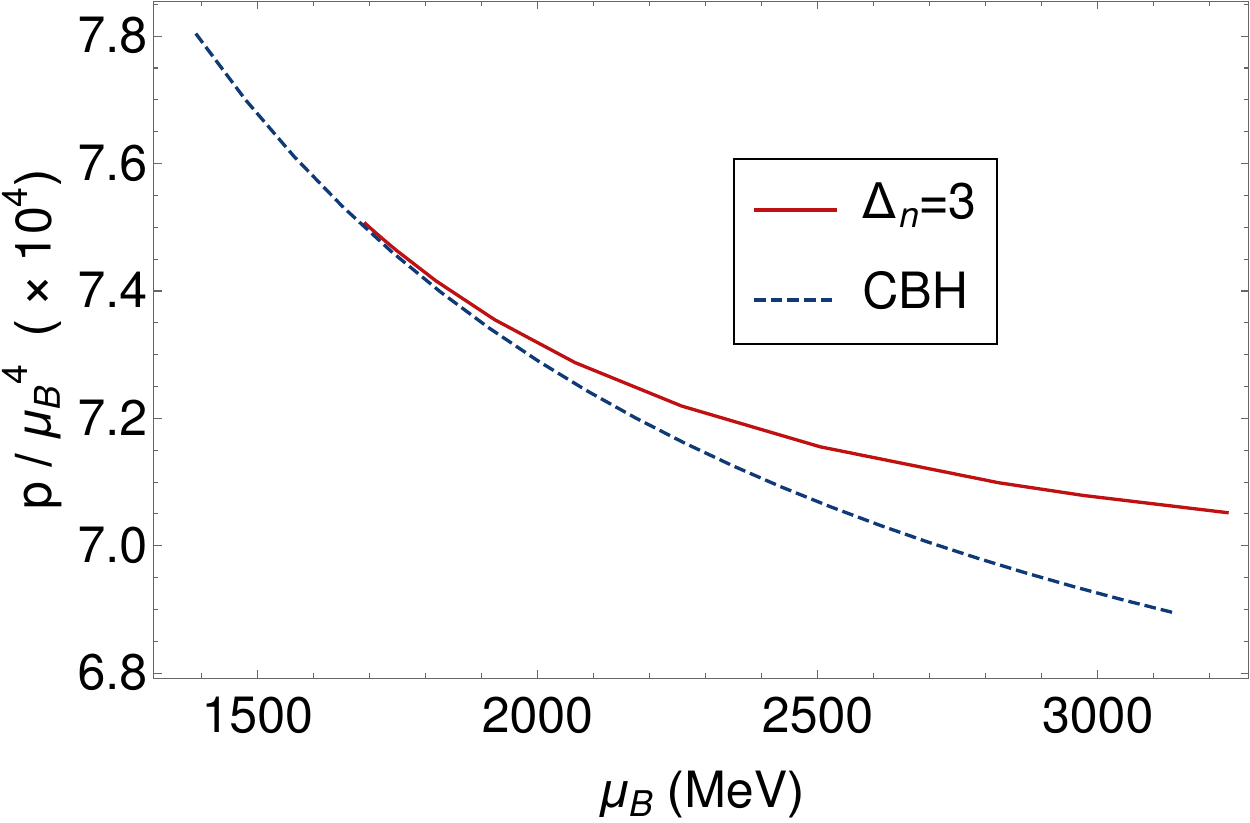}}\hfill
    \subfloat[Reduced entropy density]{\includegraphics[width=0.3\linewidth]{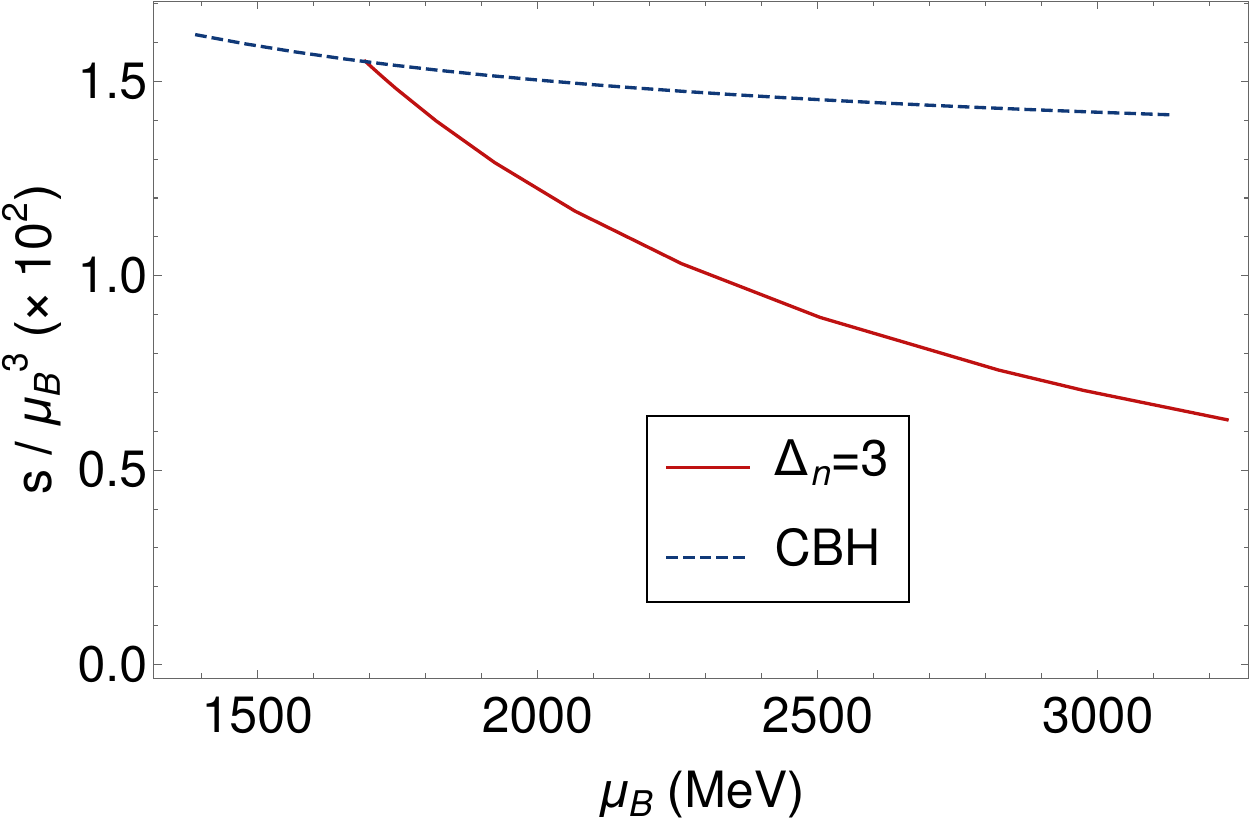}}\hfill
    \subfloat[Reduced baryon number density]{\includegraphics[width=0.3\linewidth]{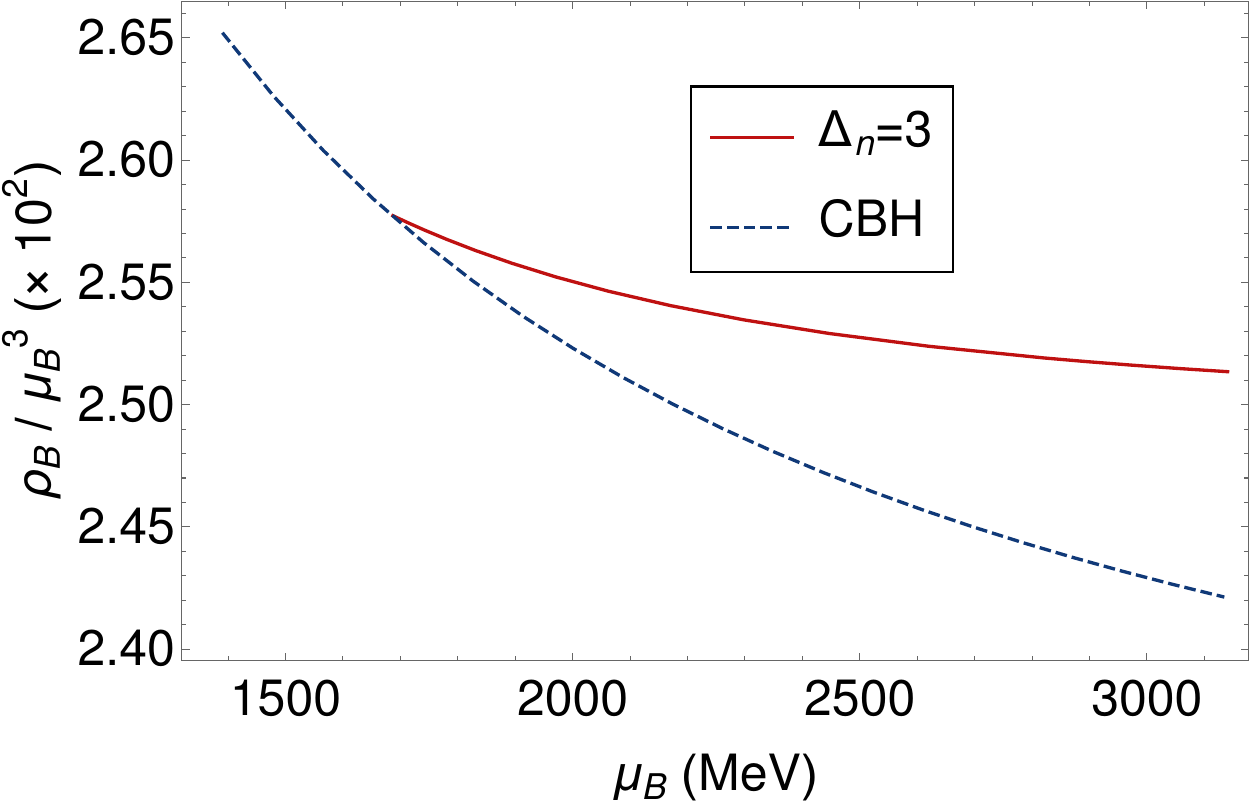}}
    \caption{Reduced thermodynamic quantities as a function of baryon chemical potential at $T=15$MeV. }
    \label{fig:CBHCondTD}
\end{figure}
As in the case of holographic superconductors in four dimensions, even in the five-dimensional model, condensed phases are preferred at sufficiently high densities, as seen in figure \ref{fig:CBHCondTD} - that is to say, condensation leads to increased pressure.

The range for $\mu_B$ shown in the figure was chosen in anticipation that these phases will be relevant at high density. Entropy is decreased due to condensation and the total density is correspondingly increased and stored in the condensate. 

\subsection{Charged AdS}
We now turn to the CAdS backgrounds that describe the confined phase. We will undertake a detailed study of the solutions with a non-trivial scalar field using the boundary condition $\f(z_0)=0$. In particular, we focus on the qualitative effects of varying the parameters $g_0,\z,q$ in our model. This will inform our expectations when we study the system using the phenomenological boundary conditions of the previous section.

\begin{figure}[h]
\centering
\subfloat[Scalar field]{
\includegraphics[width=0.45\linewidth]{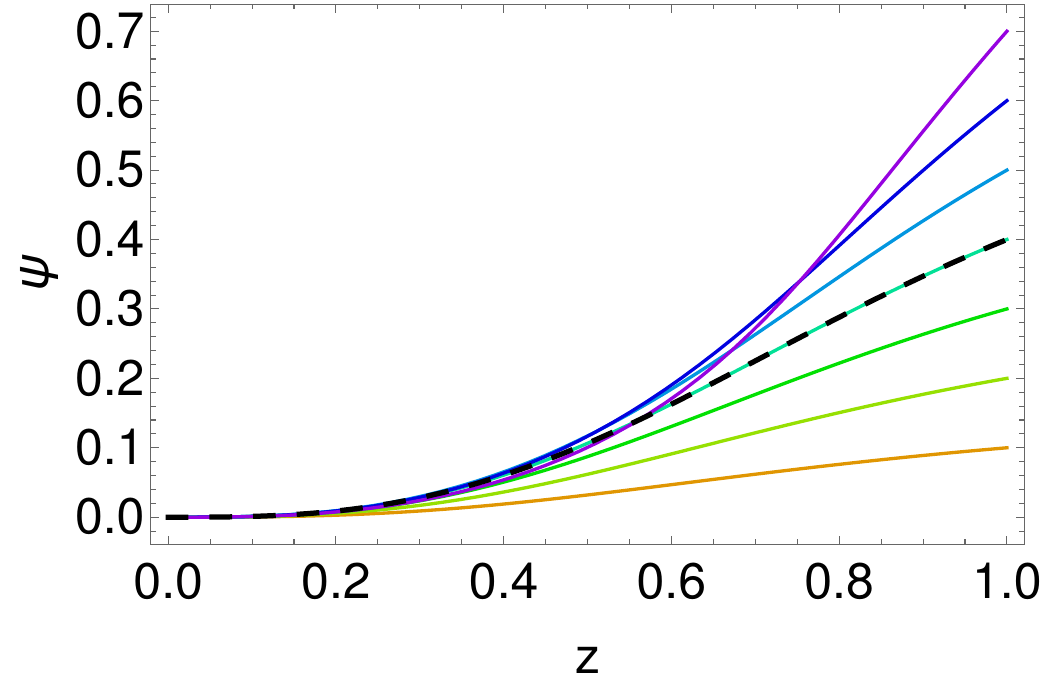}}\hfill
\subfloat[Electric field]{\includegraphics[width=0.47\linewidth]{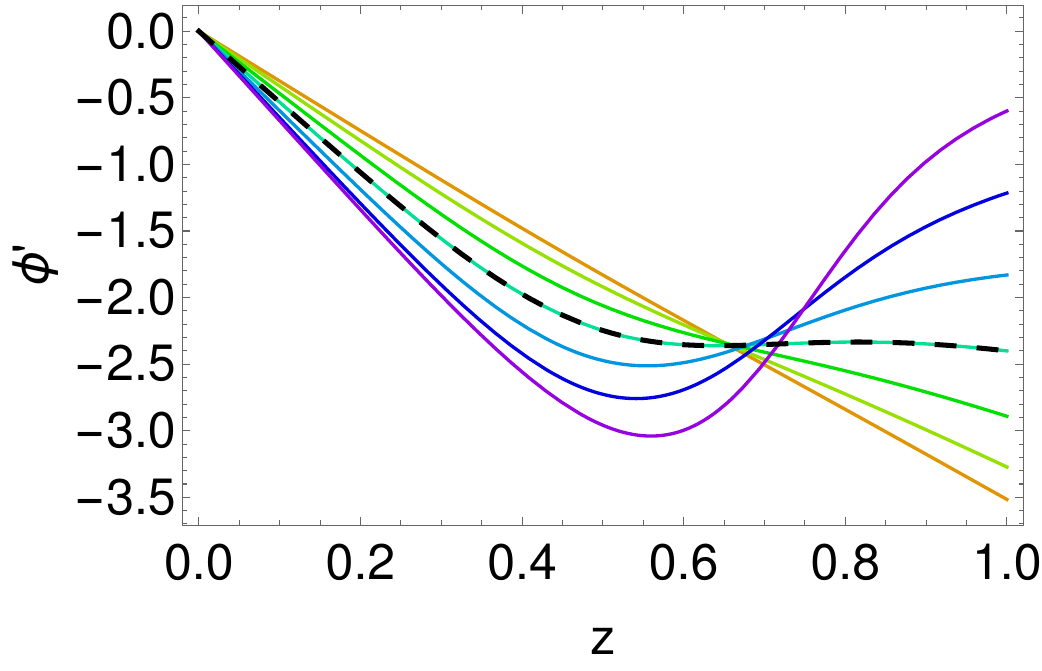}}\\
\subfloat[$h$]{\includegraphics[width=0.45\linewidth]{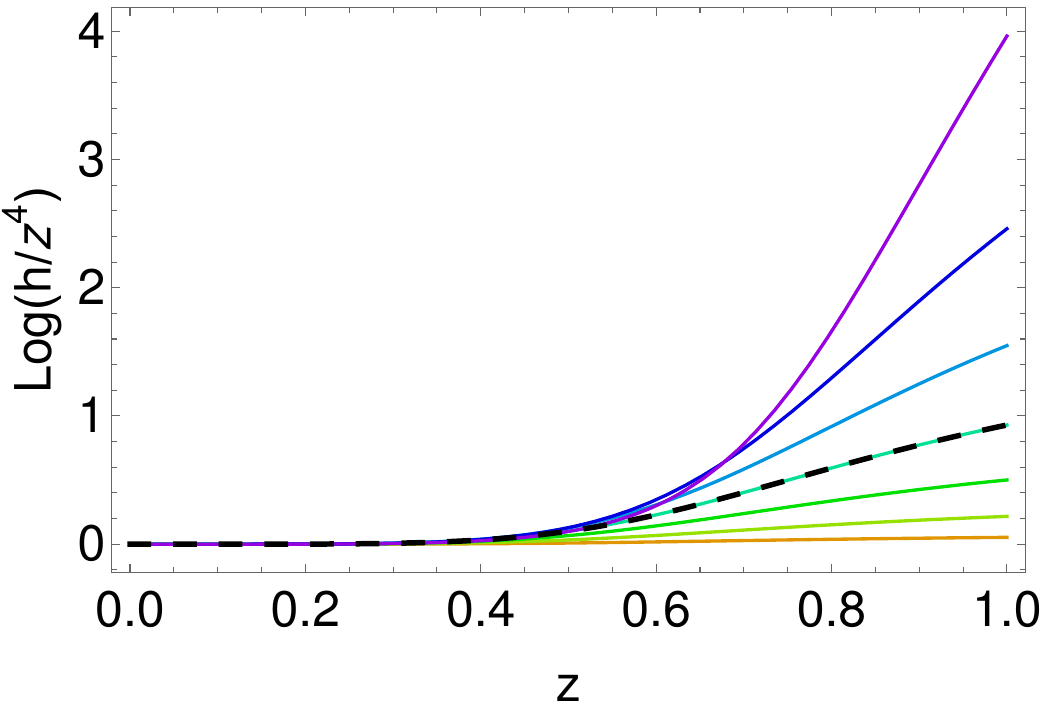}\label{fig:SOL_phi0_H}}\hfill
\subfloat[$g_{tt}$]{\includegraphics[width=0.45\linewidth]{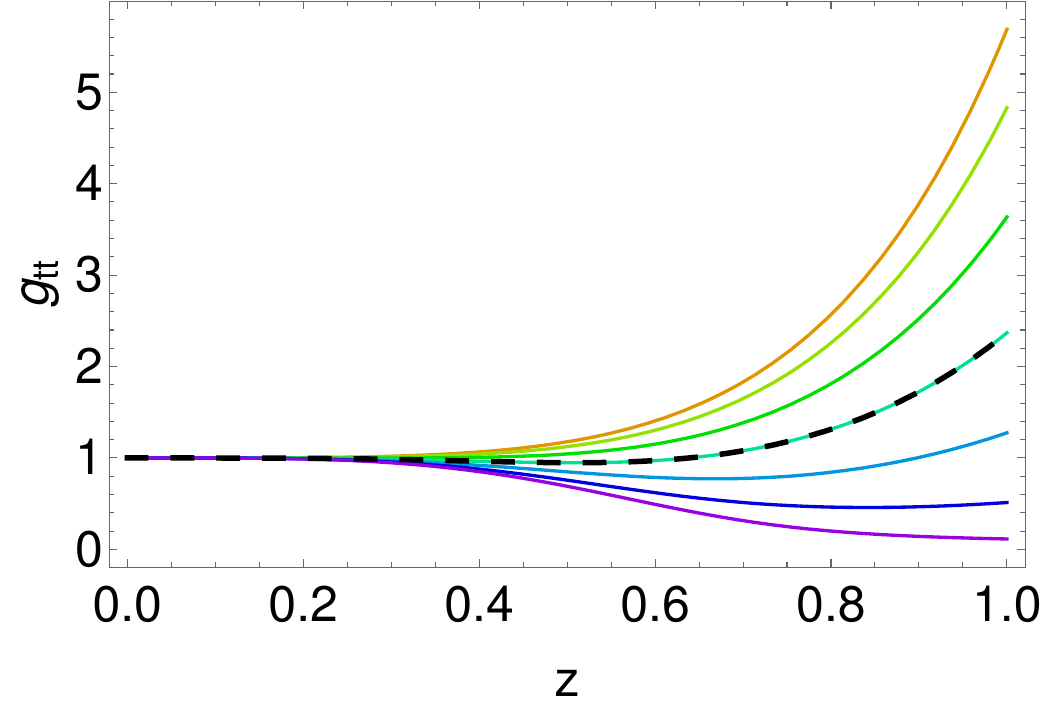}}
\caption{Numerical solution for the system with parameters $q=2$, $g_0=6$, $\Delta_n=3$, $\m_B=1566$ MeV, and $\phi_0=0$ boundary condition.}
\label{fig:SOL_phi0}
\end{figure}

Given an initial value $\psi(z_0)$, we can numerically solve the differential equation for the scalar field to obtain a solution without a boundary source term. Figure \ref{fig:SOL_phi0} shows numerical solutions for $\z=0.77, q=2, \D_n=3, g_0=6$, and $\m_B=1566$ MeV. Different colors represent various values of $\y_0$. The black dashed curve denotes the solution that maximizes the pressure (maximum pressure solutions will be discussed below).

For these cases, the metric component $g_{tt}=z^2\frac{g}{h}$ shown in figure \ref{fig:SOL_phi0} reveals that these solutions do not exhibit any pathology that would invalidate them. It may be noted that the maximum variation in $\psi$ occurs around the cutoff region $z_0=1$.

Interestingly, there appears to be an upper limit to $\y_0$ beyond which there are no solutions to the scalar field (at the least, these are numerically hard to find). This could be attributed to the appearance of a singularity $h\to \infty $ at the IR cutoff as seen in figure \ref{fig:SOL_phi0_H}. This singularity is a naked singularity at which $\det(g)\to 0$ because $g_{tt}$ vanishes (but $g_{zz}$ does not).

For the solution to describe a {\em ground} state, the scalar field profile should not have a node (this is because of general expectations from Sturm-Liouville theory). 
However, we will solve the equation for a range of $z$ beyond the cutoff $z>z_0$ and require that the solution must not have a node even in the extended geometry. This is under the expectation that once a node appears, it is likely to persist in an IR complete geometry because it is a topological feature and hence will not represent a valid {\em ground state} description.

\begin{figure}[h]
    \centering
    \includegraphics[width=0.4\linewidth]{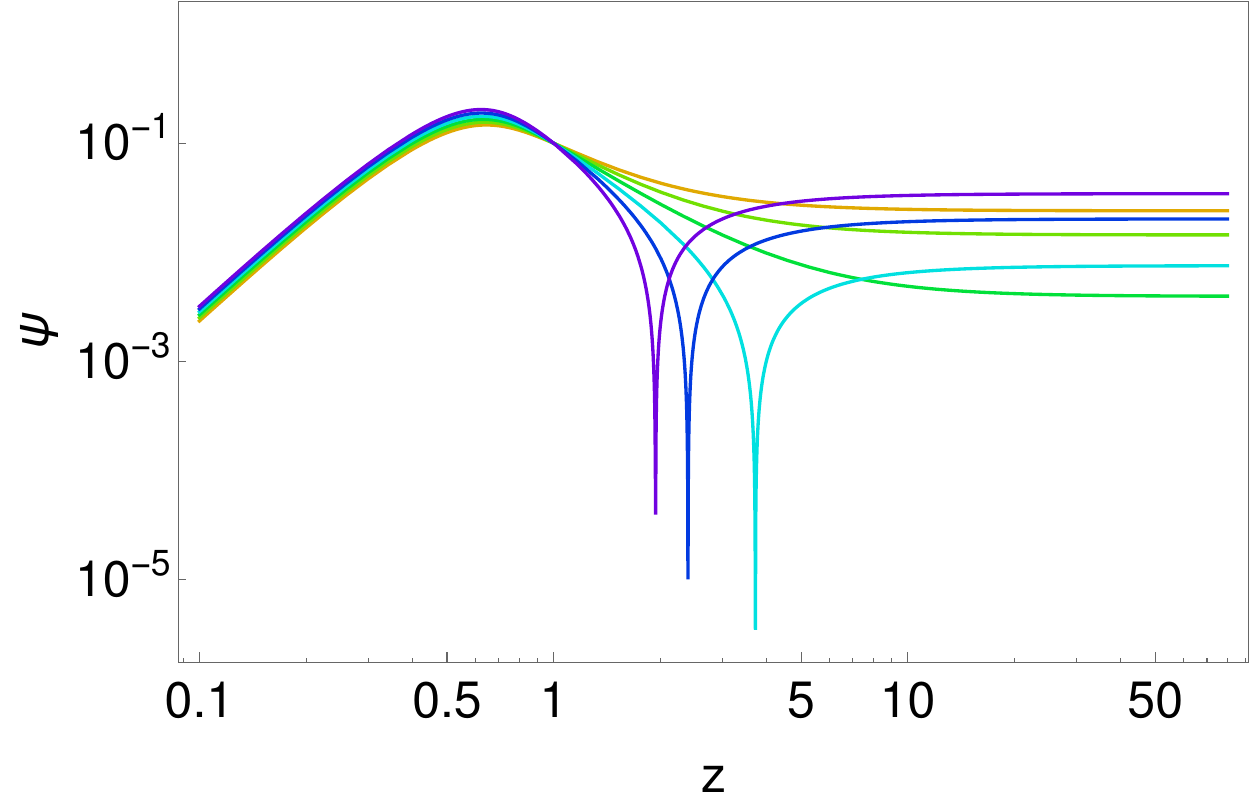} 
    \caption{The function $\y(z)$ is extended beyond the cutoff $z_0$. The log-log plot displays the function for a fixed $\y_0$ and different values of $\m$, with the smallest value of $\m_B=2436$ MeV represented by the orange color.}
    \label{fig:psi}
\end{figure}

In figure \ref{fig:psi}, we show that scalar field solutions with a node indeed exist at large enough chemical potential. 
The log-log plot of the scalar field shows that beyond the IR cutoff, for a fixed value of $\y_0$, as we increase $\m_B$, a node appears in the scalar field as shown by curves that sharply dip downwards. These solutions share the same parameter values as those in figure \ref{fig:SOL_phi0}, except for \(\m_B\), which corresponds to 2436 MeV for the orange curve and 2523 MeV for the purple curve. Clearly, for a given value of $\mu_B$, there are several values of $\y_0$ which give acceptable profiles of the condensate. 

We can fix the $\y_0$ in two ways. We can take it to be a constant (which will be fixed by phenomenological boundary conditions at the IR). This will lead to the offset of the condensate, signaled by the node in the scalar field profile. A more realistic physical possibility is to allow the IR value to depend on $\m_B.$ This requires additional physics input, such as the dependence of the pole mass of the meson on the density. 

To motivate our choice, let us consider the form \eqref{IFormS} of the on-shell free energy presented in the appendix. If we regard the IR region $z>z_0$ as being described by another set of gauge fields and scalar fields, then $z_0$ is like a UV cutoff for this region, and hence the on-shell action can be written in one of these forms. This shows that the pressure of the matter in the region $z>z_0$ can be expressed in terms of only the fields $g, \f, \f', h$ all evaluated at the IR surface $z_0$.  In particular, even if this region also contains a condensate, as we may expect, the on-shell action does not depend on the values of the scalar field at the IR cutoff. Consequently, the natural choice for $\y_0$ is the one that extremizes the on-shell action. Since we are working in the grand canonical ensemble, this is equivalent to maximizing the pressure.
\begin{figure}[h]
            \centering
            \subfloat[Reduced pressure]{\includegraphics[width=0.3\linewidth]{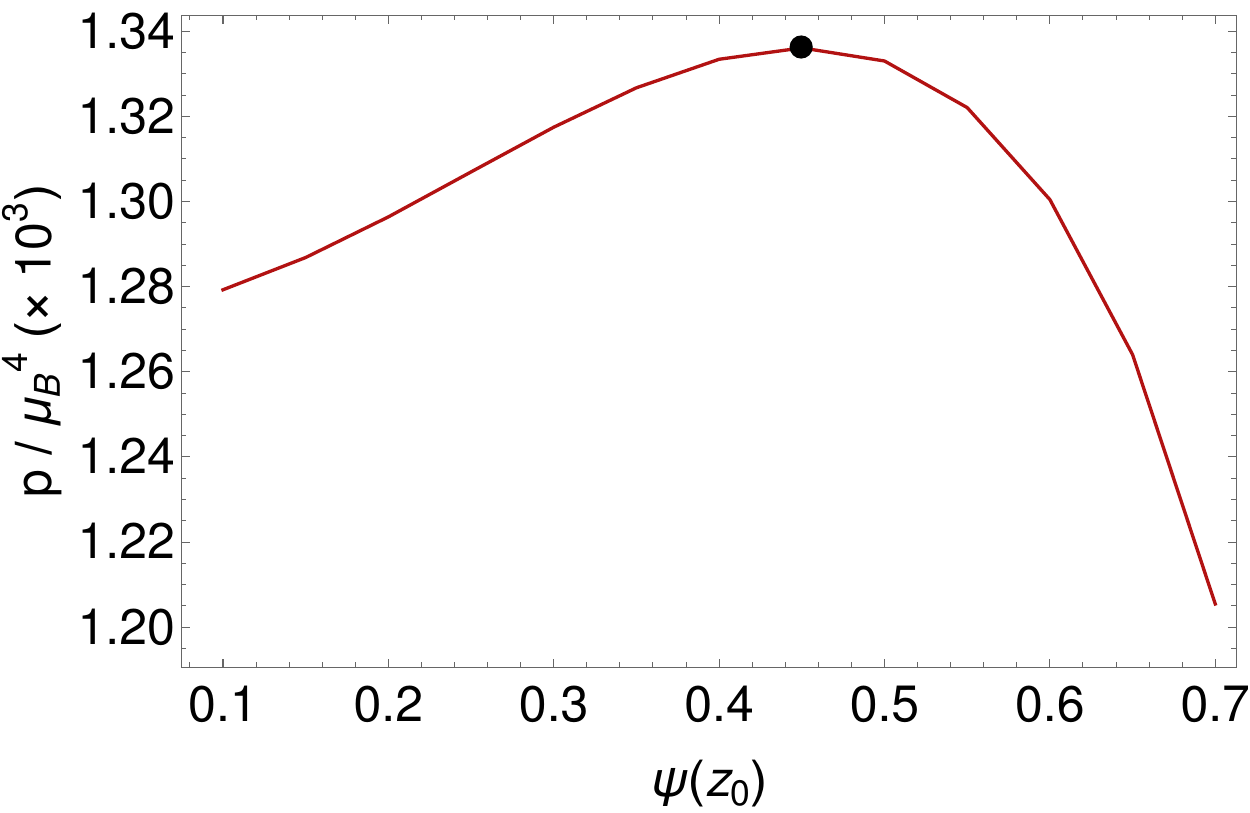}}\hfill
            \subfloat[Reduced baryon number density]{\includegraphics[width=0.3\linewidth]{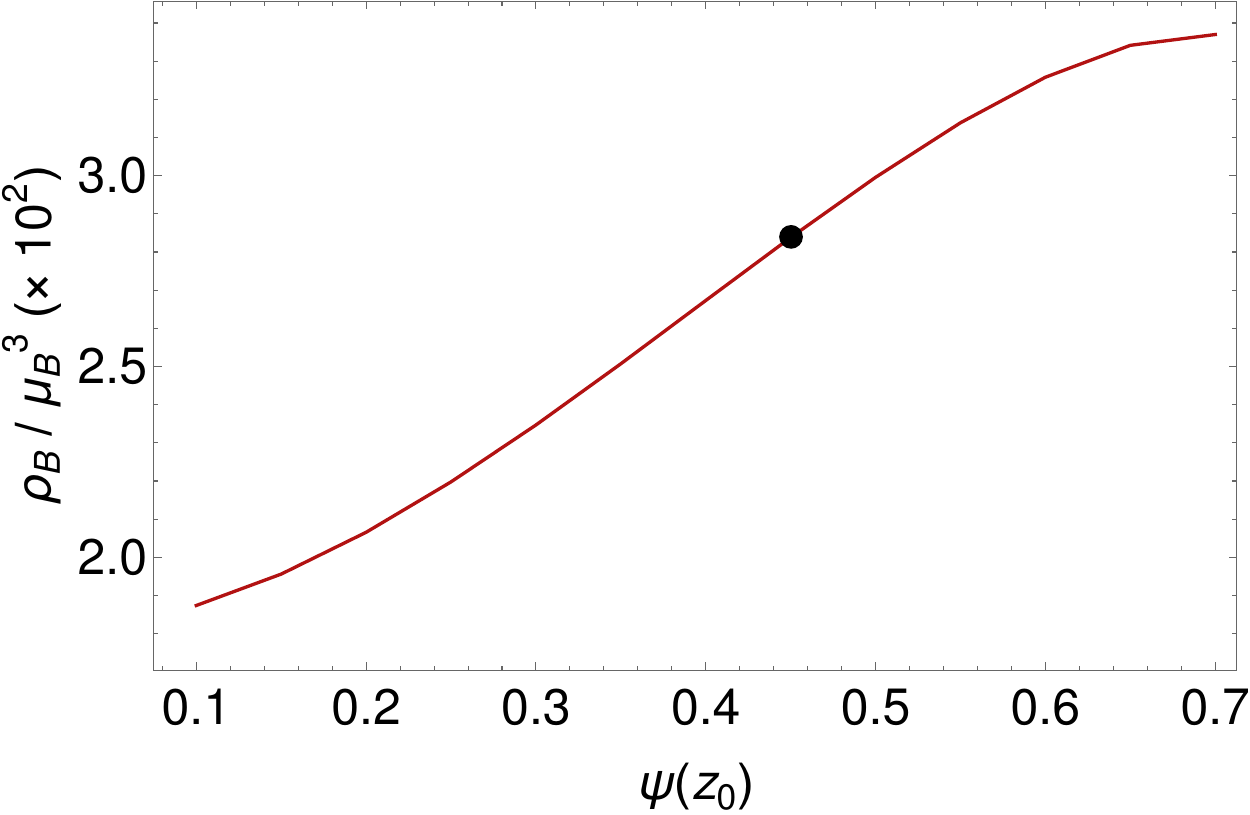}}\hfill
            \subfloat[ Reduced Condensate]{\includegraphics[width=0.3\linewidth]{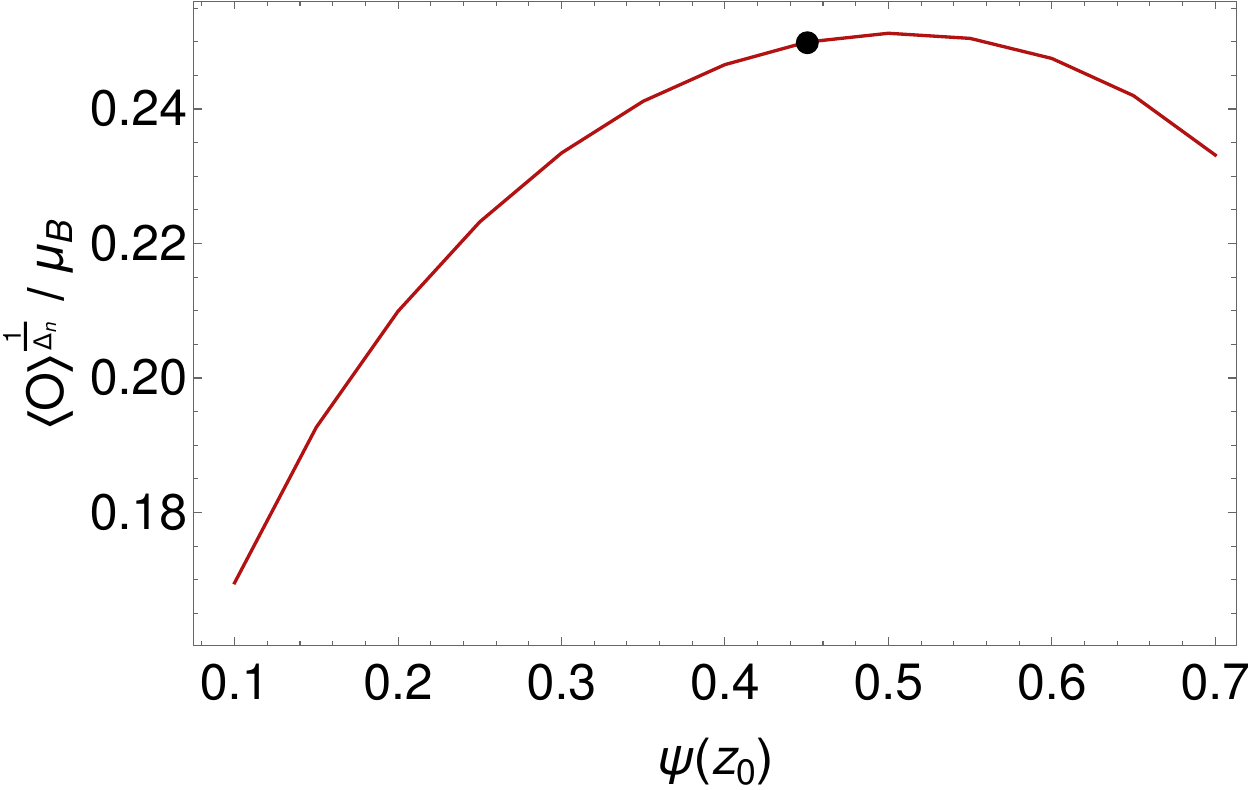}}
            \caption{Thermodynamic variables for $q=2$, $g_0=6$, $\Delta_n=3$, $\m_B=1566$ MeV, and $\phi_0=0$ boundary condition.}
            \label{pvspsi0}
\end{figure}
The first panel of figure \ref{pvspsi0} shows that the pressure attains a maximum (represented by a black dot) at an intermediate value of the cutoff value $\psi_0.$ In figure \ref{fig:SOL_phi0}, the gravity solutions corresponding to maximum pressure (shown by dashed lines) are also seen to be intermediate and are hence sans any feature that will restrict their validity. 
Note that the condensate also has a maximum as a function of $\y_0$, which may not be the same as the pressure maximum. 
\begin{figure}[h]
    \centering
    \includegraphics[width=0.4\linewidth]{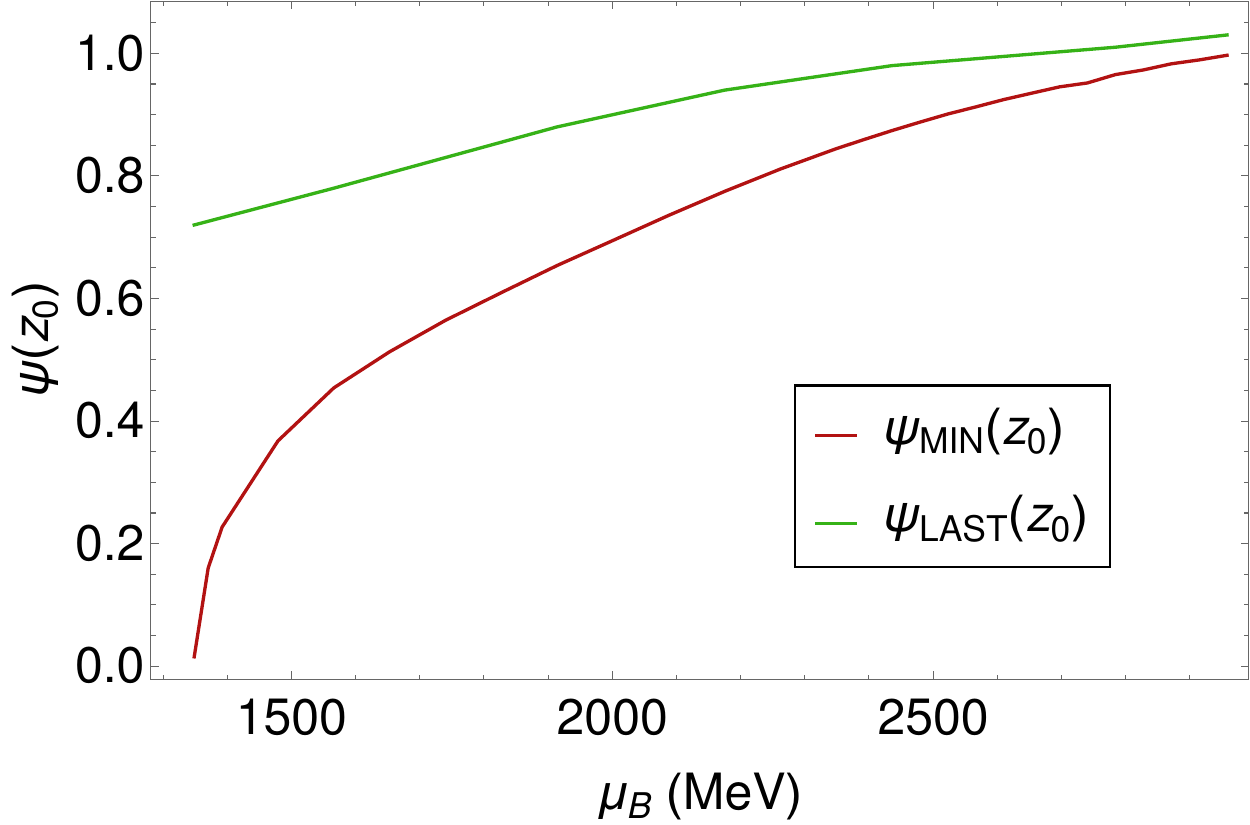}
    \caption{$\y(z_0)$ as function of $\m_B$ for $q=2$, $g_0=6$, $\Delta_n=3$, and $\phi_0=0$ boundary condition. 
    }
    \label{fig:YMIN_Phi0_vs_MU.pdf}
\end{figure}
As we increase the chemical potential, the $\psi_0$ corresponding to maximum pressure (or minimum free energy) shifts to larger values as illustrated by the red curve in \ref{fig:YMIN_Phi0_vs_MU.pdf}. The green curve represents the initial condition corresponding to the last possible solution which leads to $g_{tt}(z_0) \to 0$. Clearly, at larger densities, the maximum pressure tends to this singular geometry. 
However, as we will see below, the CAdS condensate solution will make a transition to condensate solutions in CBH geometry at $\m_B^*$. Thus, we can conclude that condensate solutions exist at all values of $\m_B>\mu_c.$ 

\begin{figure}[h]
            \centering
            \subfloat[Reduced pressure]{\includegraphics[width=0.45\linewidth]{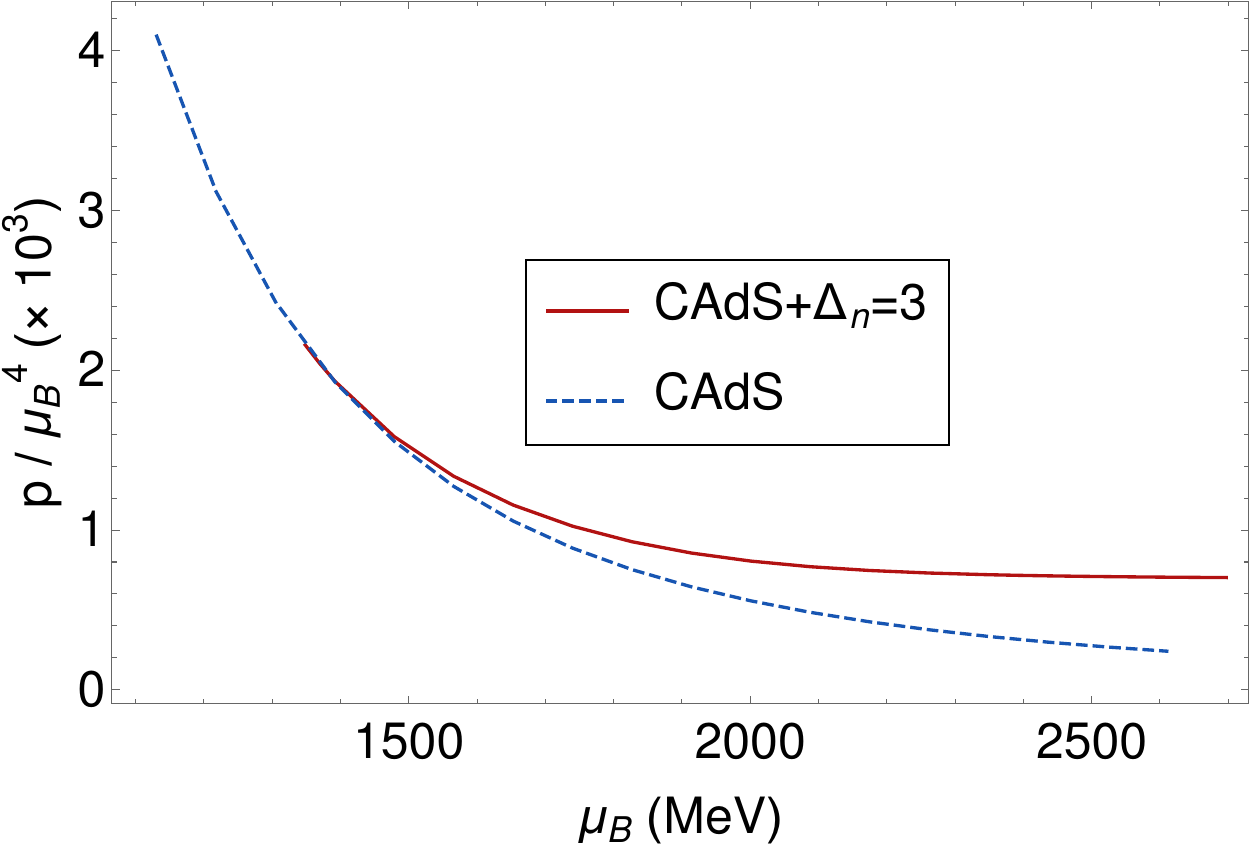}}\hfill
            \subfloat[Reduced baryon number density]{\includegraphics[width=0.45\linewidth]{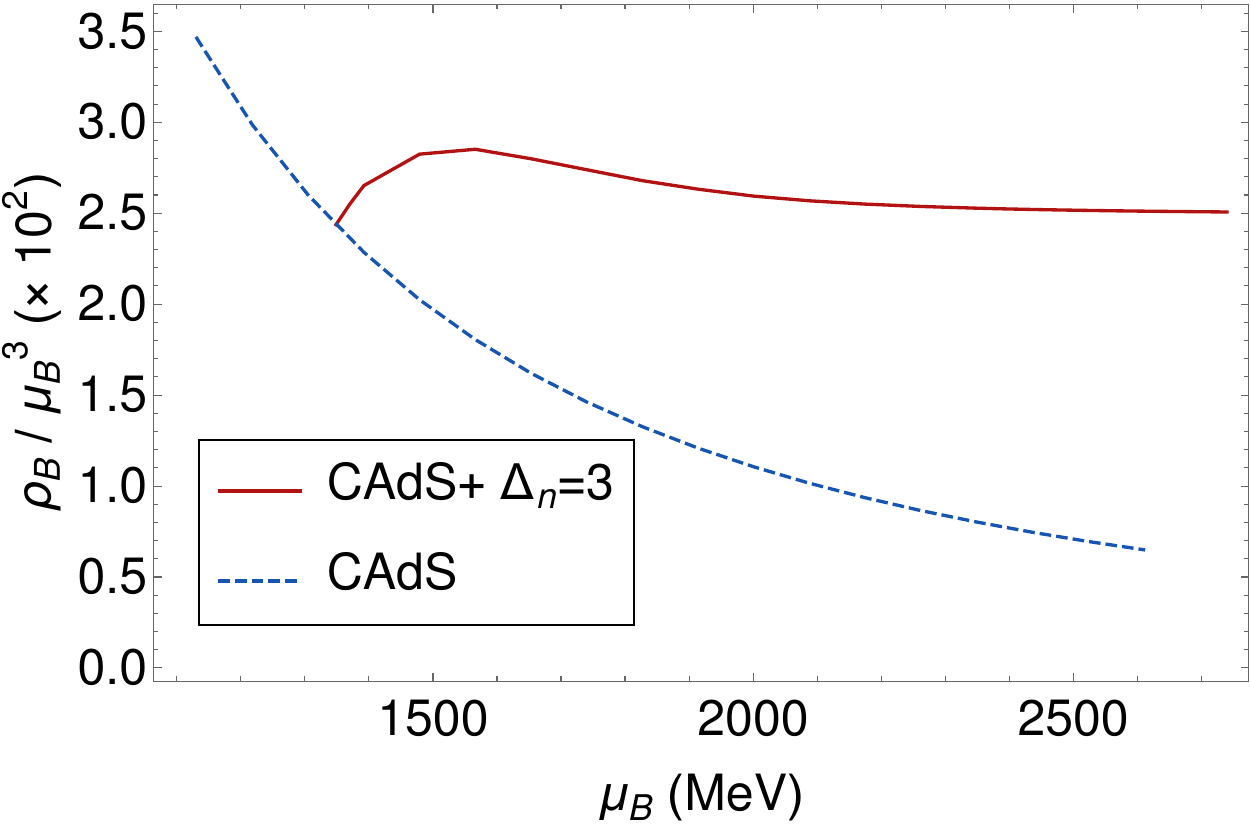}}\hfill
            \caption{Thermodynamic variables for $\f_0=0,g_0=6$ and $q=2.$}
            \label{fig:condp}
\end{figure}

We can now compare the pressure of the CAdS geometry with and without condensation, as shown in figure \ref{fig:condp}. The dashed blue line represents the pressure of CAdS without the scalar condensate. We observe that the presence of condensation leads to an increase in pressure, indicated by the solid red curve. As we increase the chemical potential, the condensate solutions continue to exhibit lower free energy. 
For the value $g_0=6$, and $\z=0.77$, the transition to a condensed phase occurs at around $\m_B=1400$ MeV. The second panel shows that baryon density increases sharply upon condensation. 

Increasing the value of $\z$ results in a rise in both pressure and baryon density. Furthermore, we note that the onset of condensation occurs earlier for larger values of $\z$. 
\begin{figure}[h]
    \centering
    \subfloat[Reduced pressure]{\includegraphics[width=0.3\linewidth]{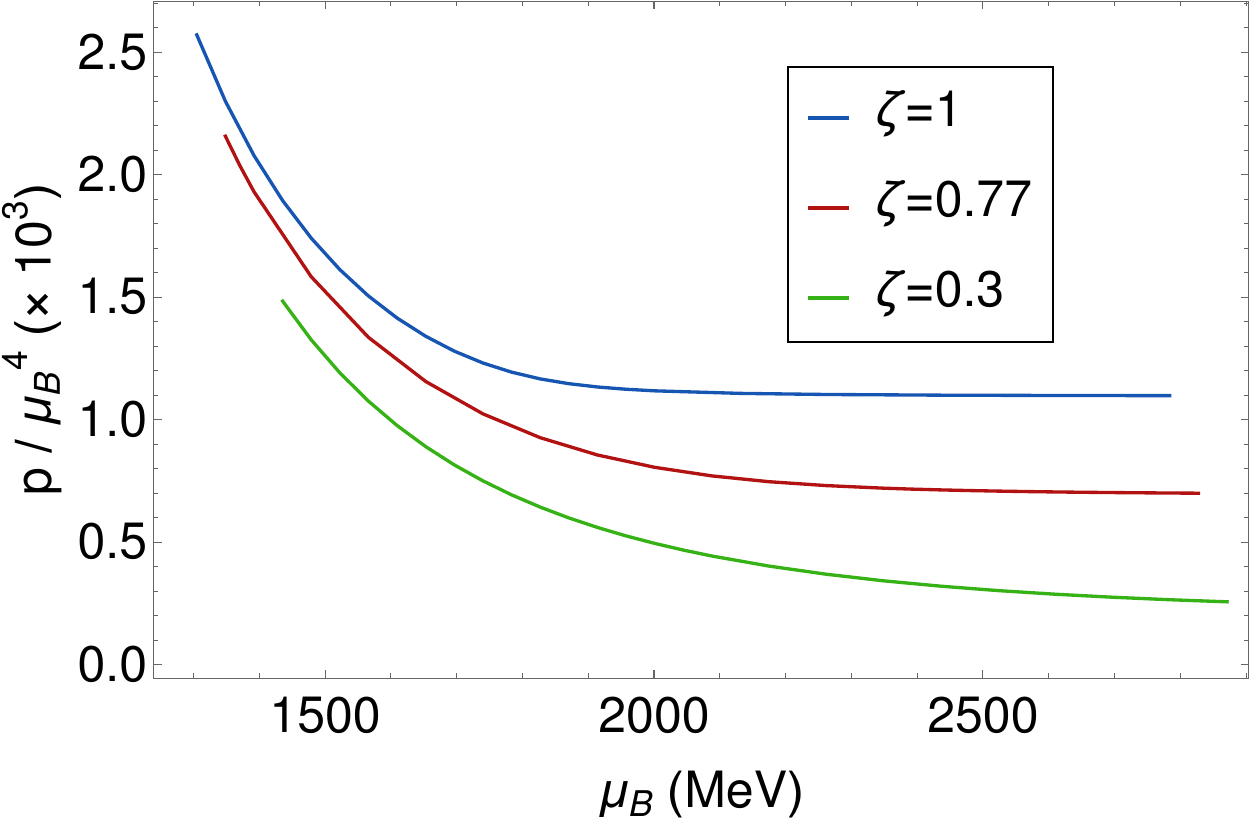}}\hfill
    \subfloat[Reduced Baryon number density]{\includegraphics[width=0.3\linewidth]{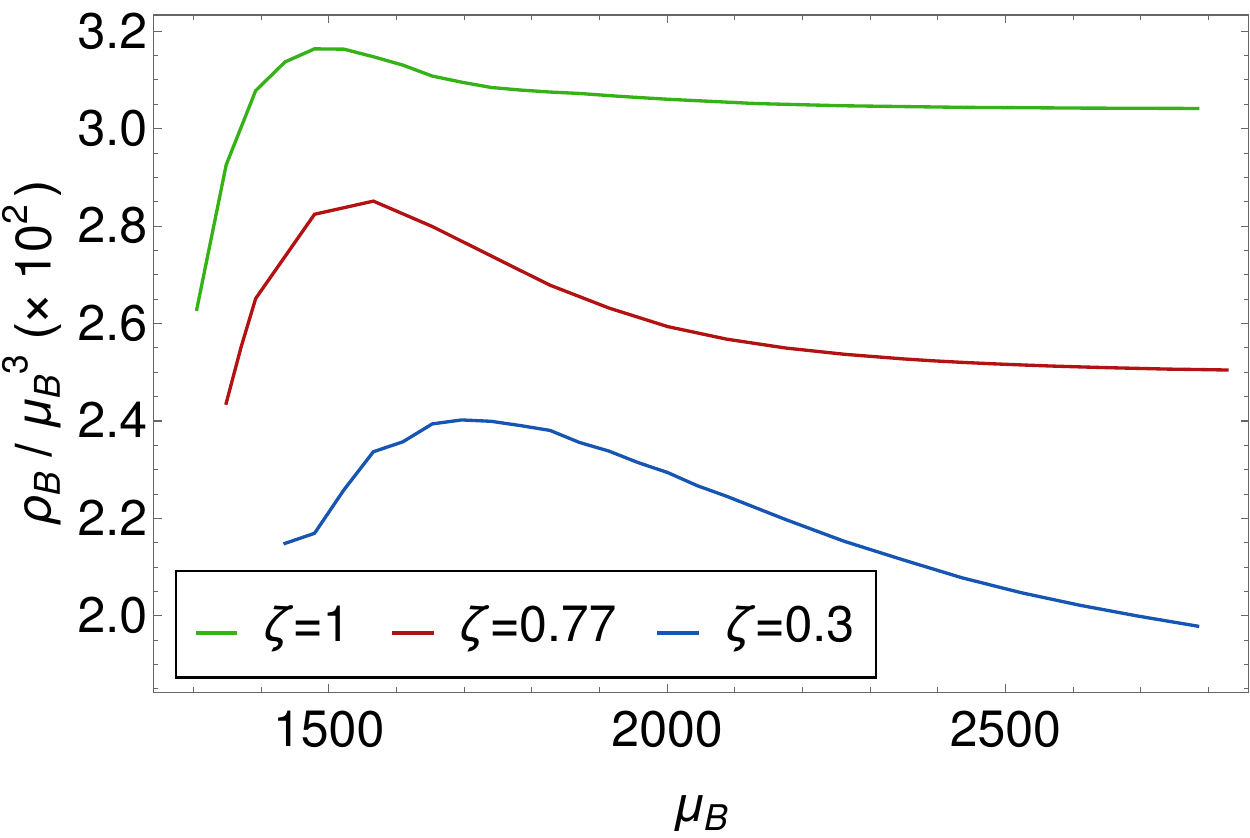}}\hfill
    \subfloat[Reduced Condensate]{\includegraphics[width=0.3\linewidth]{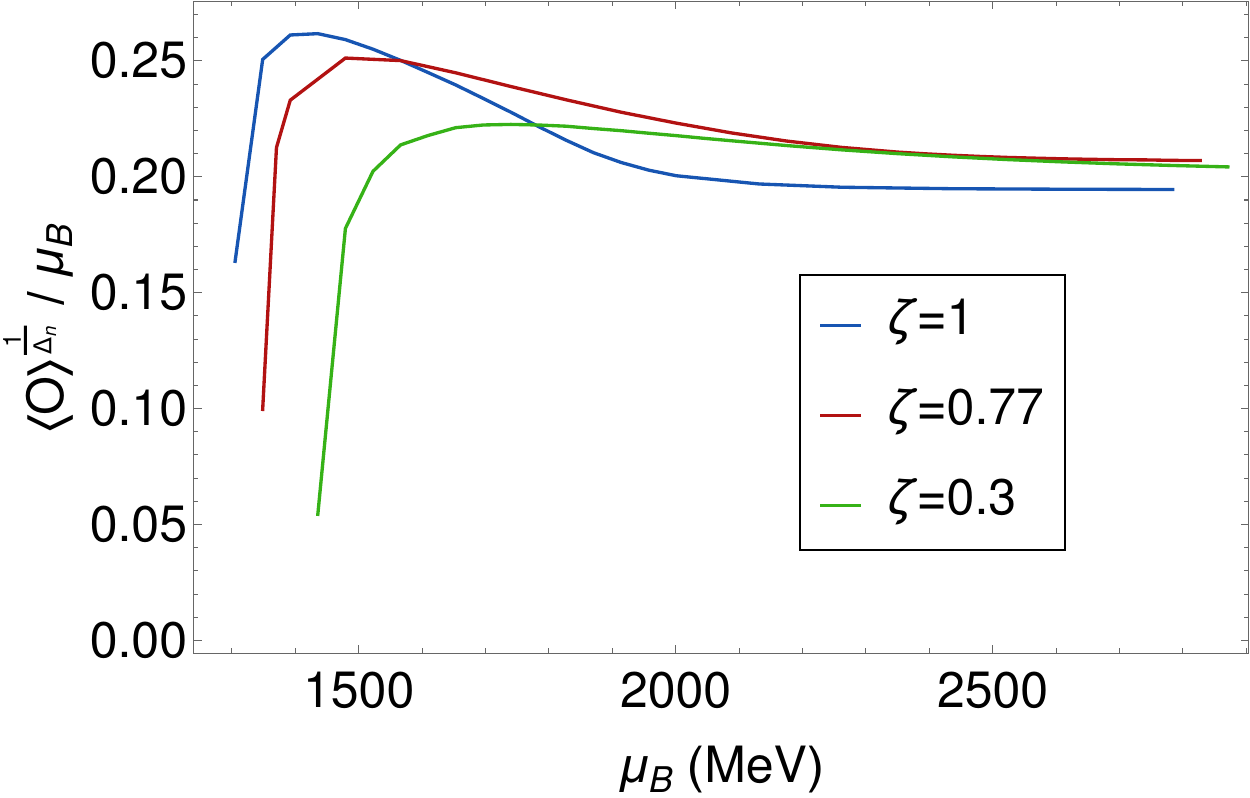}}
    \caption{Thermodynamics quantities as a function of $\m_B$ for $\f_0=0,g_0=6,\D=3,$ and fixed $q=2.$}
    \label{fig:phi0_pressure}
\end{figure}
Additionally, the condensate saturates more rapidly as a function of $\m_B$ for higher values of $\z$. These observations are illustrated in figure \ref{fig:phi0_pressure}.
Since $\z$ controls the interaction between the quarks (gauge fields) and the glue (metric fields), we see that increasing this interaction strength actually leads to higher pressure and densities.

\begin{figure}[h]
    \centering
    \subfloat[Reduced condensate]{\includegraphics[width=0.45\linewidth]{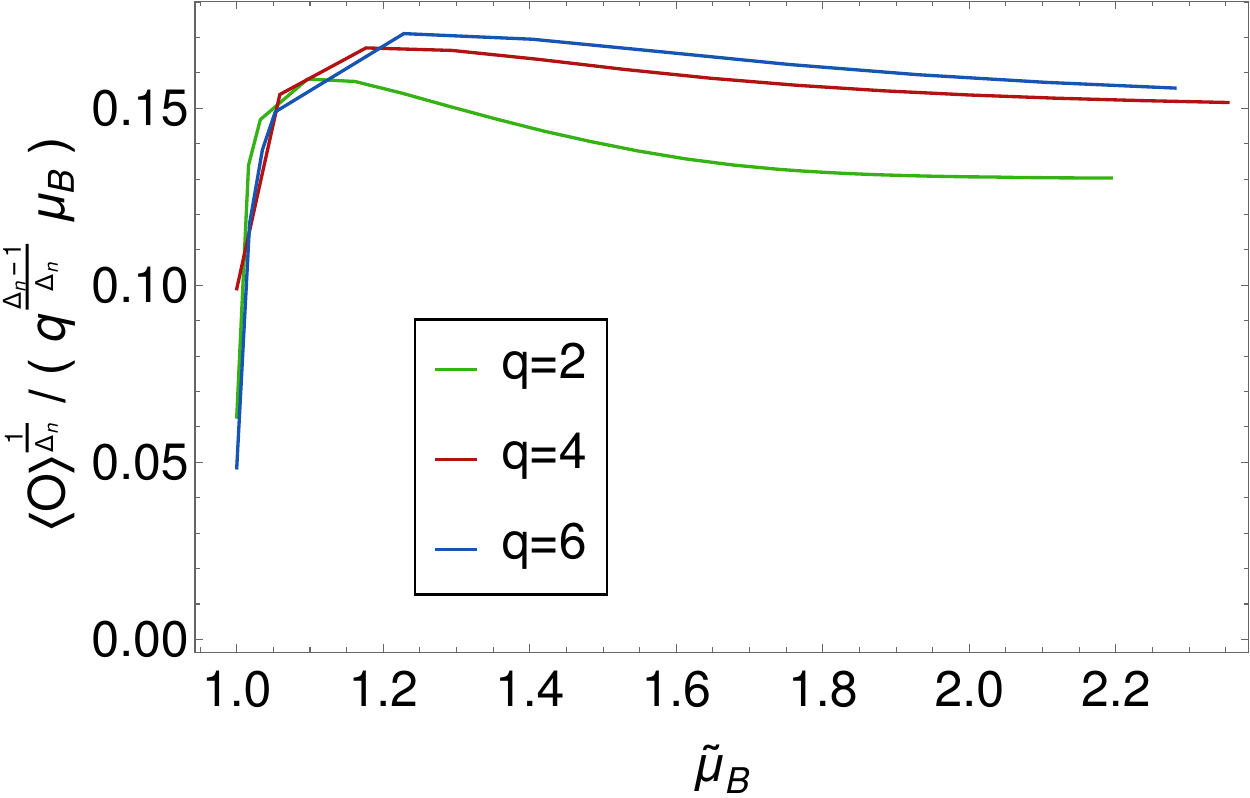}}\hfill
    \subfloat[Reduced Baryon number density]{\includegraphics[width=0.45\linewidth]{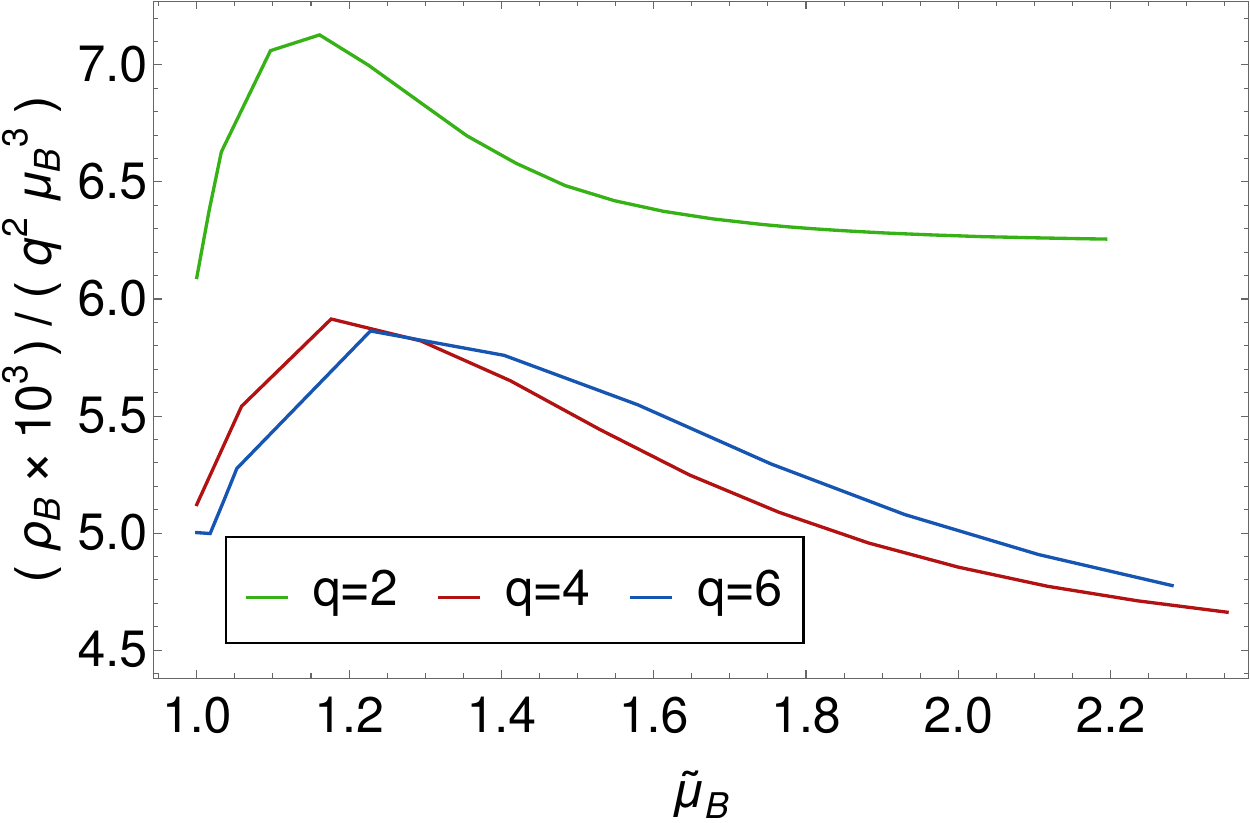}}
    \caption{Thermodynamics quantities as function of $\Tilde{\m}_B=\frac{\m_B}{\m_B^c}$ for different value of $q.$}
    \label{fig:phi0_pressure_varyq}
\end{figure}

The effects of varying the baryonic charge of the condensate is shown in figure \ref{fig:phi0_pressure_varyq} where we see that both baryon density and the condensate depend on the baryon charge of the condensate in a non-linear manner. 
We also see that operators with higher baryonic charge
condense significantly less, as might be expected. 

\subsection{Phase diagrams}
We can now identify the solutions with the largest pressure and thereby construct the phase diagram in the $\mu_B-T$ plane.  We will present our results for a few choices of the parameters with a view toward the more phenomenological case to be discussed in the upcoming section. The dependence of the phase diagram on the parameters $g_0$ and $\z$ can be interpreted in terms of phenomenological quantities that these parameters determine \cite{Erlich:2005qh}. 
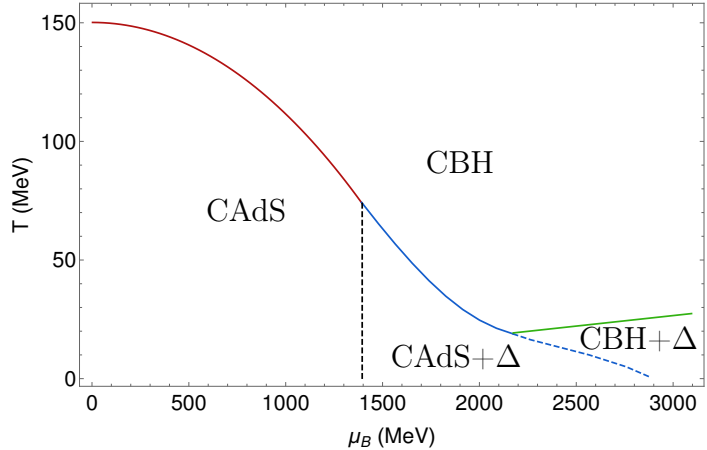
\begin{figure}[h]
    \centering
    \begin{overpic}[scale=0.4, unit=1mm]
{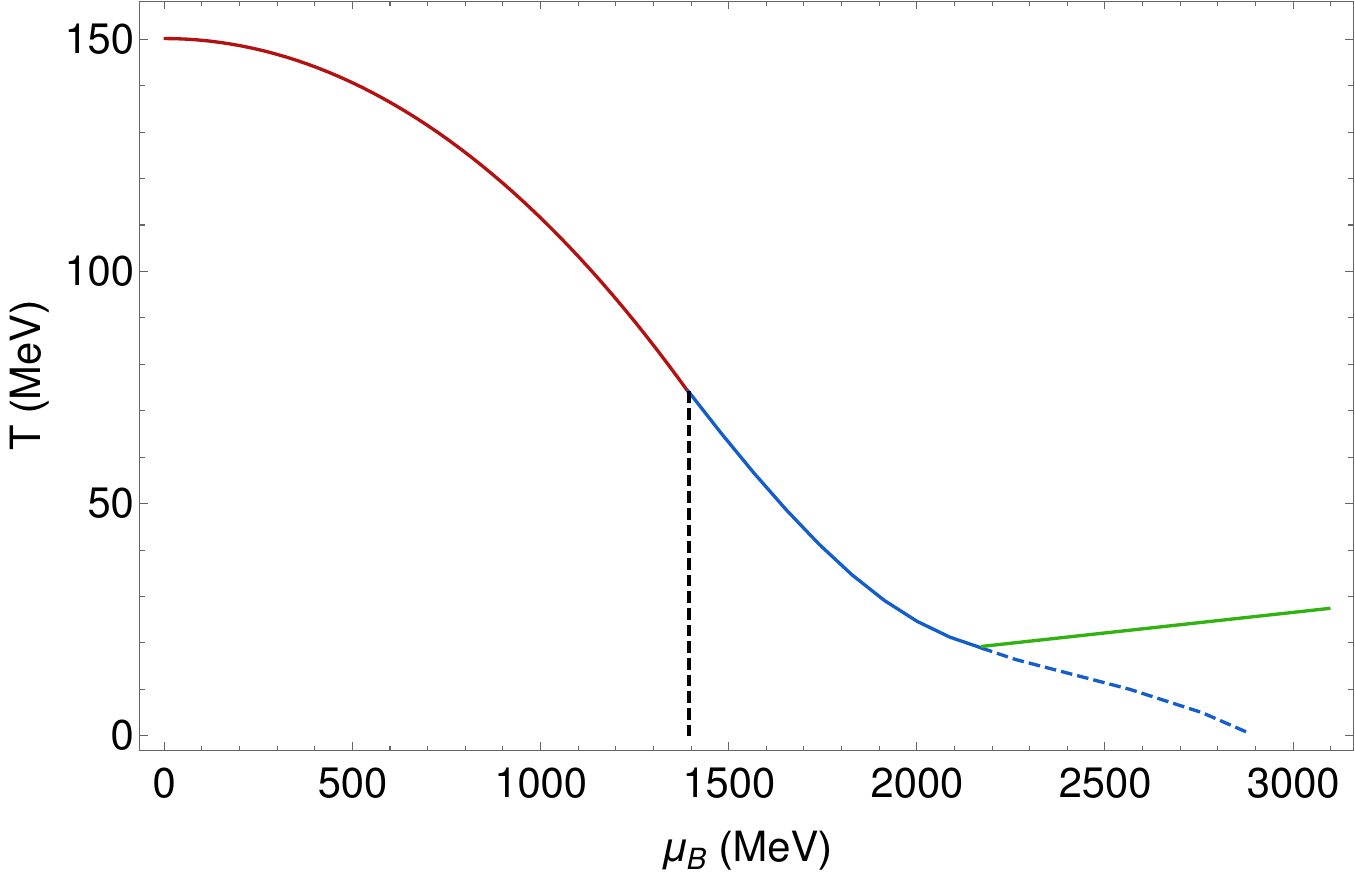}
\put(28.5,33){CAdS}
\put(55,12){CAdS+$\D$}
\put(60,40){CBH}
\put(82,14.5){CBH+$\D$}
\end{overpic}
    \caption{Phase Diagram}
    \label{fig:PD_Phi0}
\end{figure}
As mentioned in the previous section, once the condensate appears in the CAdS geometry, it always remains the lower free energy solution. Since the pressure for the horizonless geometries with the boundary condition $\f_0=0$ does not depend on temperature, the transition line is independent of temperature and is shown by a black dotted curve in figure \ref{fig:PD_Phi0}. As the chemical potential increases, the CAdS with condensate makes a first-order phase transition to the charged black hole with condensate, represented by the blue dashed line. As the temperature increases sufficiently high, the charged black hole phase becomes favored. These transitions are illustrated by the red, blue, and green solid curves, which represent the CAdS, CAdS with condensate, and charged black hole with condensate phases, respectively.
\begin{figure}[h]
            \centering
            \subfloat[$q=2$]{\includegraphics[width=0.3\linewidth]{Paper3_FIG/PD_Phi0_q2_zeta0p7.pdf}}\hfill
            \subfloat[$q=4$]{\includegraphics[width=0.3\linewidth]{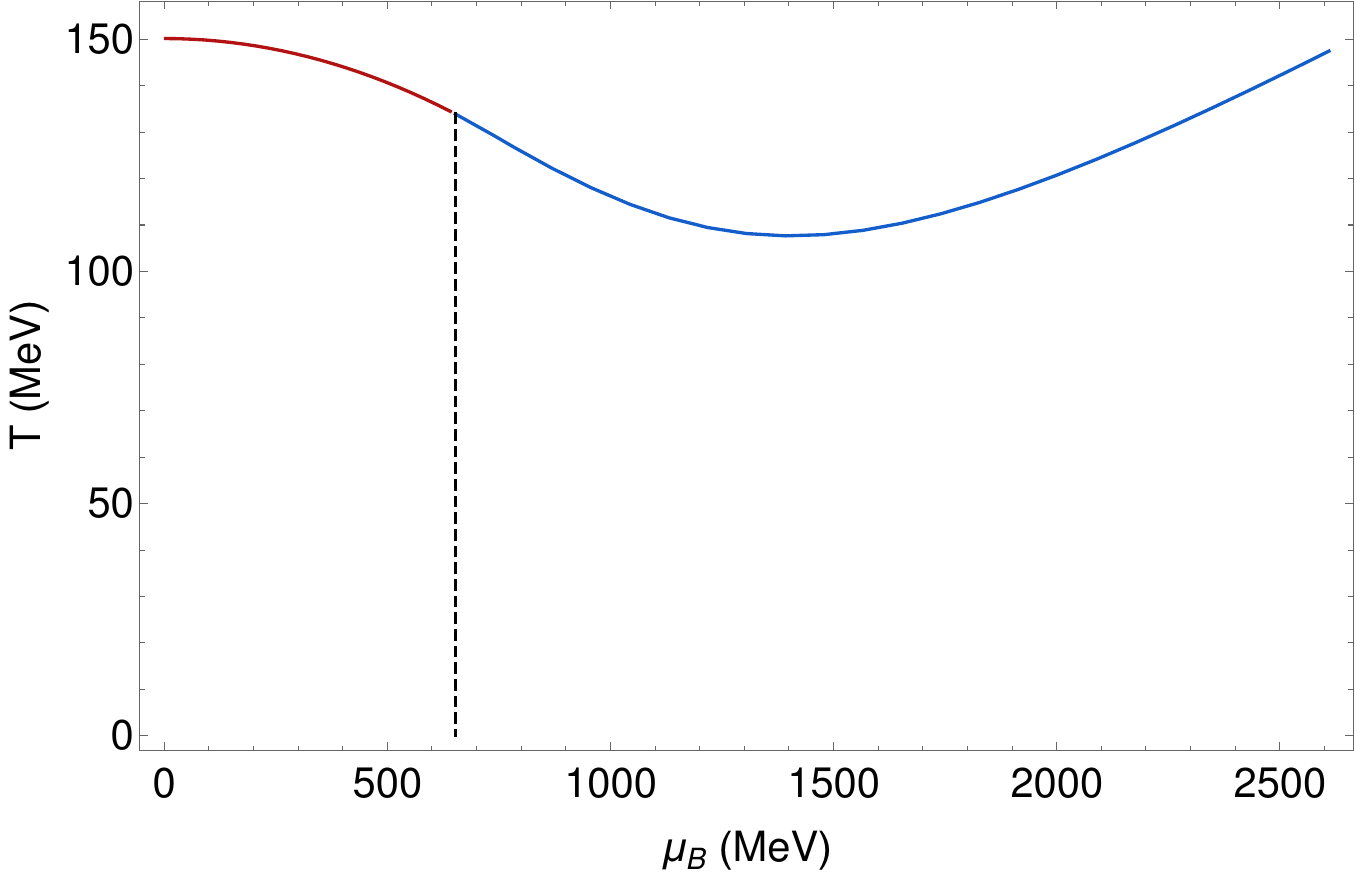}}\hfill
            \subfloat[$q=6$]{\includegraphics[width=0.3\linewidth]{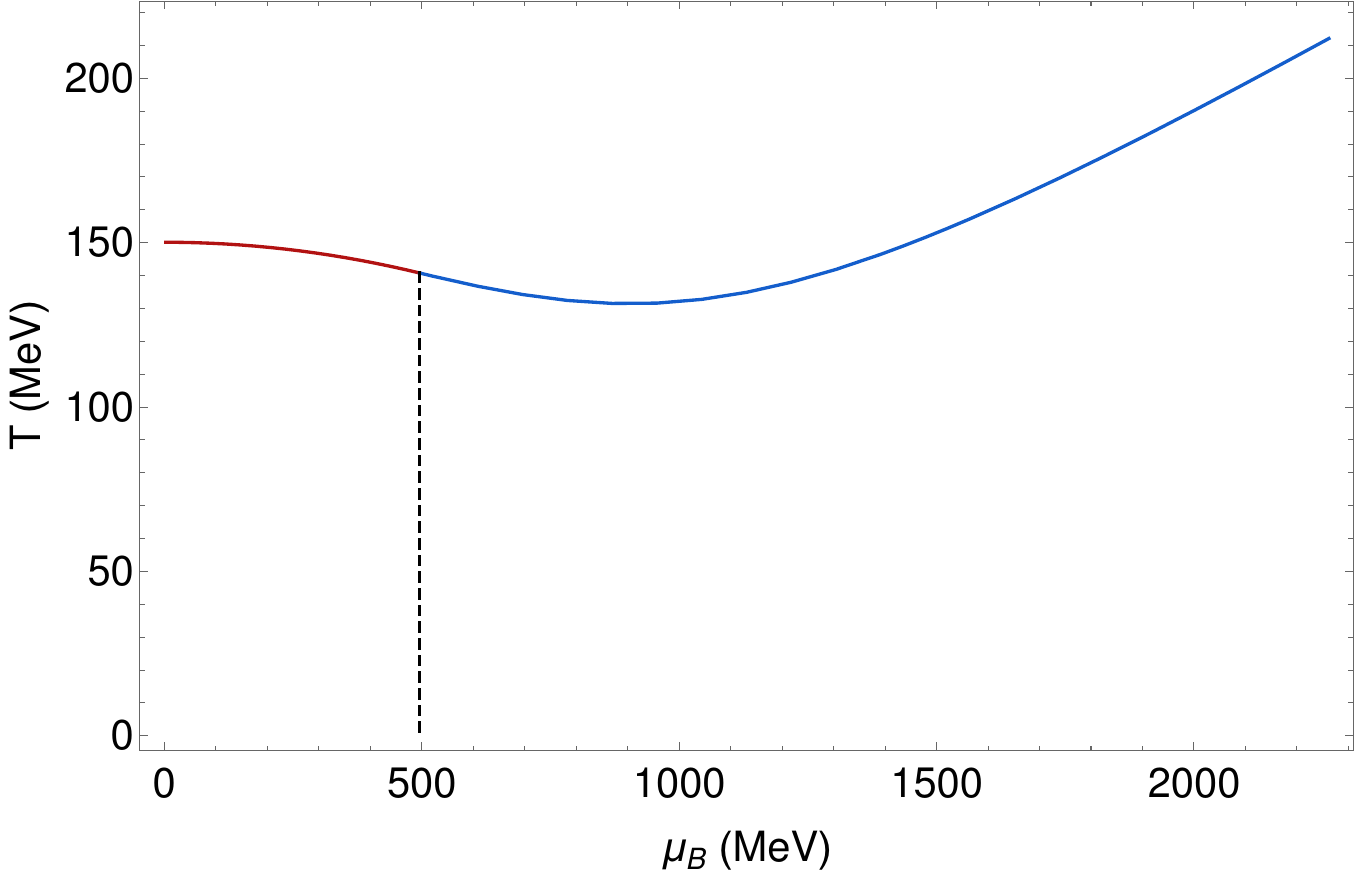}}
            \caption{Phase diagram for $\f_0=0,g_0=6$ and $\z=0.77$}
            \label{fig:PDq}
        \end{figure}

An important question in the context of condensed phases is the nature of the condensate. This may be studied as the effect of the baryonic charge of the condensate $q$ on the phase diagram, as illustrated in figure \ref{fig:PDq}. In the phase diagrams presented here, the charge of the condensate operator in the black hole background is consistently set at $q = 2$ because we can only expect a quark condensate in the deconfined phase. The nature of the condensing operator in the confined phases can, however, vary significantly from baryonic pairs to quark pairs. Figure \ref{fig:PDq} suggests that operators with a greater charge (such as baryon pairs) condense before those with a smaller charge. It also indicates that, for larger values of $q$, the condensate in CAdS will dominate the entire large density region of the phase diagram. We will explore the variation of the scaling dimension in the subsequent sections with phenomenological boundary conditions. 
\begin{figure}[h]
    \centering
            \subfloat[$\z=0.3$]{\includegraphics[width=0.3\linewidth]{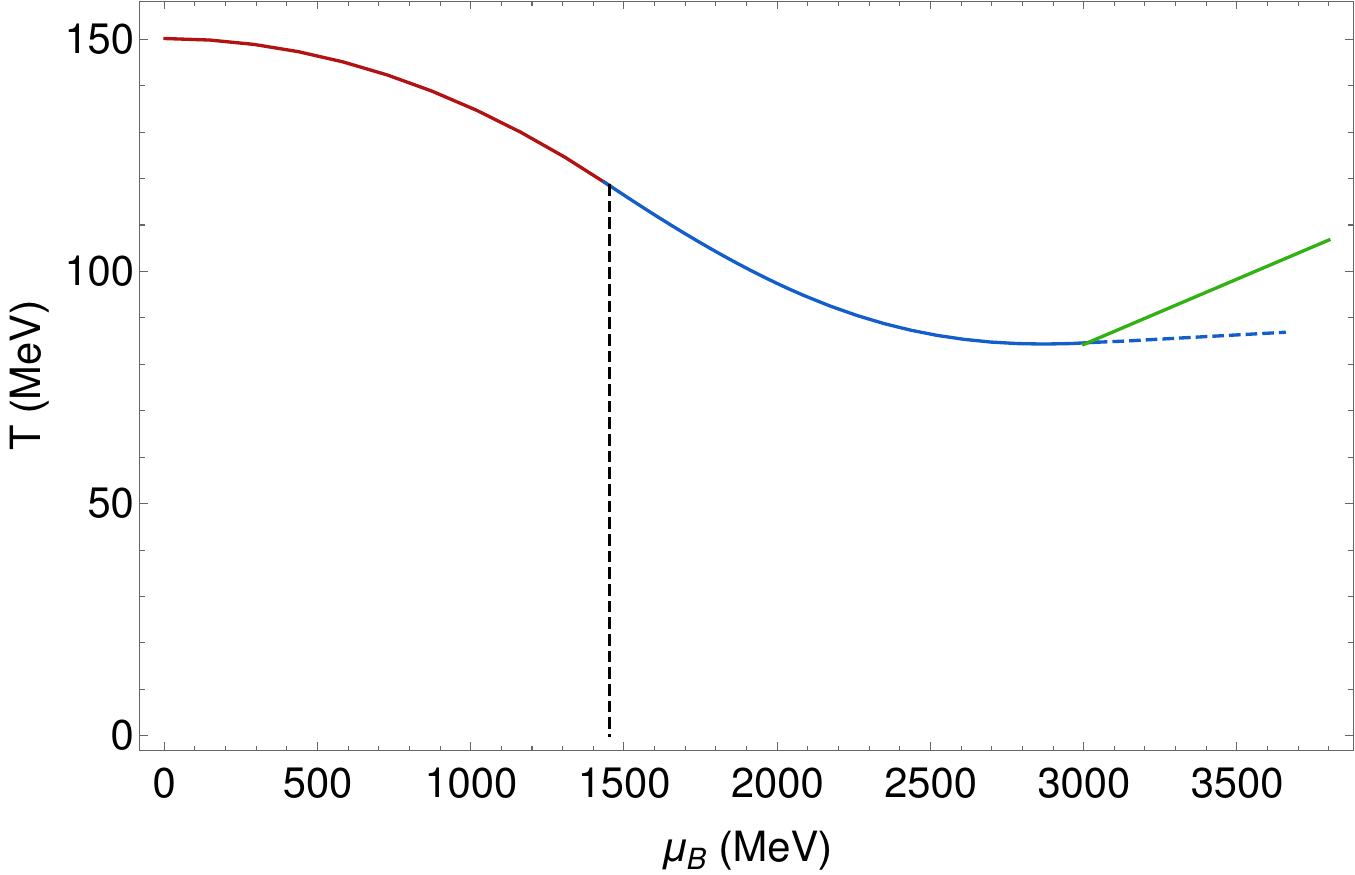}}\hfill
            \subfloat[$\z=0.77$]{\includegraphics[width=0.3\linewidth]{Paper3_FIG/PD_Phi0_q2_zeta0p7.pdf}}\hfill
            \subfloat[$\z=1$]{\includegraphics[width=0.3\linewidth]{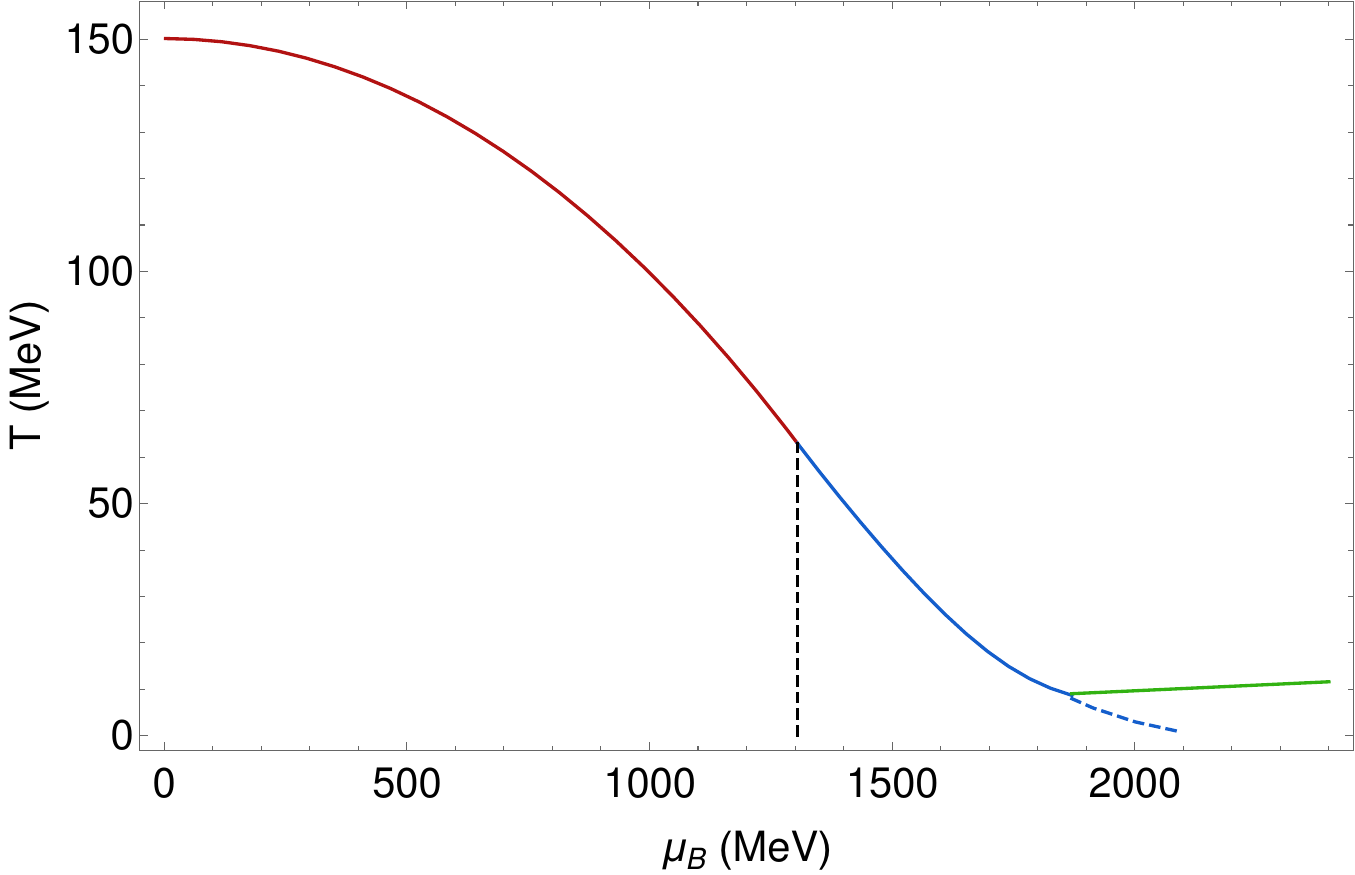}}
            \caption{Phase diagram for $\f_0=0,g_0=6$ and $q=2.$}
    \label{fig:PDzeta}
\end{figure}
We illustrate the effect of the coupling parameter $\z$ on the phase diagram in figure \ref{fig:PDzeta}. The first observation from this figure is that the onset of the condensate in CAdS backgrounds shifts to smaller values of $\m_B$ as $\z$ increases. Conversely, the transition temperature from the condensate to the uncondensed phase in CBH decreases with increasing $\z$. Furthermore, we find that for large values of $\z$, charged black hole geometries are favored over charged AdS-type geometries. One may note that the phase diagram for $\z=0.3$ resembles the phase diagram presented in \cite{Basu:2016mol}.
\begin{figure}[h]
            \centering
            \subfloat[$g_0=6$]{\includegraphics[width=0.4\linewidth]{Paper3_FIG/PD_Phi0_q2_zeta0p7.pdf}}\hfill
            \subfloat[$g_0=27$]{\includegraphics[width=0.4\linewidth]{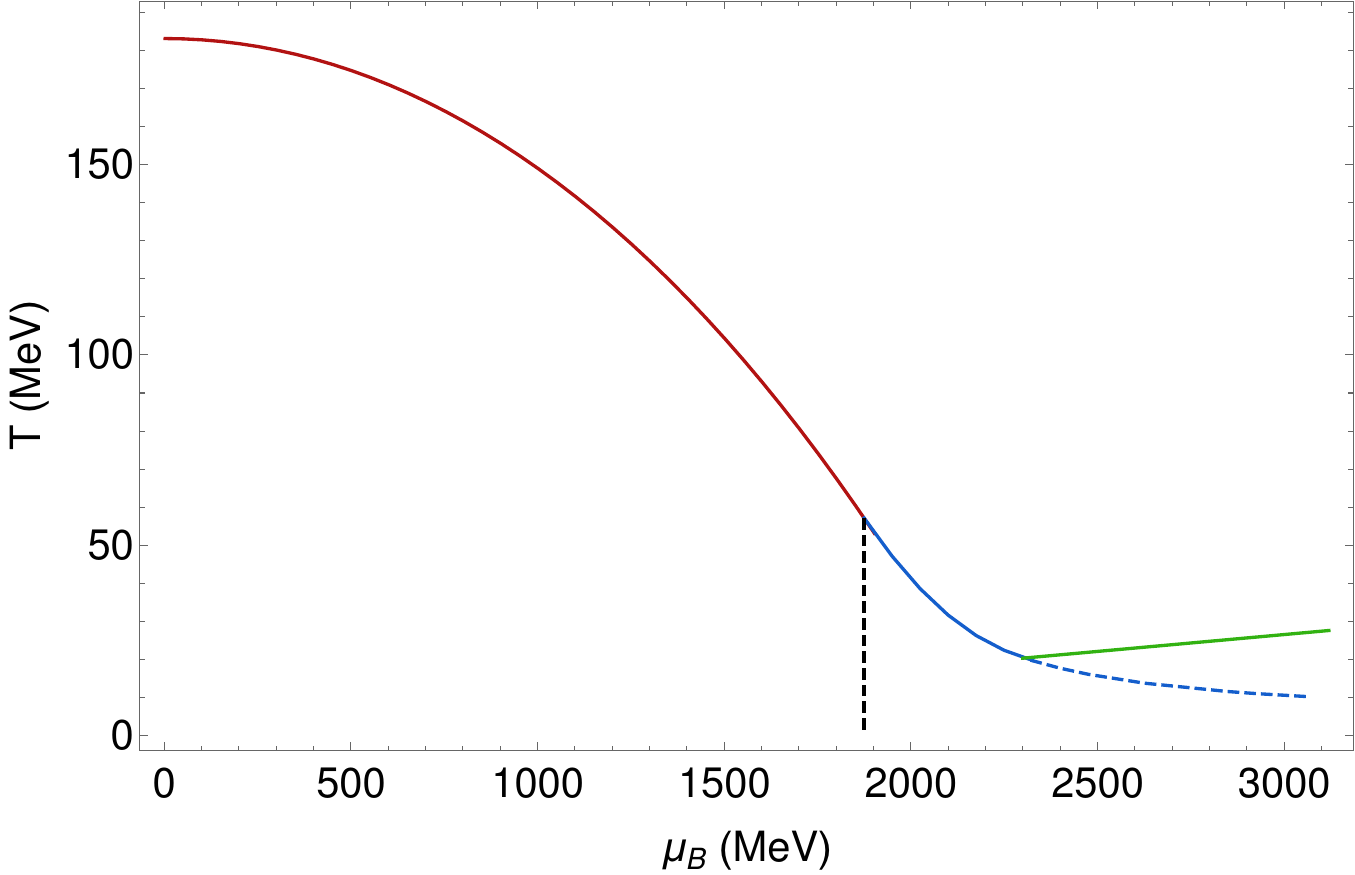}}
            \caption{Phase diagram for $\f_0=0, q=2$ and $\z=0.77$}
            \label{fig:PDg0}
        \end{figure}
Changing the values of the parameter $g_0$ seems to only alter the onset of condensation in the confined phase - increasing $g_0$ does increase the pressure of the CAdS background as expected from its relation to the mass of the mesons. The increase in the onset of condensation can then be interpreted to mean that $g_0$ does not affect the scalar field profile significantly.

\section{Pheno boundary conditions}

In this section, we will study the possibility of symmetry breaking condensates when the baryonic matter is described by the phenomenology of chiral perturbation theory and the NJL model. As we have shown, the baryonic charge and scaling dimensions of the scalar field play a crucial role in determining the possibility of condensates. In this section, we will restrict our attention to physically determined value $g_0=6$ since the previous section showed this parameter could be expected to play a relatively benign role.  We will discuss the values of $q$ and $\D$ in the sections below and present our results for a few values of $\z.$

\subsection{van der Waals phase}\label{sec-vdw}
When baryon matter is modeled by the vdW equation of state, we expect the condensates to be baryon pairs (ie, protons or neutrons) leading to superconductivity or superfluidity. This implies that the quark charge of the dual scalar field is $q=6$. Since this gas of baryons is strongly interacting, the operator describing this pair will be highly renormalized and so its scaling dimension $\D$ is likely to be far from the perturbative value $\D=9.$ A value close to $\D=3$ suggests a fermionic pair, but since the baryons are made of $N_c$ quarks, this also means that we are assuming that the medium does not destabilize the baryons themselves. On the other hand, values closer to $\D=3N_c=9$ are indicative of loosely bound quark pairs and correlate with nearly dissolved baryons. From the bulk viewpoint, the large charge of the complex scalar field leads to considerable energy costs since it generates large electric fields. Therefore, it seems reasonable to require that $m^2>0$ for the charged scalar, which will disfavor large amounts of condensation.  We will, therefore, explore a few values of $\D$.

\begin{figure}[h]
\centering
\includegraphics[width=0.45\linewidth]{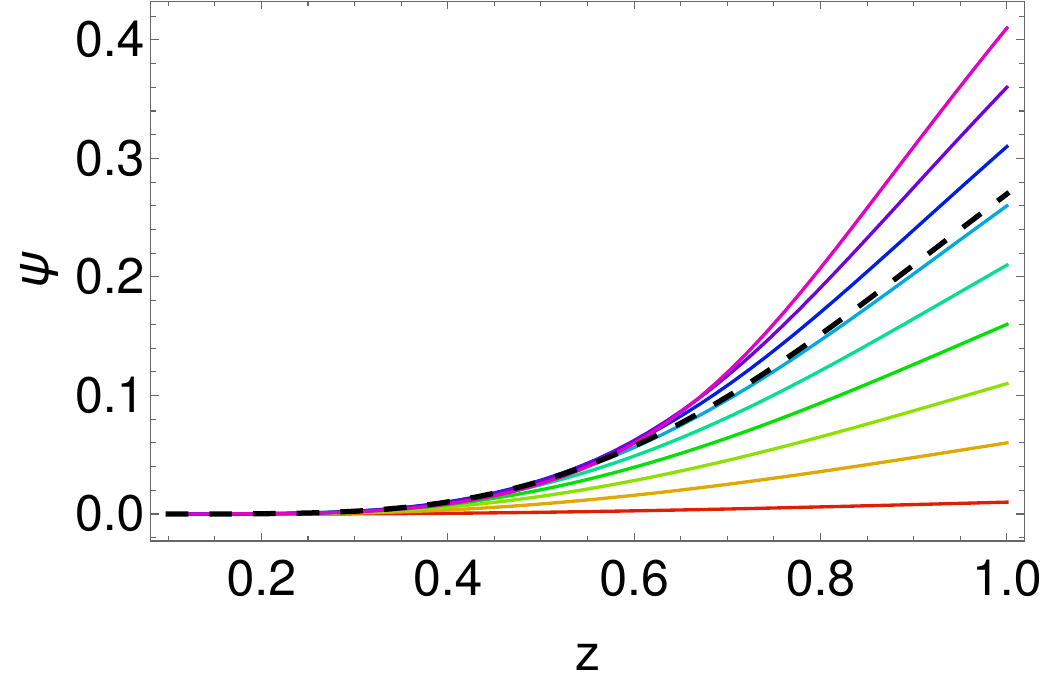}\hfill
\includegraphics[width=0.47\linewidth]{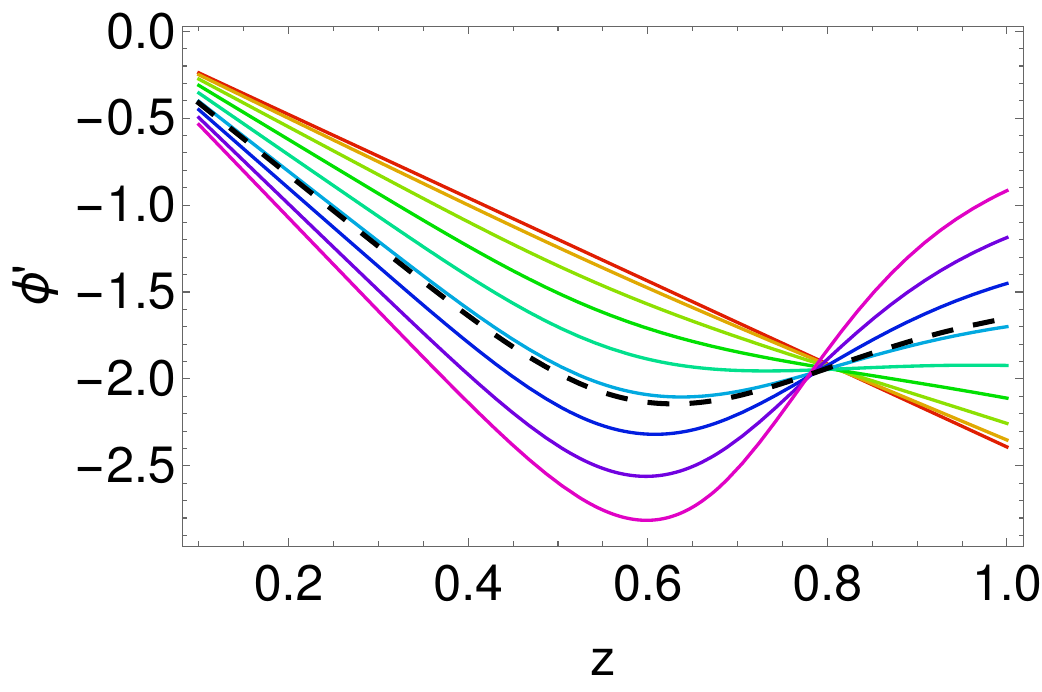}\\
\includegraphics[width=0.45\linewidth]{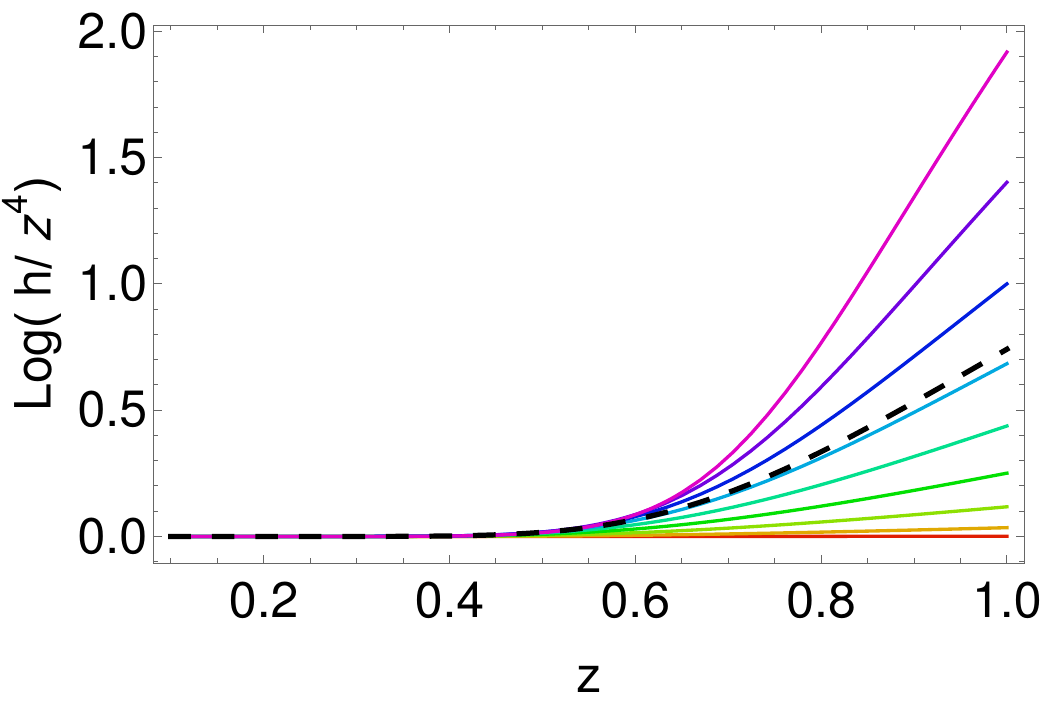}\hfill
\includegraphics[width=0.45\linewidth]{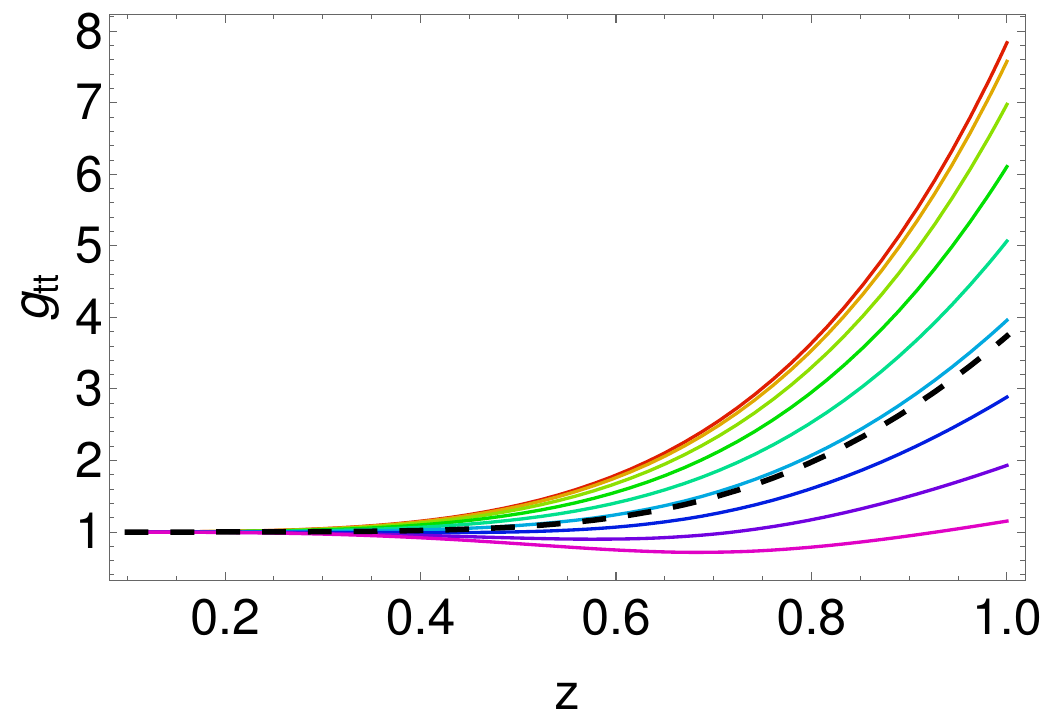}
\caption{Numerical solution for the system with parameters $q=6$, $g_0=6$, $\Delta_n=6.3$, $\m_B=1566$ MeV, and vdW-type boundary condition.}
\label{fig:SOL_vdW}
\end{figure}

To begin with, figure \ref{fig:SOL_vdW} shows that even while using phenomenologically motivated boundary conditions, condensate-type solutions do exist. The solutions are shown for the parameter values $q=6$, $g_0=6$, $\Delta_n=6.3$, $\m_B=1566$ MeV and zero temperature. Finite temperature solutions exhibit similar behavior; therefore, we will not present them here. 

It is important to mention that for large values of scaling dimensions, controlling the numerical calculations while ensuring that the non-normalizable mode remains zero using the shooting method is quite challenging. To manage the numerics effectively, we set the UV cutoff to a larger value. This challenge arises from our approach to solving the system of differential equations by transforming the boundary value problem into an initial value problem. Although there may be alternative approaches to address this issue, we expect that they will not significantly affect the thermodynamics of the system or the resulting phase diagram.
Again, the solutions obtained by maximizing the pressure over the boundary value $\psi_0$ are seen to be in the middle of a range of solutions with bulk profiles that are free of any pathology. Once again, there exists an upper limit to the value of the scalar field at the IR cutoff where the metric component $g_{tt}\to 0.$ However, we will demonstrate that the condensate undergoes a phase transition to NJL-type phases before reaching this point.

\begin{figure}[h]
    \centering
    \includegraphics[width=0.5\linewidth]{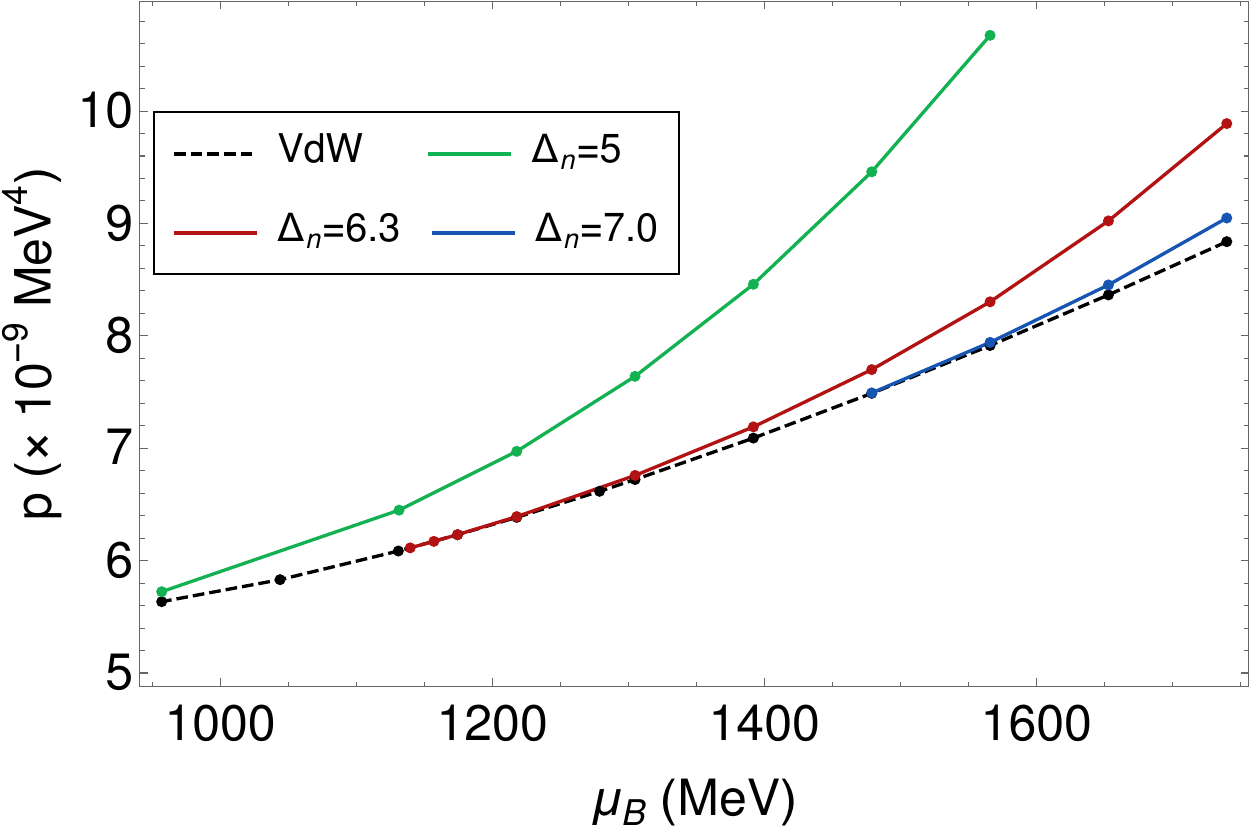}
    \caption{vdW: Maximum pressure solutions for different values of $\Delta$.}
    \label{vdWcondDelta}
\end{figure}

We can now compare the pressure of condensed solutions with that of solutions that do not contain condensates. As illustrated in figure \ref{vdWcondDelta}, the presence of a condensate increases the pressure. The figure also indicates that for larger values of $\D$, condensation onset occurs at higher chemical potential values. This has an interesting consequence: it limits the range of scaling dimensions of the condensing operators in the boundary theory. For instance, when $\D \gtrsim 7$, condensation will only occur after the NJL phase has taken over the vdW phase, meaning that the vdW with condensate never becomes the preferred phase. Conversely, for $\D \sim 5$, condensation begins at $\m_B \lessapprox M_B$, which is not a reasonable scenario. 
It should be mentioned that the range of $\D$ depends on the value of the coupling parameter $\z$.

\begin{figure}[h]
    \centering
    \subfloat[Reduced pressure]{\includegraphics[width=0.3\linewidth]{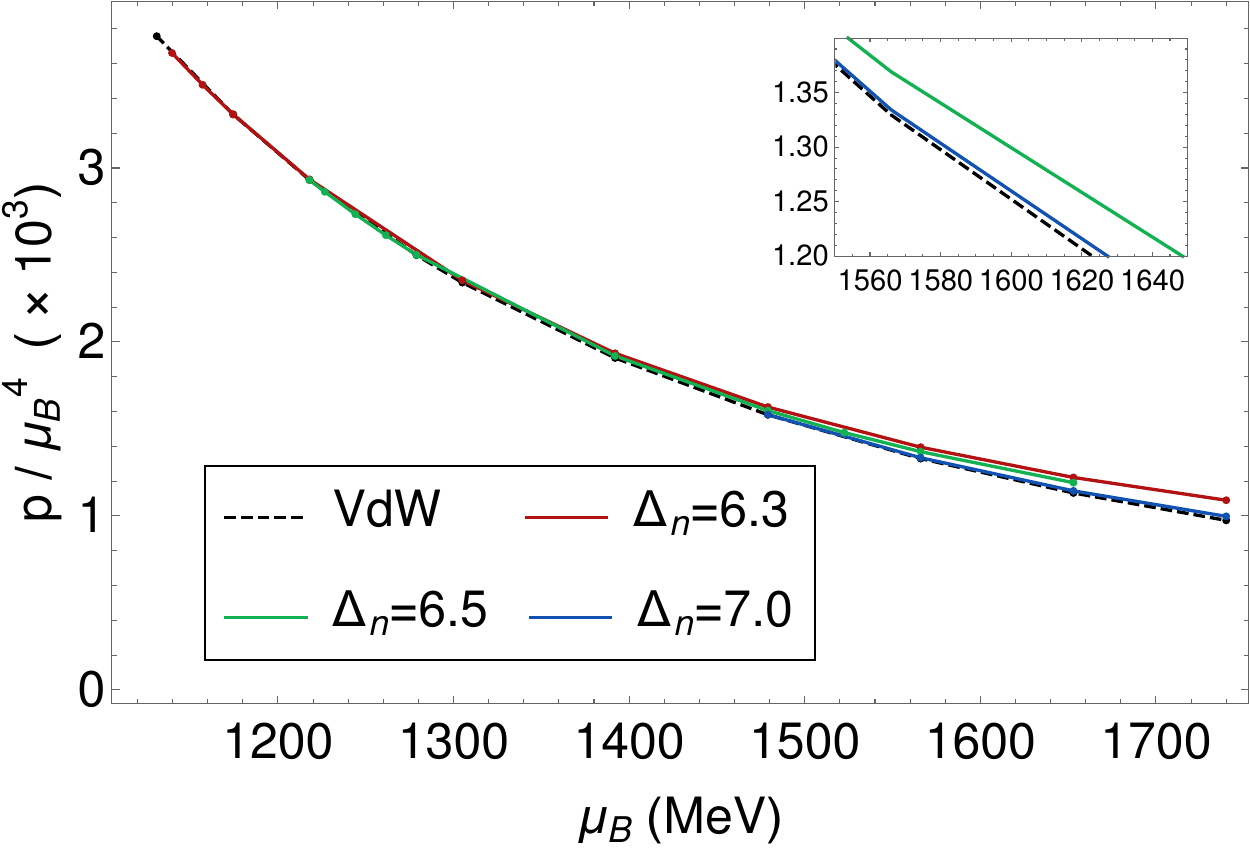}}\hfill
    \subfloat[Reduced Baryon number density]{\includegraphics[width=0.3\linewidth]{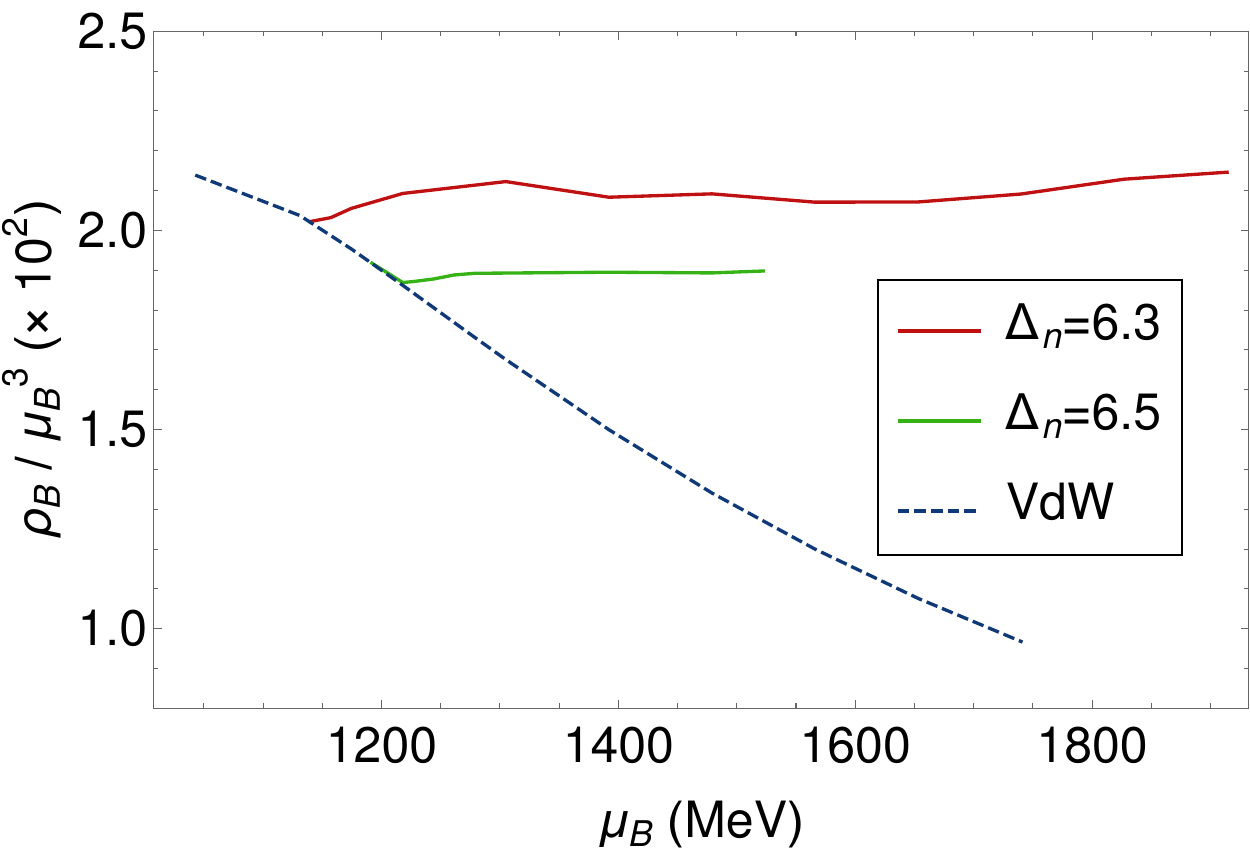}}\hfill
    \subfloat[Reduced Condensate]{\includegraphics[width=0.32\linewidth]{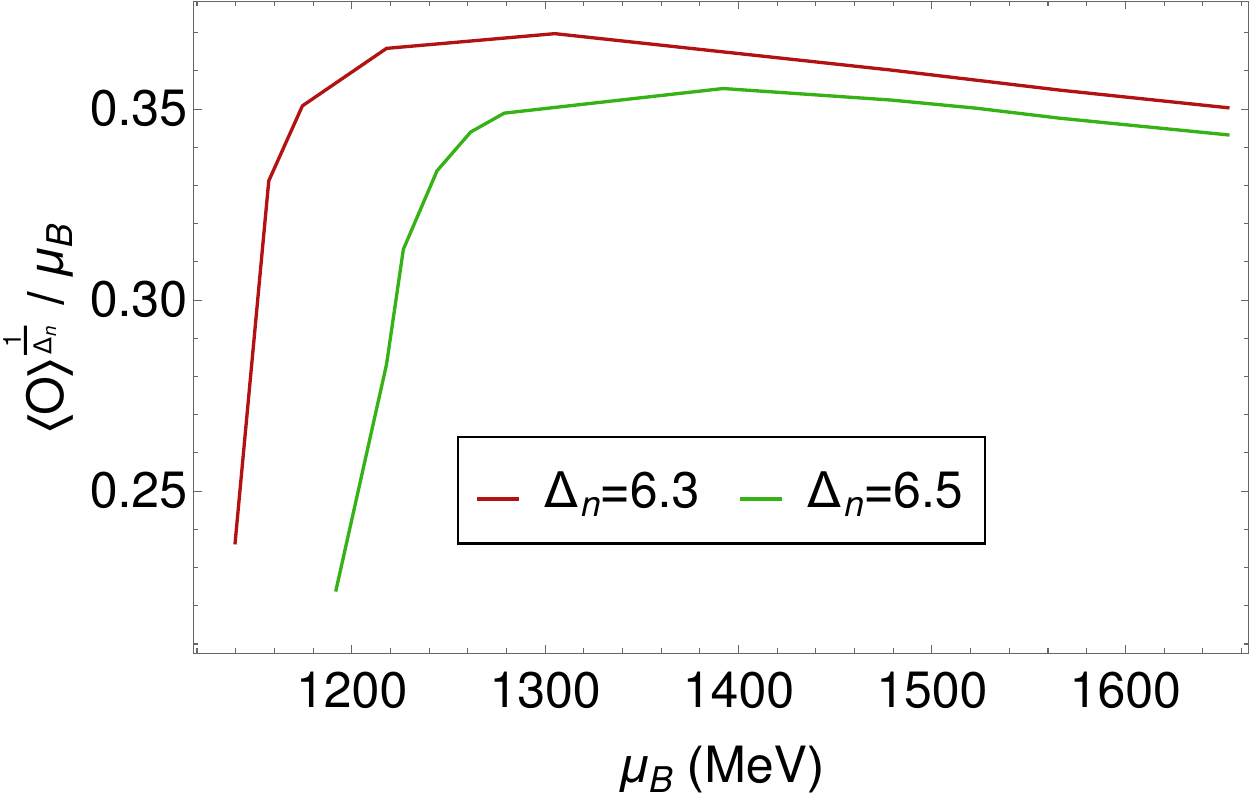}}
    \caption{Thermodynamics quantities as a function of $\m_B$ for $g_0=6,q=6$ and fixed $\z=0.77.$}
    \label{fig:vdWC-TD}
\end{figure}

Because we are extremizing over $\psi_0$, the condensation transition is continuous. Some thermodynamic properties extracted from the holographic solution at zero temperature are shown in figure \ref{fig:vdWC-TD}. The blue dashed line represents the vdW phase without the condensate. The red and green solid curves indicate the condensate phase with scaling dimensions $\D=6.5$ and $\D=6.5.$ As stated previously, operators with larger scaling dimensions tend to form condensates at high chemical potential values. A noteworthy feature is the saturation of the baryonic density. 

\begin{figure}[h]
    \centering
    \subfloat[Reduced pressure]{
    \includegraphics[width=0.3\linewidth]{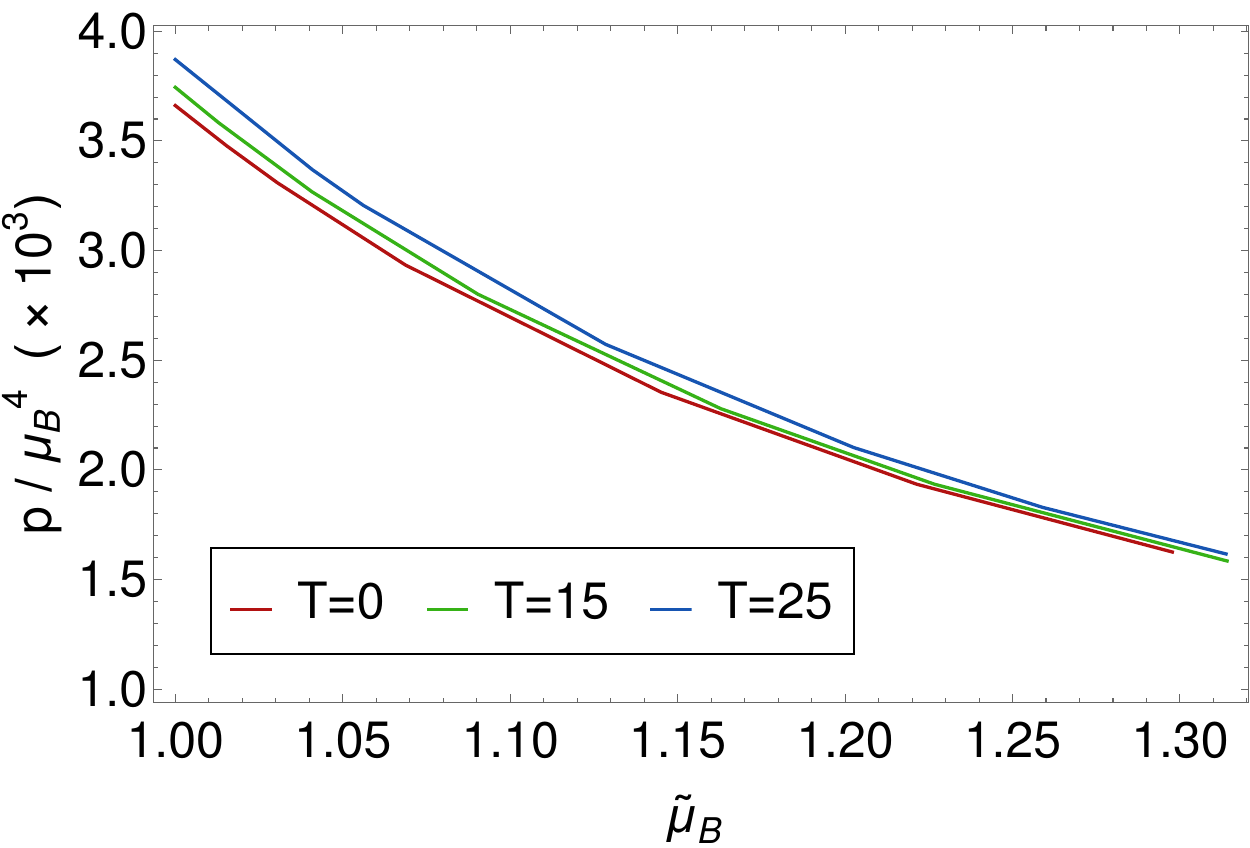}}\hfill
    \subfloat[Reduced condensate]{
    \includegraphics[width=0.3\linewidth]{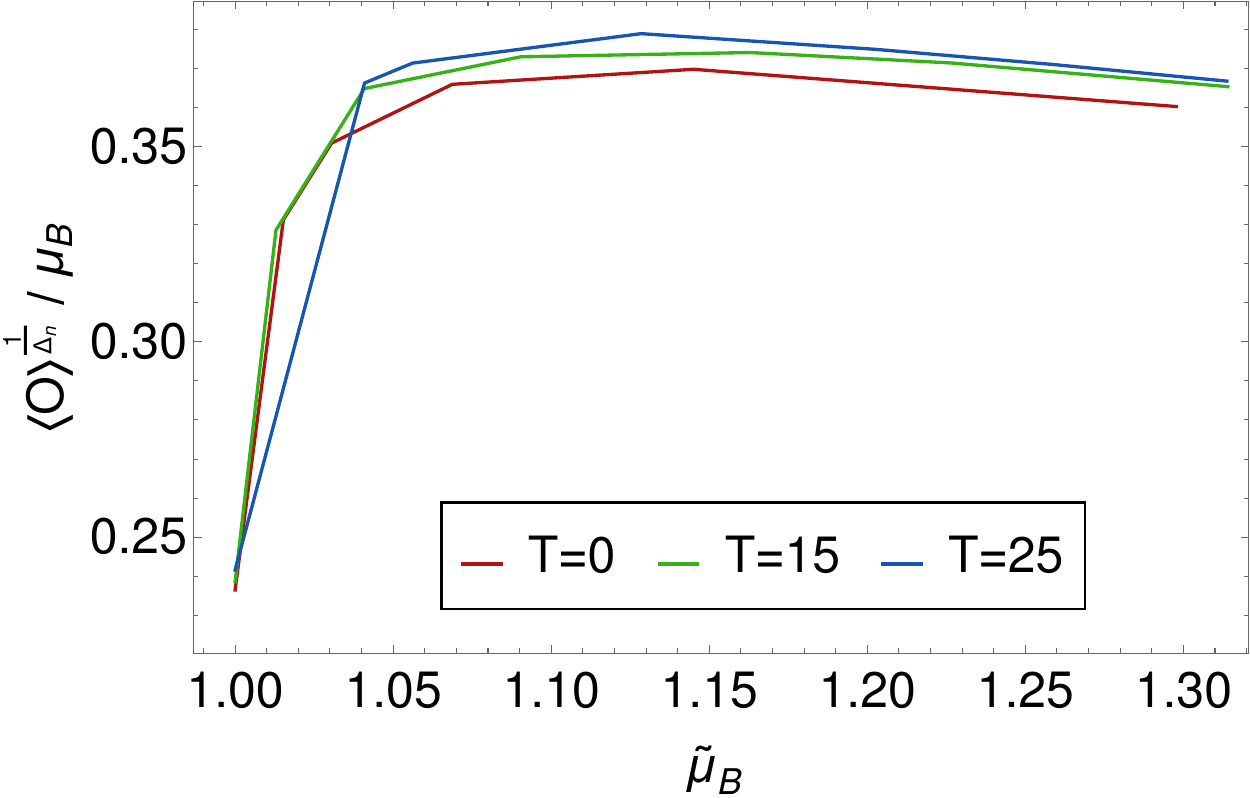}}\hfill
    \subfloat[Relative condensate number density]{
    \includegraphics[width=0.3\linewidth]{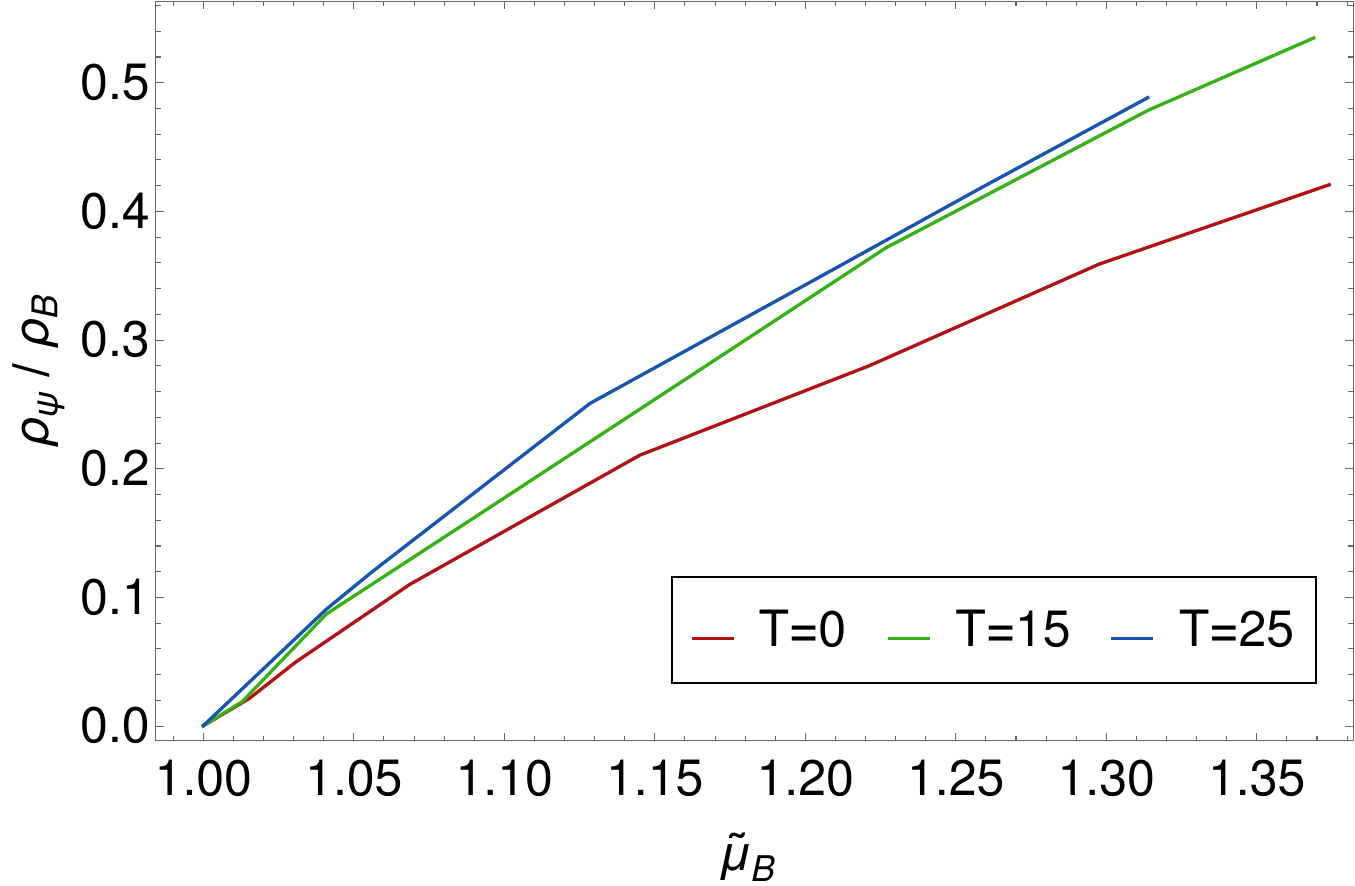}}
    \caption{Thermodynamic quantities as a function of $\Tilde{\m}_B=\frac{\m_B}{\m_B^c}$ for $g_0=6,q=6, \D=6.3$ and fixed $\z=0.77.$}
    \label{fig:vdWC-Cond}
\end{figure}

Finally, figure \ref{fig:vdWC-Cond} shows that the condensate does not alter appreciably at the low temperatures of interest to the interiors of neutron stars. It can be concluded from the fraction of density that for larger temperatures, the amount of matter that condenses increases. 

\subsection{NJL}
At somewhat higher densities,  we have shown that the vdW phase of the preceding section gives way to an NJL gas of quarks with a gradually vanishing chiral condensate.
Since the NJL baryon density is in the form of quarks, the baryon charge of the scalar field will be set to $q=2$ as befits a Cooper pair of quarks. In this sense, these phases can be thought of as either a superconducting/superfluid phase of quarks or even a CFL phase. Correspondingly, we will assume that the scaling dimension of the dual operator is $\D=3$, which implies that $M^2L^2=-3.$ The negative mass of the scalar field suggests that condensation will be energetically favored in contrast with the vdW phase where the mass was positive. 
\begin{figure}[h]
\centering
\includegraphics[width=0.45\linewidth]{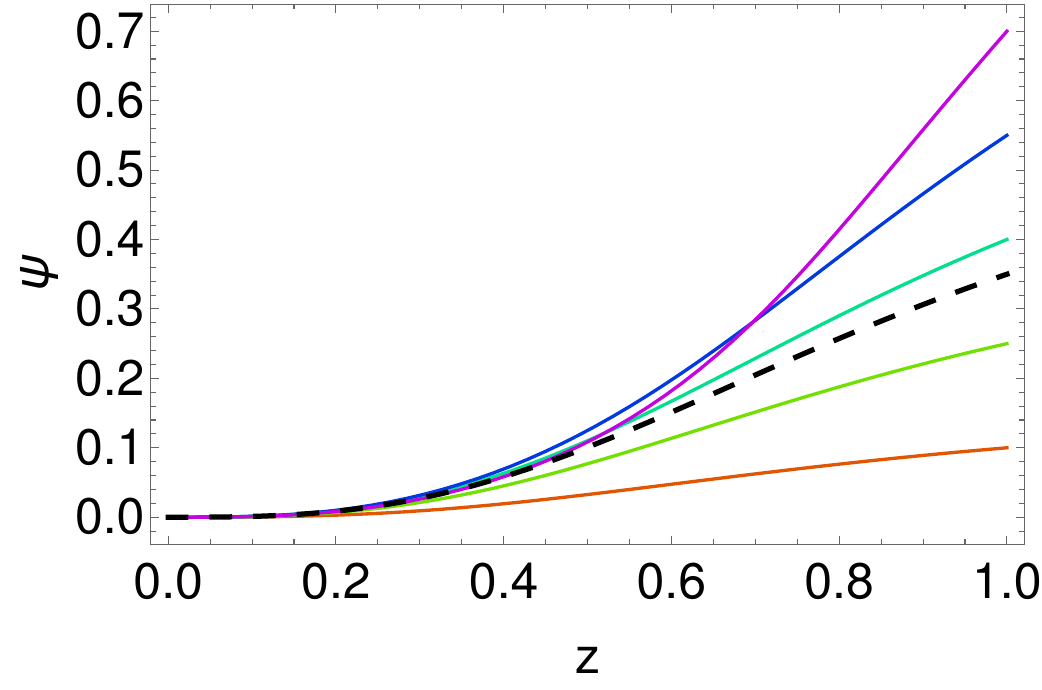}\hfill
\includegraphics[width=0.47\linewidth]{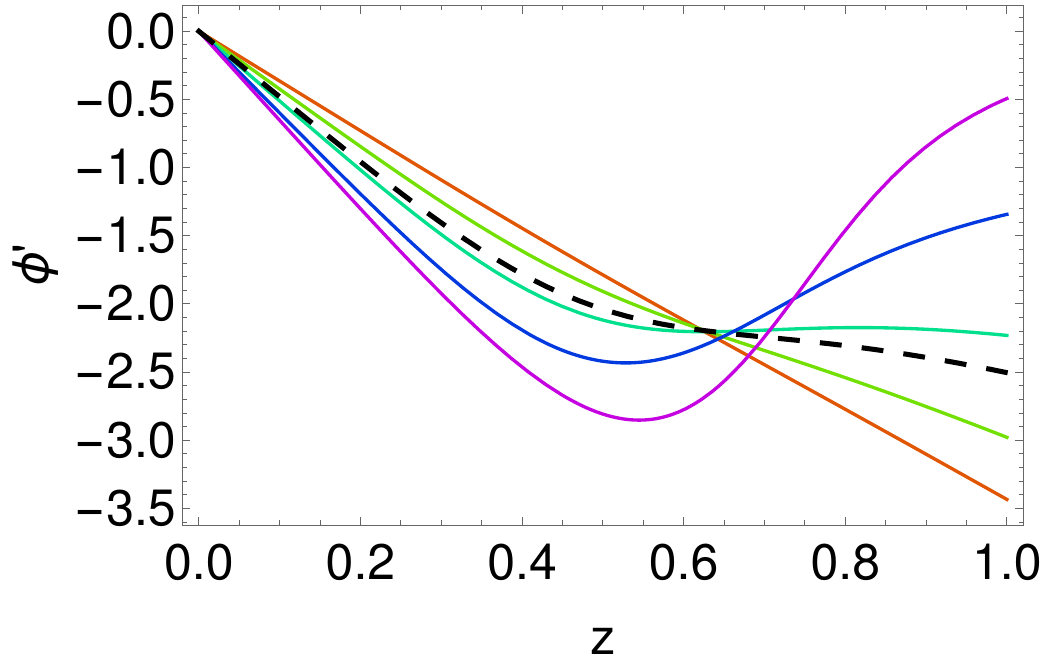}\\
\includegraphics[width=0.45\linewidth]{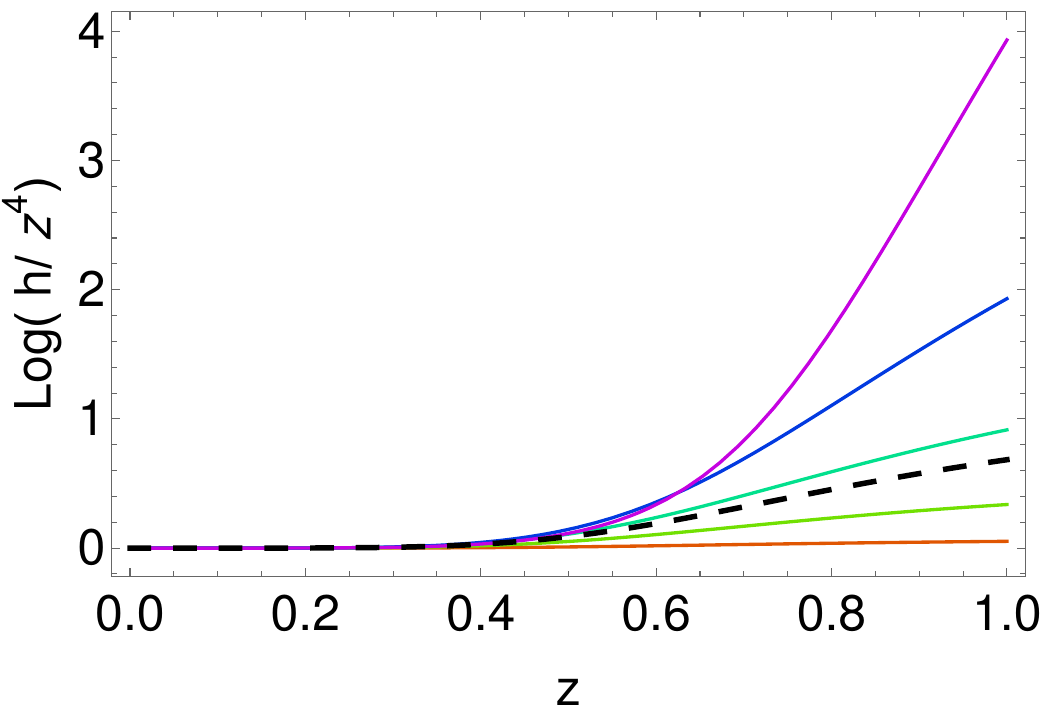}\hfill
\includegraphics[width=0.45\linewidth]{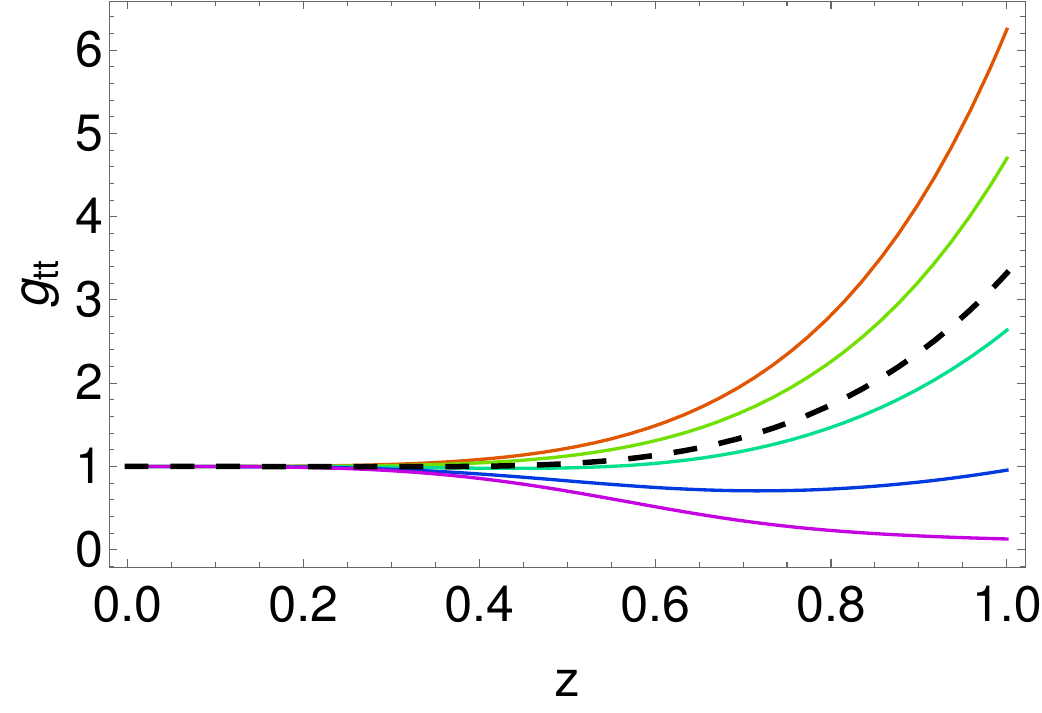}
\caption{Numerical solutions for fixed $\m_B=1566$ MeV and $T=1$ MeV in the NJL phase. Different color shows solutions with different values of $\y(z_0)$. Black dashed curves represent the minimum energy solution.}
    \label{NJLProfiles}
\end{figure}
Using the pressure and baryon density computed from the NJL model to set the IR boundary conditions, we can numerically solve the equations to obtain solutions with nonzero scalar field profiles.

\begin{figure}[h]
    \centering
        \subfloat[Reduced pressure]{\includegraphics[width=0.3\linewidth]{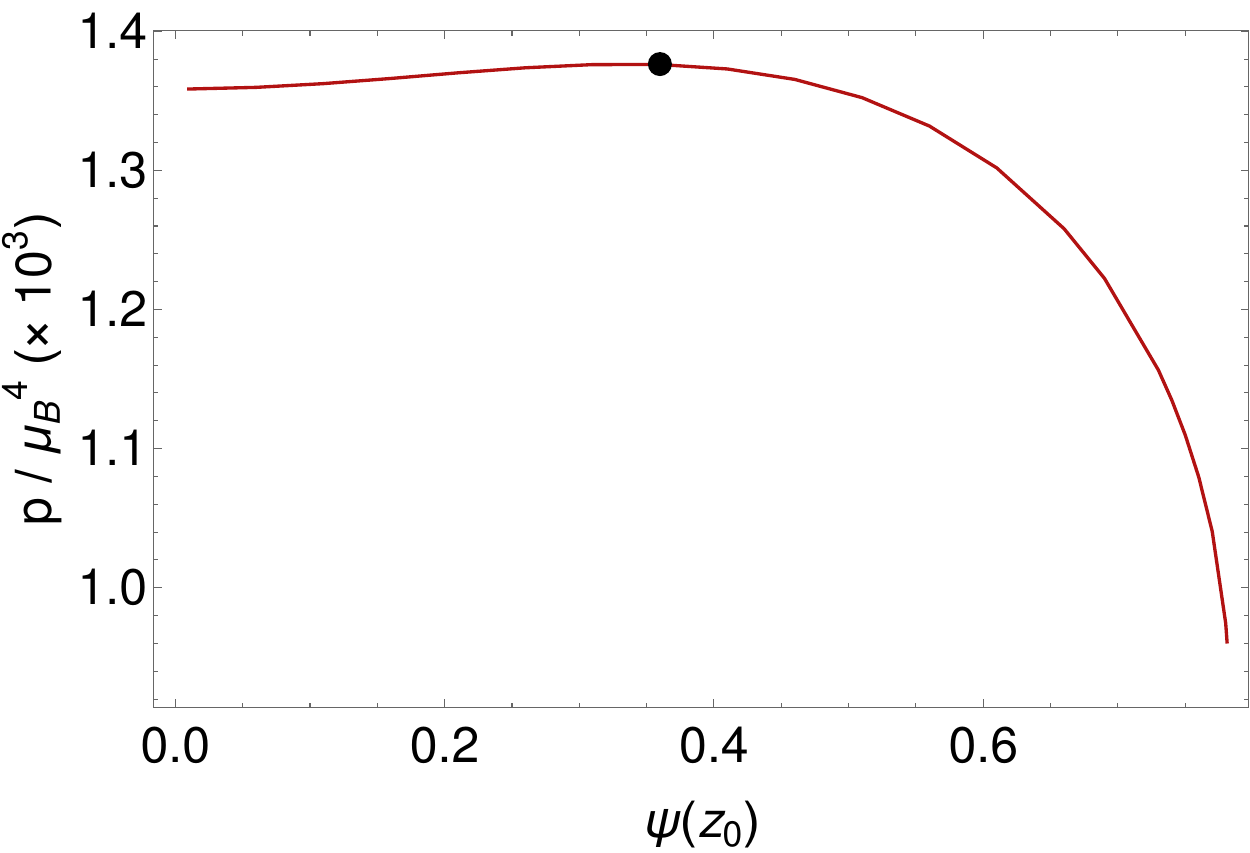}}\hfill
        \subfloat[Reduced baryon number density]{\includegraphics[width=0.3\linewidth]{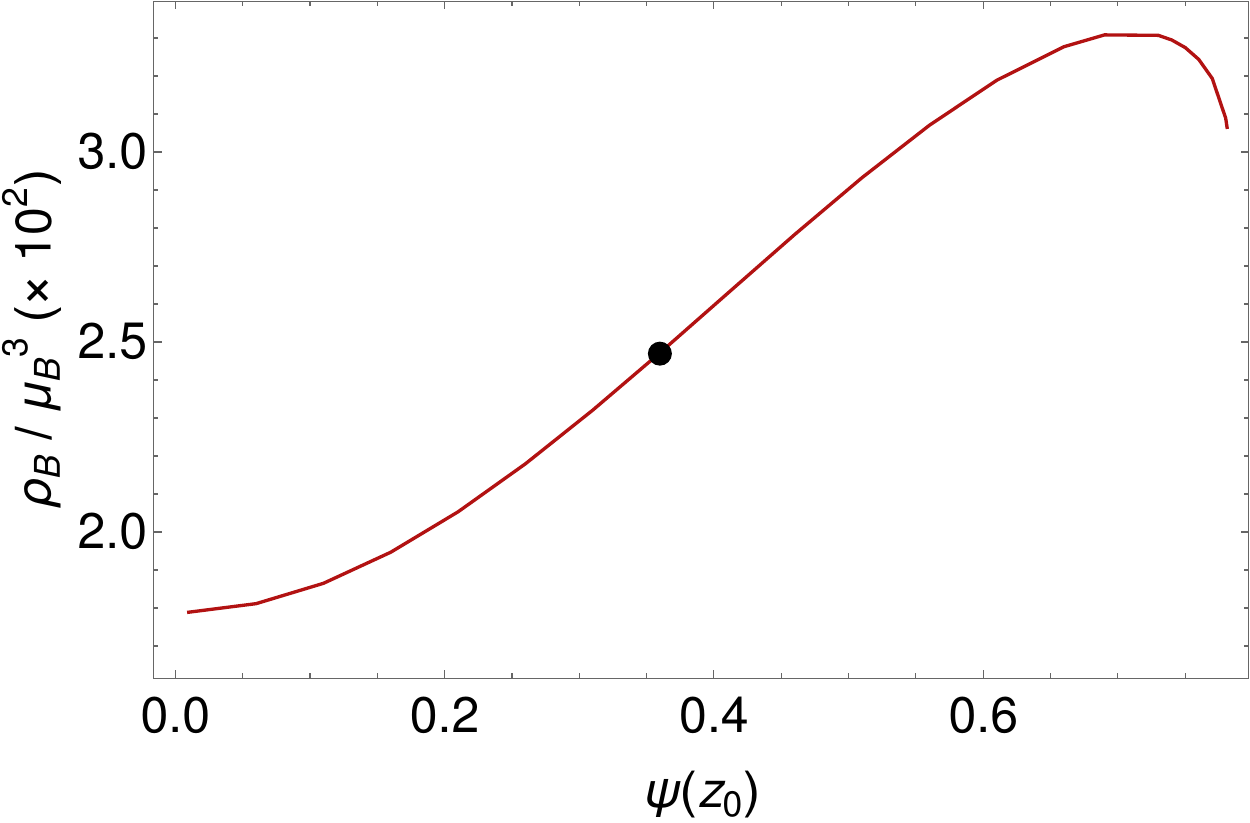}}\hfill
        \subfloat[Reduced Condensate]{\includegraphics[width=0.3\linewidth]{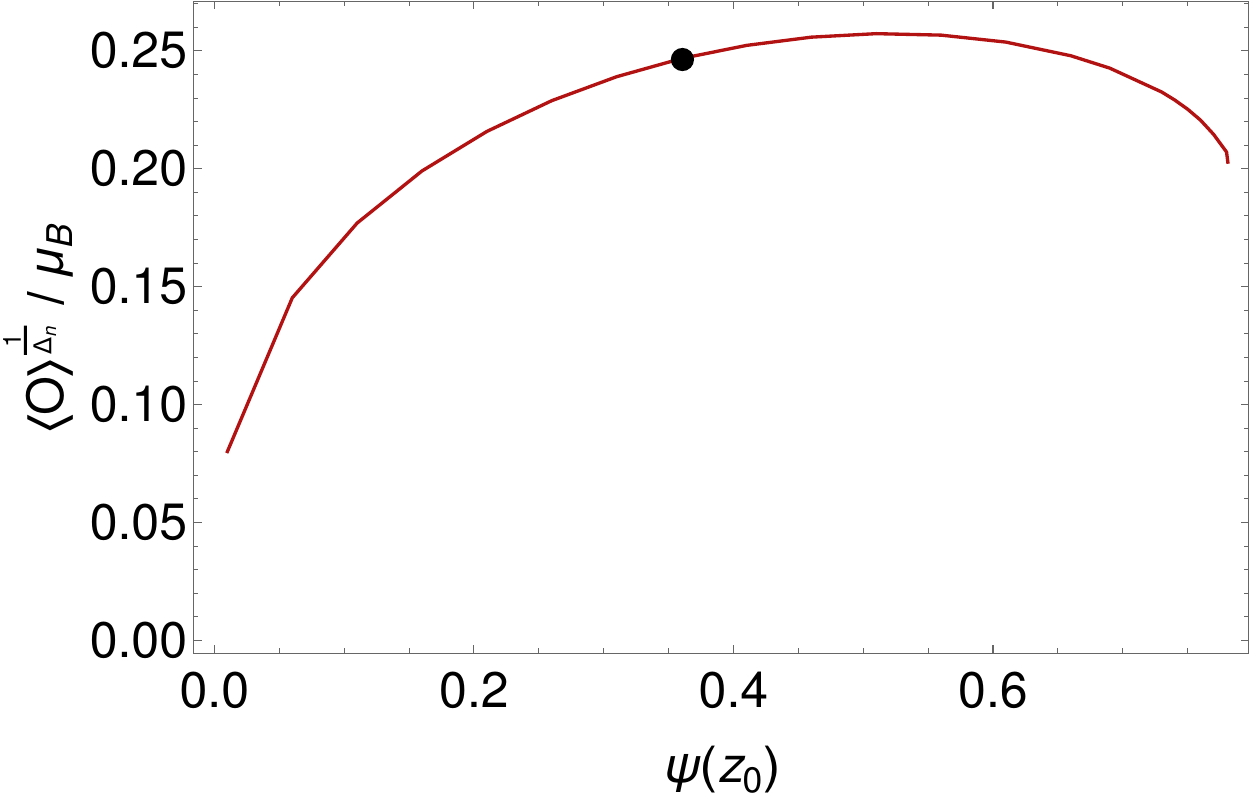}}
        \caption{Thermodynamic variables for $q=6$, $g_0=6$, $\Delta_n=3$, $\m=1566$ MeV. The minimum energy solution is represented by the black dot.}
        \label{NJLMaxP}
\end{figure}

These are shown in figure \ref{NJLProfiles} and clearly show that the solutions are well-behaved and under control. As before, the maximum pressure solution occurs at an intermediate value of the initial condition $\psi_0$ as shown in figure \ref{NJLMaxP}. The maximum pressure solution once again approaches the last solution as $g_{tt}$ approaches 0 for large values of $m_B$.
\subsection{Thermodynamic properties}
We first discuss the thermodynamics of the hard wall framework.
The baryon density $\r_B$ and entropy $s$ are defined as 
\begin{equation}\label{thermoCons}
    \r_B=\frac{\partial p}{\partial \mu_B} \; ; \qquad s=\frac{\partial p}{\partial T}
\end{equation} 
As discussed in section \ref{Confined}, the pressure contribution from matter behind the IR cutoff is encapsulated in the boundary condition. Consequently, the total pressure, along with any derived quantities, should be separated into two components: the contribution of the boundary condition and the portion within the region defined by $\e < z < z_0$.
Thus, the total baryon number density can be expressed as $\rho = \rho_B + \rho_\y$, where $\rho_B$ is the number density derived from the NJL model, which is used in the boundary condition to determine $\phi'(z_0)$ as shown in \eqref{fprime}. We have verified that the sum matches with the total number density obtained from \eqref{thermoCons}. We will then use the Euler relation $\ve + p = \m_B \rho_B + sT$ to compute the energy density $\ve.$

We will now present the thermodynamic properties of these configurations as we vary the chemical potential and temperature.
\begin{figure}[h]
    \centering
    \subfloat[Reduced pressure]{\includegraphics[width=0.3\linewidth]{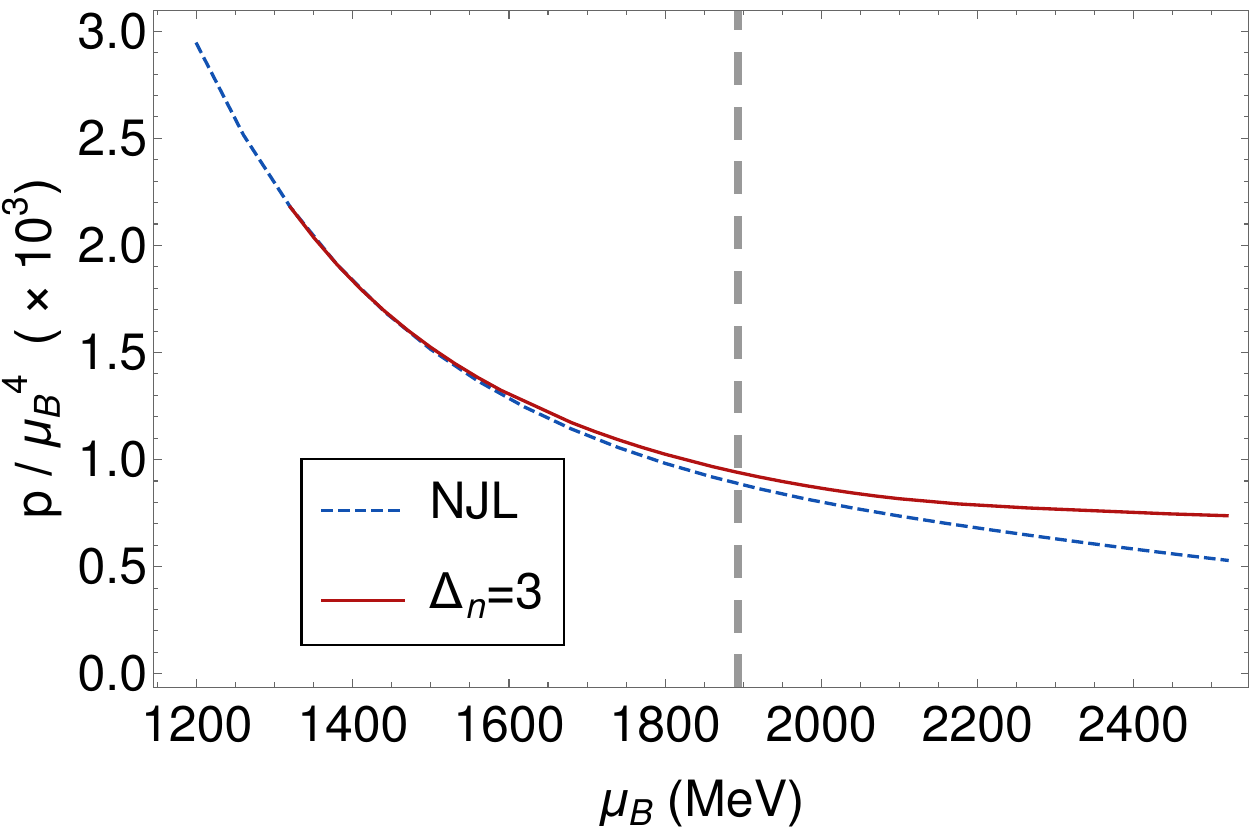}}\hfill
    \subfloat[Reduced Baryon number density]{\includegraphics[width=0.3\linewidth]{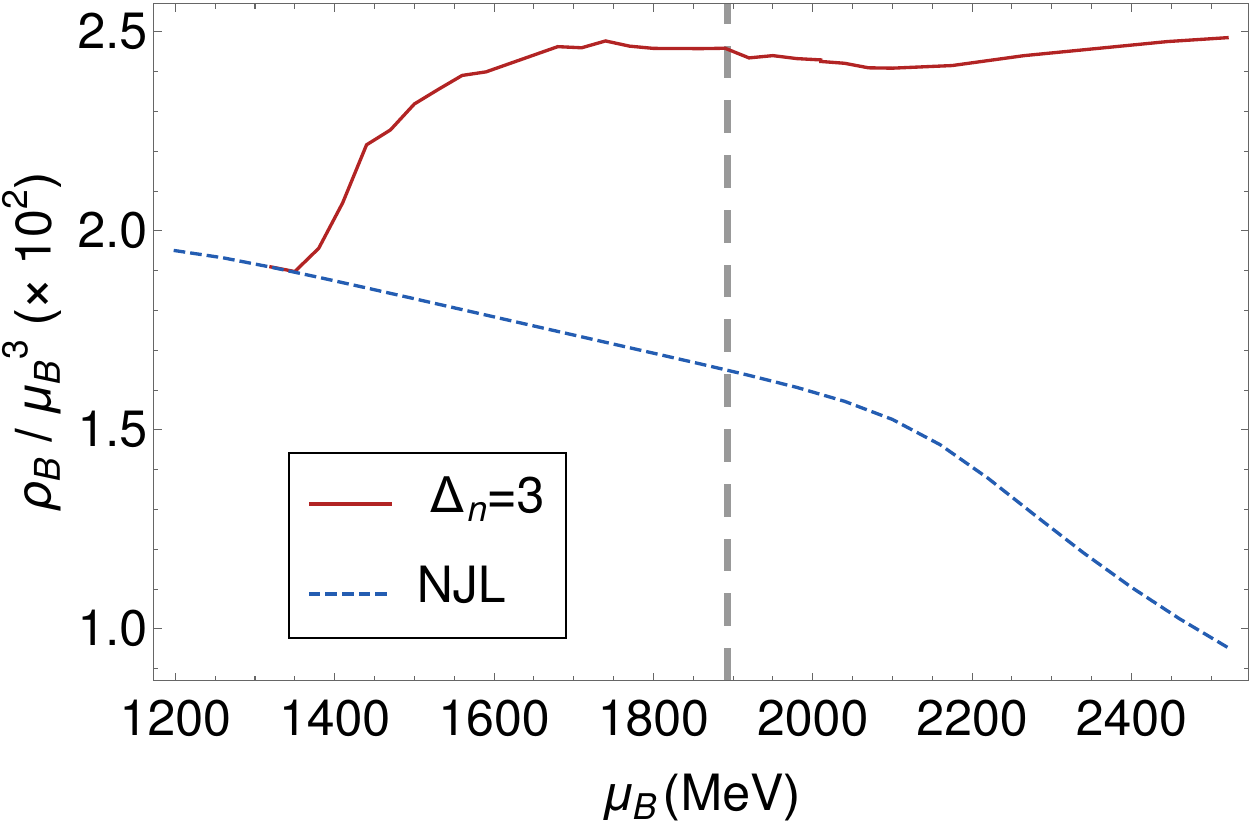}}\hfill
    \subfloat[Reduced Condensate]{\includegraphics[width=0.3\linewidth]{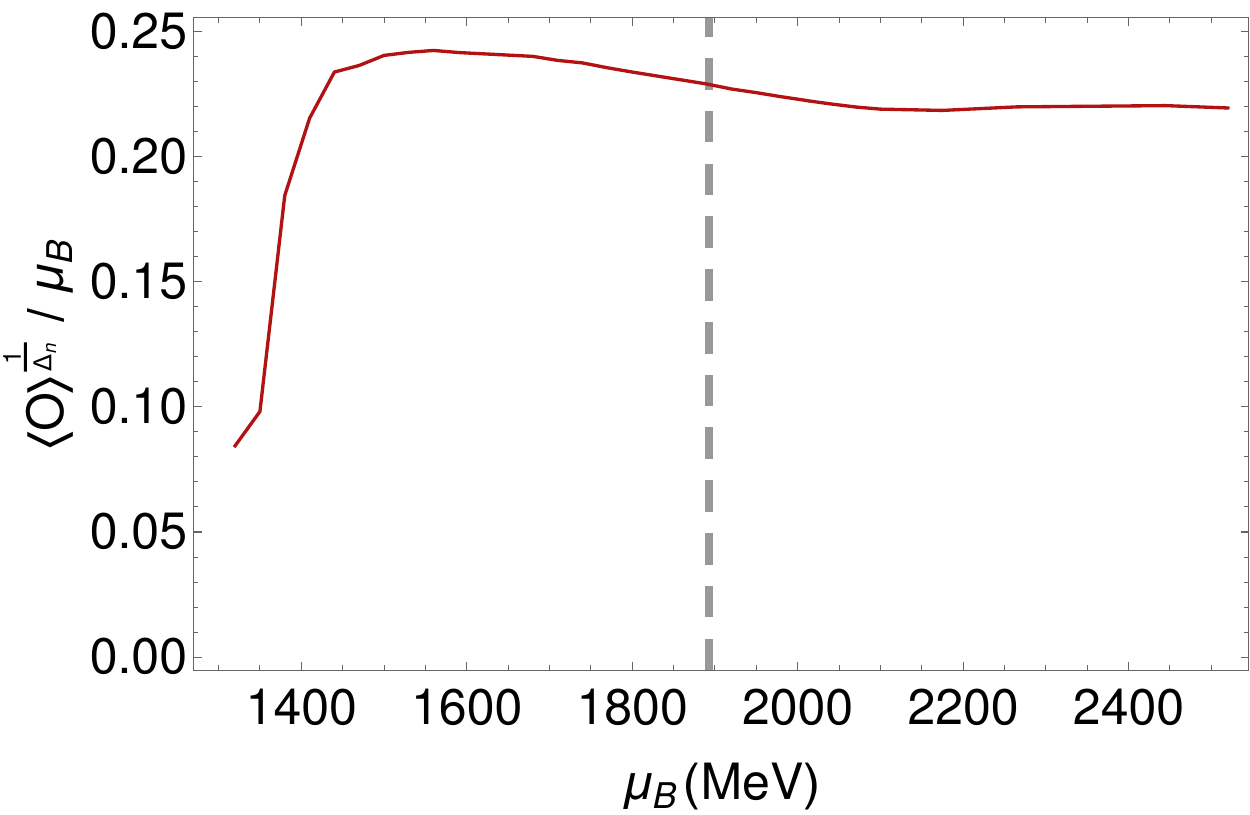}}
    \caption{Thermodynamics quantities as a function of $\m_B$ for $g_0=6,q=6$ and fixed $\z=0.77.$ }
    \label{fig:NJLC-TD}
\end{figure}
The first panel in figure \ref{fig:NJLC-TD} shows that condensation occurs at $\m_B\approx 1300$ MeV. A comparison of the pressure between the NJL model with and without condensate shows that condensate phases always have higher pressure. In particular, this indicates the absence of a first-order transition, which would have indicated that pairs are formed early on and condensation occurs later.  However, we know from our review of confined phases in section \ref{Confined} that the vdW to NJL transition will occur at around $\m_B\sim 1500$. Thus, we will not observe this second-order phase transition in the phase diagram at sufficiently low temperatures. The second panel shows the variation of the condensate and the total baryon density as a function of $\mu_B$
in comparison with the situation sans condensate. Concomitant with the increase in the condensate, the total baryon density increases sharply with condensation. The figure above is not reliable beyond $\m\approx 1900$ MeV (represented by the grey dashed vertical line) since we are close to the NJL cutoff $\L=631$ MeV. 

\begin{figure}[h]
    \centering
    \subfloat[Variation with $q$]{
    \includegraphics[width=0.45\linewidth]{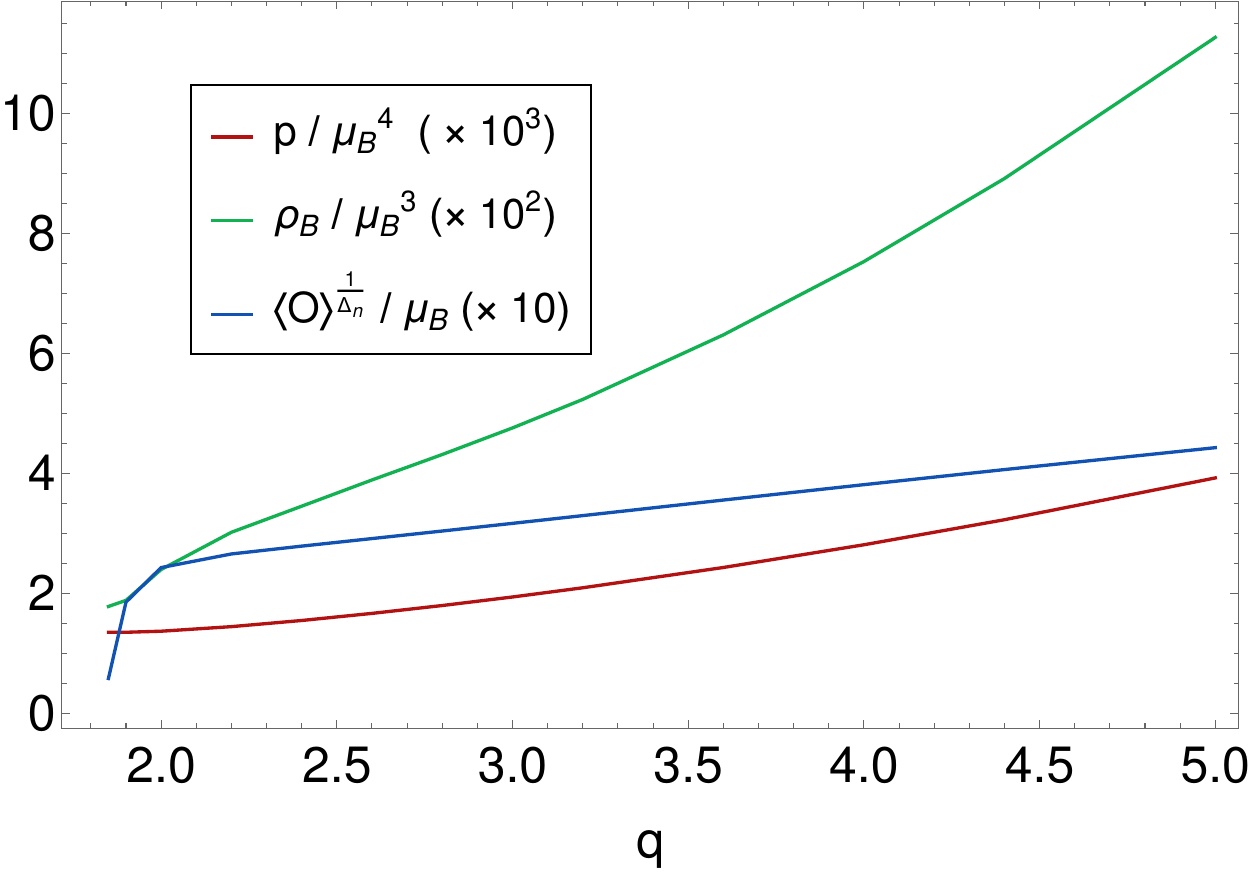}}\hfill
    \subfloat[Variation with $\D$]{
    \includegraphics[width=0.45\linewidth]{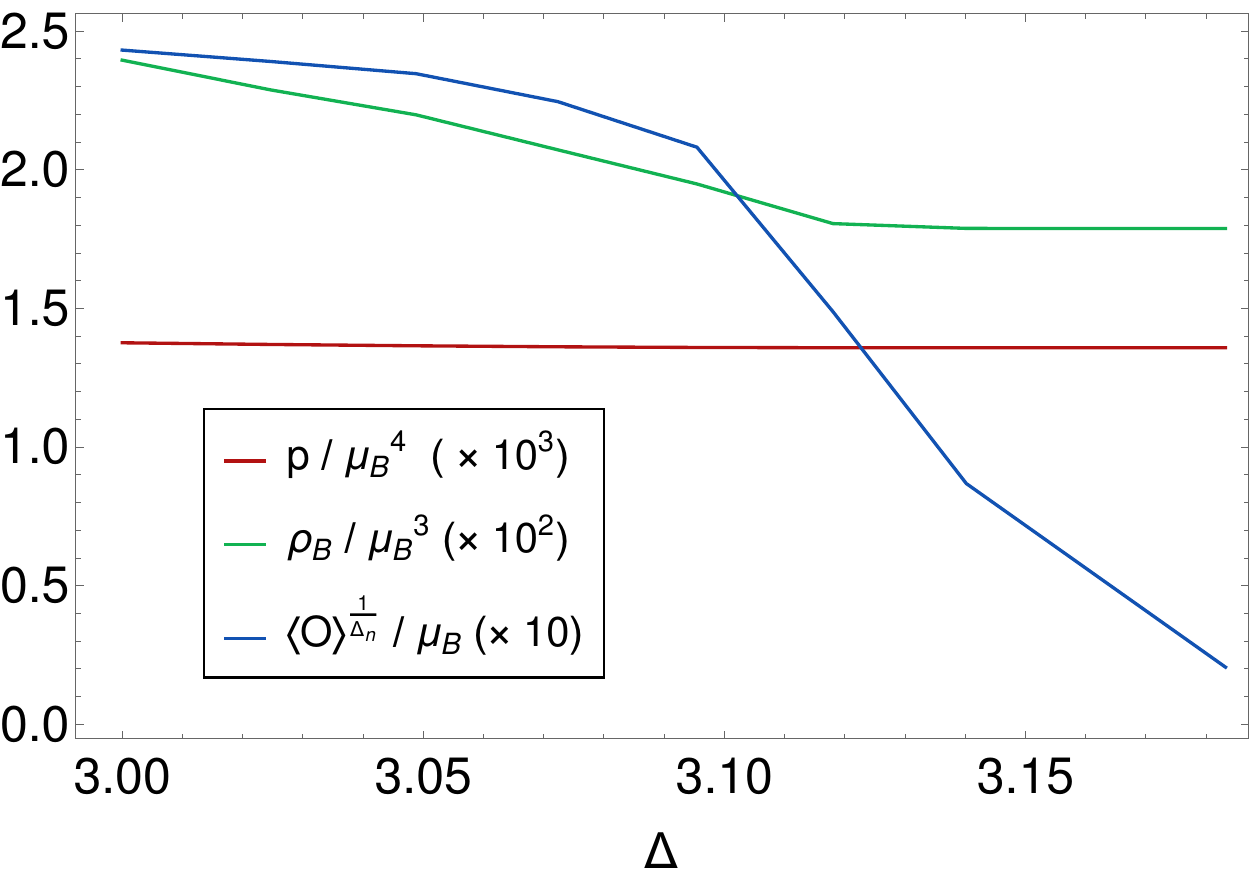}}
    \caption{Thermodynamics quantities for $g_0=6,\m_B= 1566$ MeV, $T=1$ MeV, and fixed $\z=0.77.$}
    \label{fig:NJLC_q}
\end{figure}
The charge and the scaling dimension for the NJL phase have already been discussed. However, it will be of interest to explore other systems that fall within the same universality class as QCD, but with slightly different scaling dimensions.  Therefore, we present a plot of thermodynamic quantities as a function of charge $q$ and scaling dimension $\D$ in figure \ref{fig:NJLC_q}. The quantities represented are reduced pressure (red curve), trace anomaly (green curve), reduced number density (blue curve), and condensate (magenta curve). These quantities are computed for $g_0=6,\m_B= 1566$ MeV, $T=1$ MeV, and fixed $\z=0.77.$
From this figure, we conclude that for a given value of the chemical potential and temperature, the operator with the largest charge will condense first. Condensation is more challenging for operators with larger scaling dimensions, which aligns with observations made in Van der Waals phases. Additionally, other thermodynamic quantities do not show significant variation with respect to $\D$.

\begin{figure}[h]
    \centering
    \subfloat[Reduced pressure]{\includegraphics[width=0.3\linewidth]{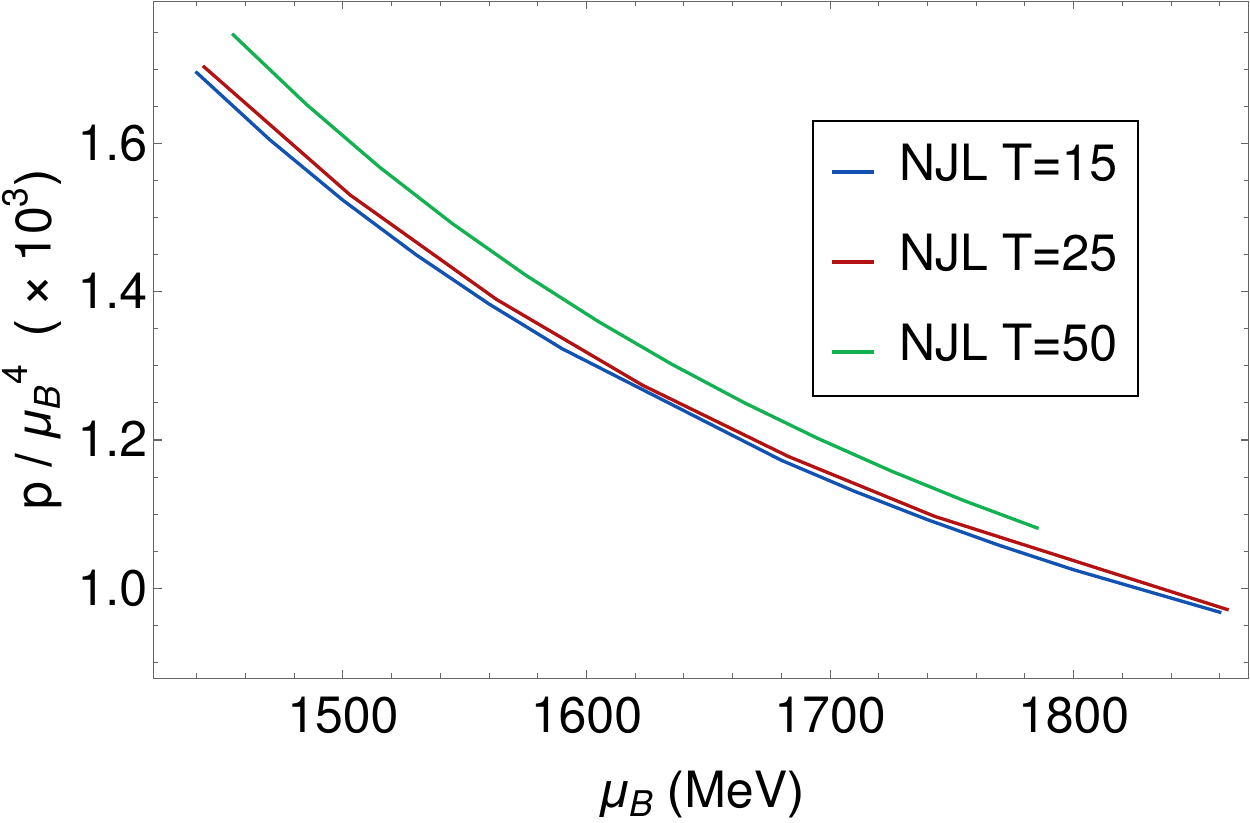}}\hfill
    \subfloat[Reduced Baryon number density]{\includegraphics[width=0.3\linewidth]{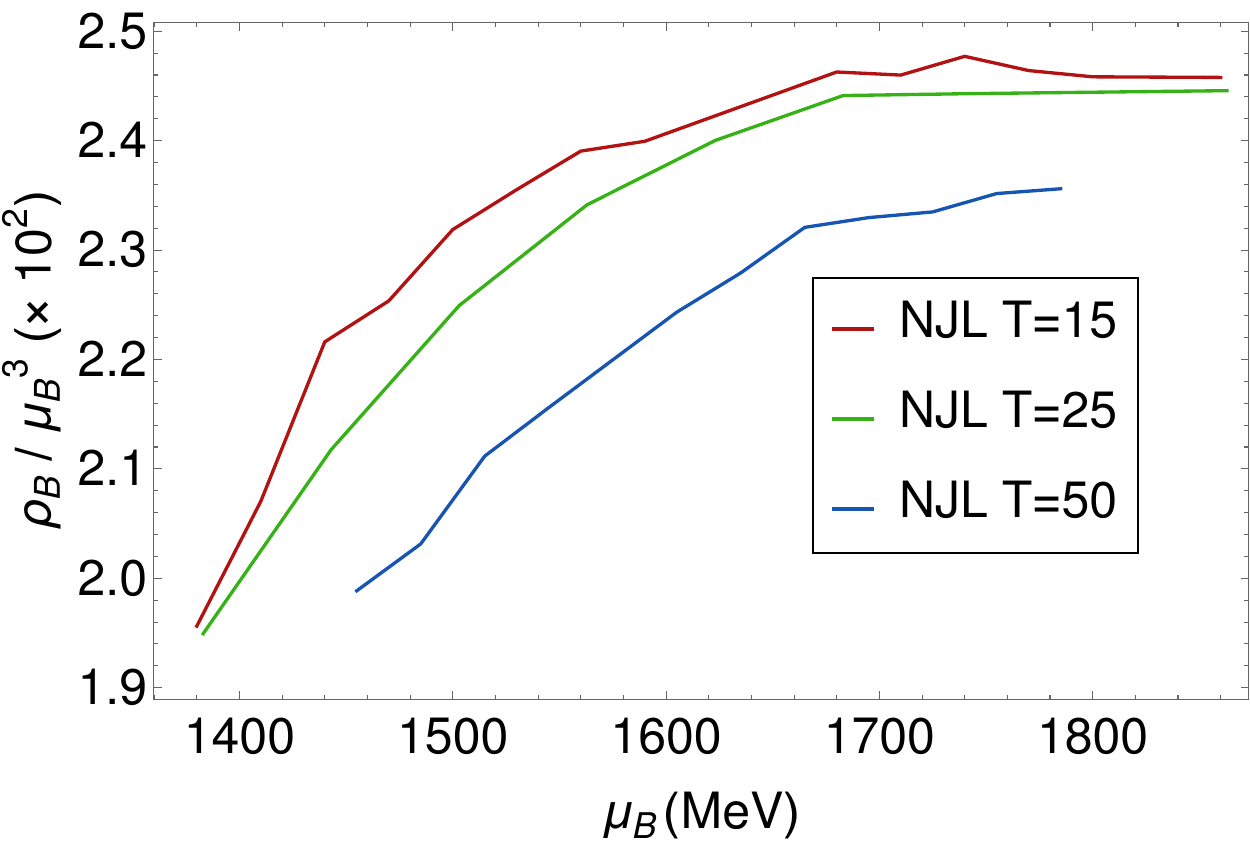}}\hfill
    \subfloat[Reduced Condensate]{\includegraphics[width=0.3\linewidth]{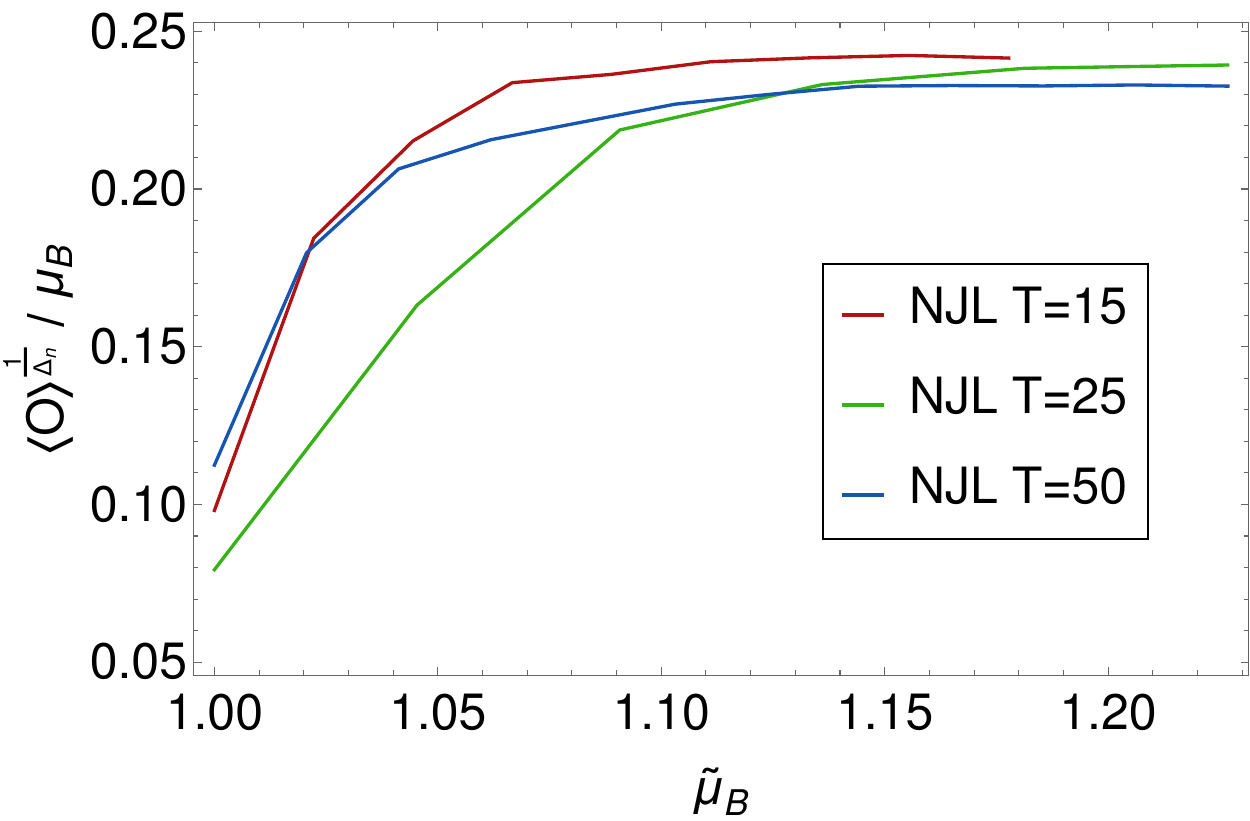}}
    \caption{Thermodynamics quantities as a function of $\m_B$ for $g_0=6,q=6$ and fixed $\z=0.77.$}
    \label{fig:NJLC-TDwithT}
\end{figure}

At finite temperatures, the dependence is shown in figure \ref{fig:NJLC-TDwithT}. The scenario is quite similar to that of a condensate in the Van der Waals phase at finite temperature. The solid red, blue, and green curves represent the thermodynamic variables for $T= 15, 25$, and $50$ MeV, respectively. As the temperature increases, the pressure increases while the baryon number density decreases as might be expected.
\begin{figure}[h]
    \centering
    \subfloat[Reduced pressure]{\includegraphics[width=0.3\linewidth]{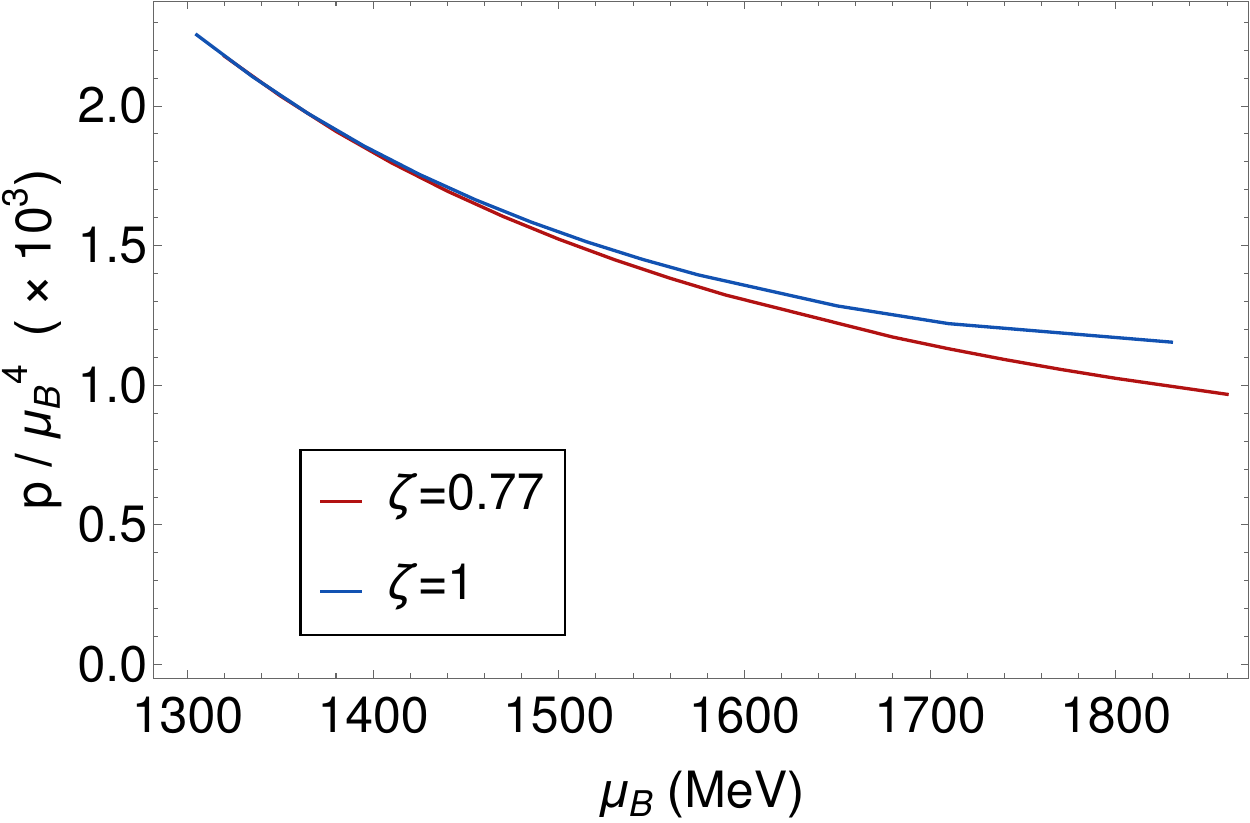}}\hfill
    \subfloat[Reduced Baryon number density]{\includegraphics[width=0.3\linewidth]{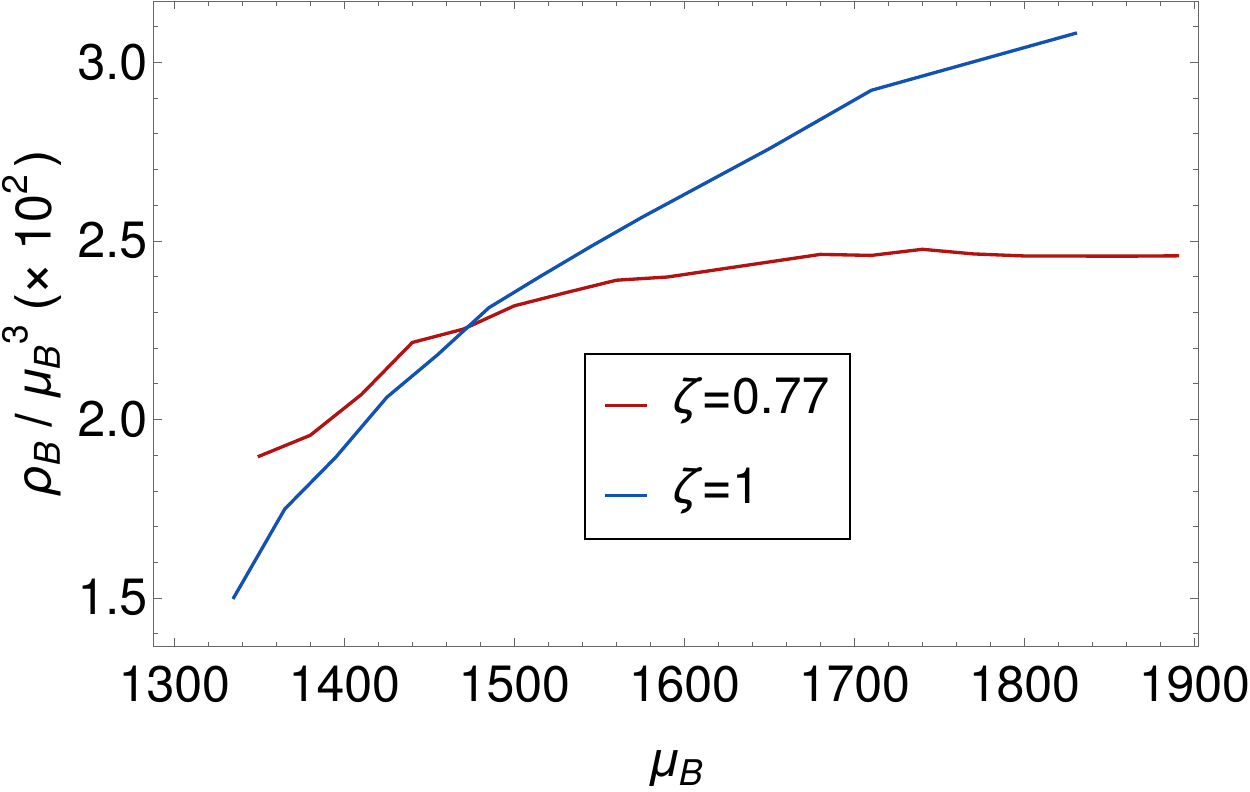}}\hfill
    \subfloat[Reduced Condensate]{\includegraphics[width=0.3\linewidth]{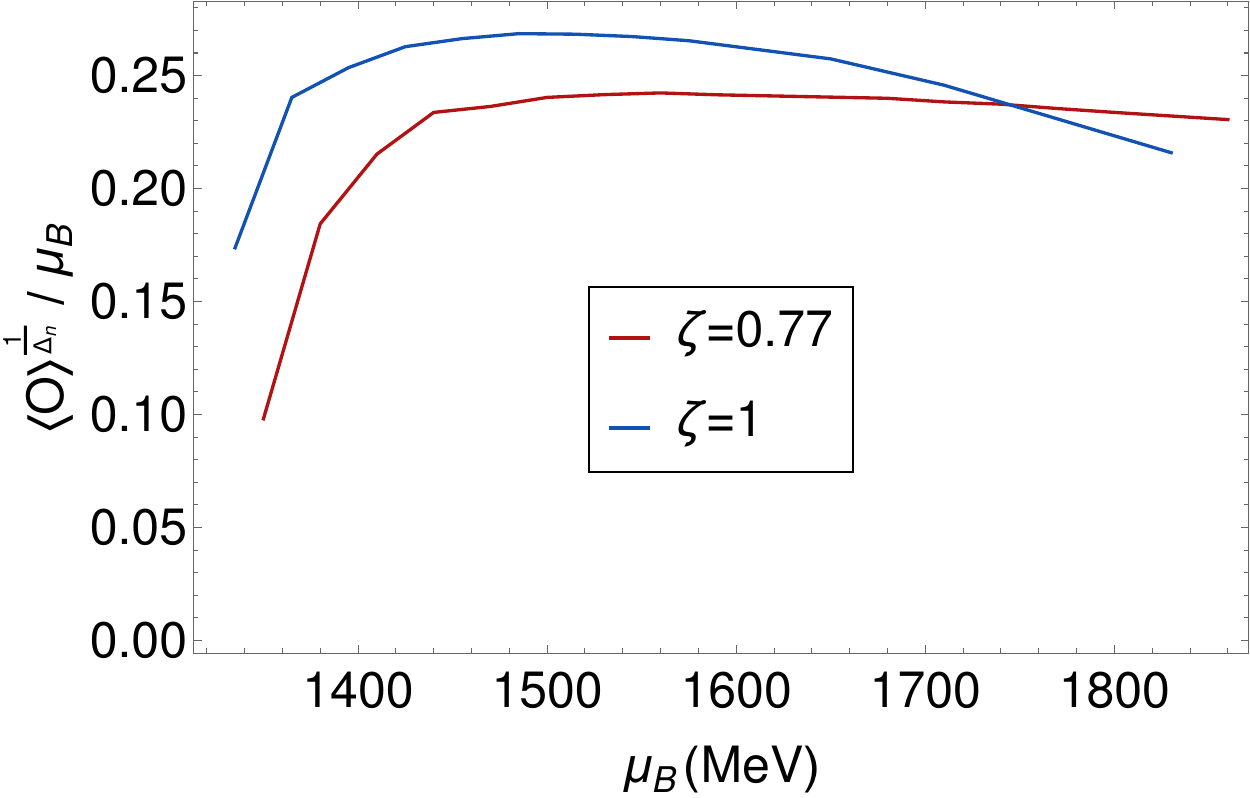}}
    \caption{Thermodynamics quantities as a function of $\m_B$ for $g_0=6,q=2, \D=3$ and different $\z$}
    \label{fig:NJLC-TD-zeta}
\end{figure}
We plot the variation of thermodynamic quantities as a function of $\m_B$ for two different values of $\z$ in figure \ref{fig:NJLC-TD-zeta}. As expected from our previous analysis with $\f_0=0$, the pressure and number density increase for larger $\z$ values. The onset of condensation occurs later for smaller $\z$.  These plots are computed for $g_0=6, q=2, \D=3$ and $T=15$ MeV.

\section{Phase diagrams}
In this section, we will construct the phase diagram of the boundary theory in the $\m_B -T$ plane, now including the condensate phases. The phase boundaries are obtained by comparing the pressures of the various solutions.

\subsection{Phase transition with NJL}
\begin{figure}[h]
    \centering
    \subfloat[$T=1$ MeV]{
    \includegraphics[width=0.45\linewidth]{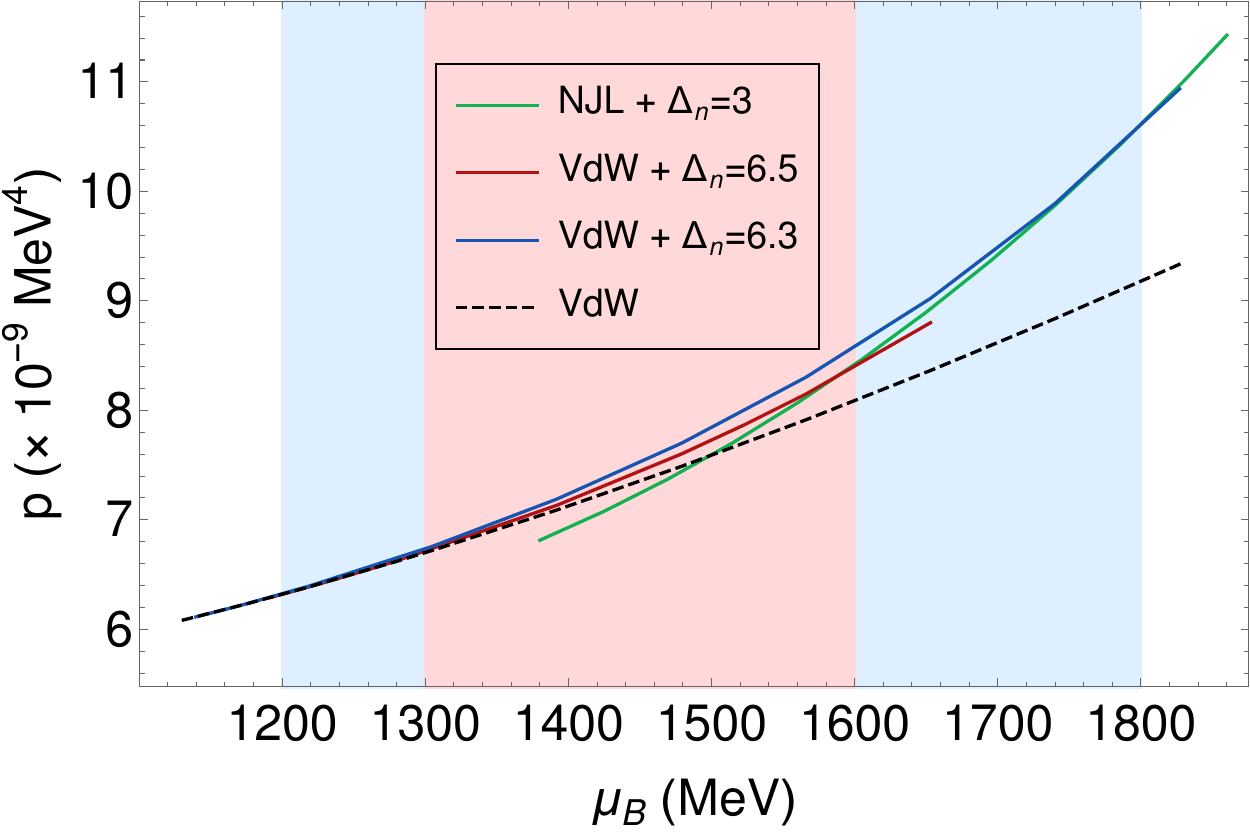}}\hfill
    \subfloat[$T=15$ MeV]{
    \includegraphics[width=0.45\linewidth]{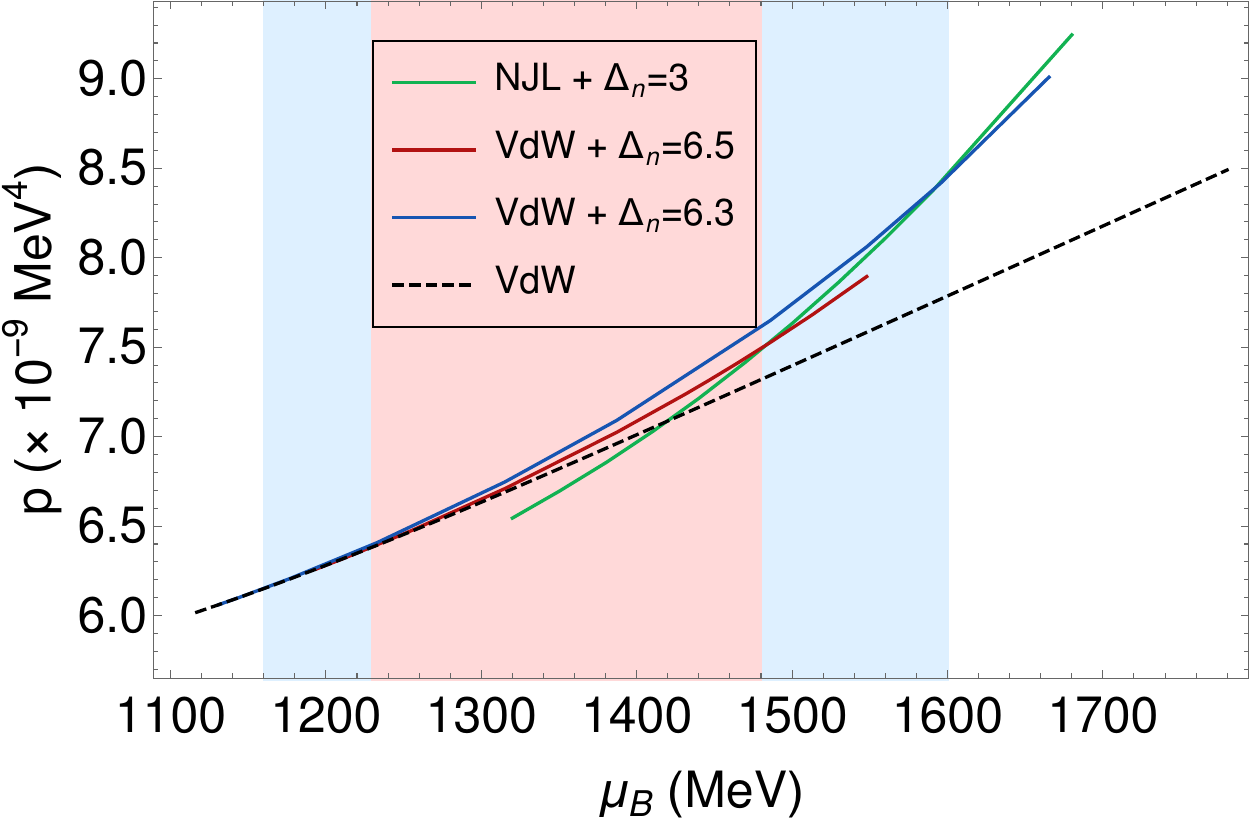}}
    \caption{The phase transition between vdW and NJL}
    \label{fig:PT_VDW_NJL}
\end{figure}

The figure \ref{fig:PT_VDW_NJL} shows a comparison of the pressure obtained with NJL boundary conditions and vdW boundary conditions.
As we have discussed in section \ref{sec-vdw}, the onset of condensation depends on the value of the scaling dimension. Therefore, we compare the vdW condensate for two values of $\D$. We see in figure \ref{fig:PT_VDW_NJL} that the vdW condensate for $\D=6.3$ (blue curve) will have higher pressure above $\m_B\sim1200$ MeV but the NJL with condensate (green curve) has even higher pressure at the slightly larger value $\m_B>1800$ MeV. For $\D=6.5$, the NJL description is preferred above $\mu_B>1580$ MeV. 

The blue-shaded region in the figure indicates the range where the vdW condensate ($\D=6.3$) is preferred compared to the vdW liquid and NJL phase. Notably, when we increase the value of $\D$ to $6.5$ (shown by the red curve), this preferred region shrinks, as represented by the red-shaded area. For $\D \gtrsim 7$, this region will completely disappear. Conversely, if we decrease the value of $\D$, the preferred region expands, and the vdW condensate will dominate the entire low-temperature phase.

In the second panel, at a slightly higher temperature $T=15$ MeV, the preferred region becomes smaller. Eventually, at a critical temperature, it will shrink to a point. This behavior is seen in the complete phase diagram that we will discuss further below.

\subsection{Phase transition with CBH phase}

\begin{figure}[h]
    \centering
    \subfloat[$\zeta=0.77, T=1$ MeV]{
    \includegraphics[width=0.45\linewidth]{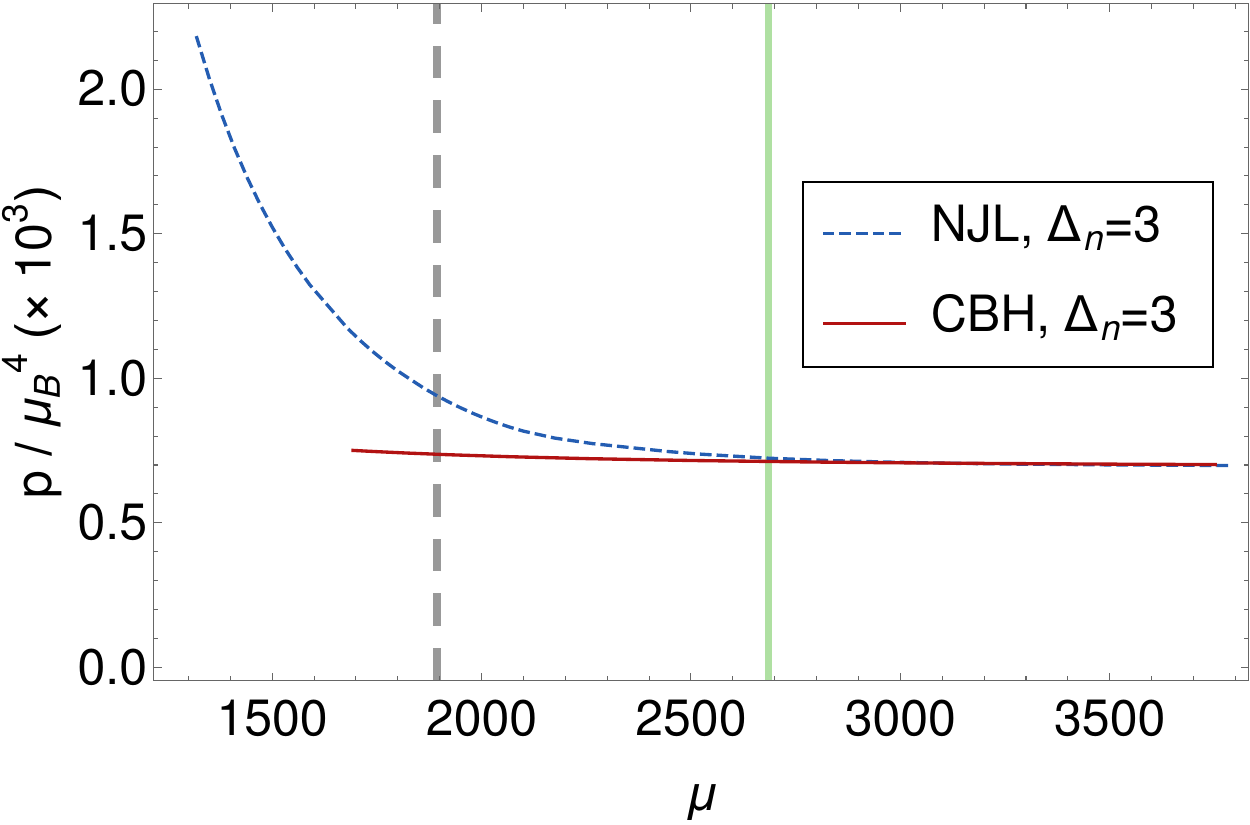}}\hfill
    \subfloat[$\zeta=1,T=15$ MeV]{
     \includegraphics[width=0.45\linewidth]{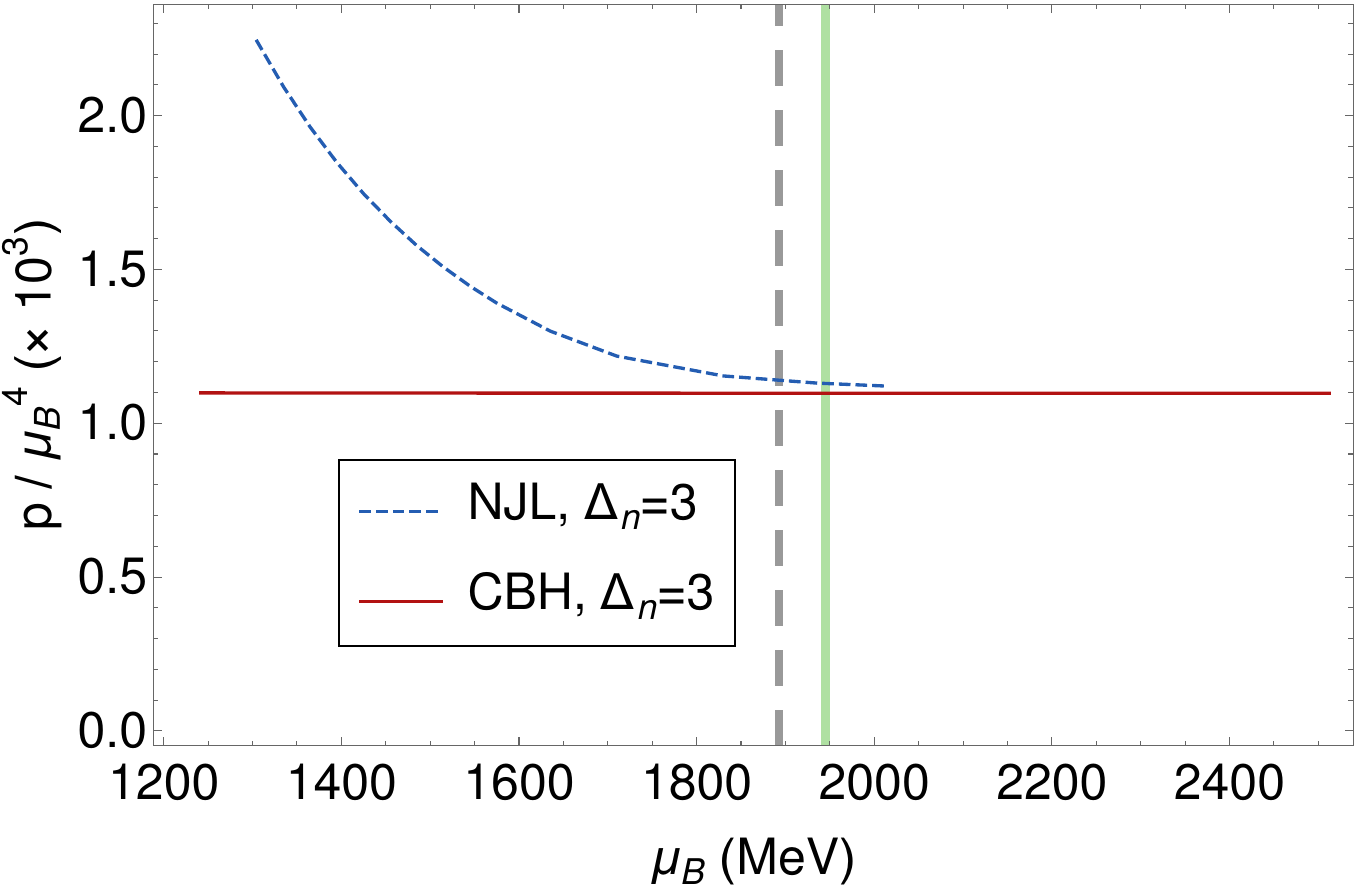}}
    \caption{Reduced pressure as a function of baryon chemical potential. The charged black hole with the scalar is shown by a red curve, and the NJL with the scalar field is shown by a blue dotted curve. The grey dashed curve represents a region of validity for the NJL equations of state. }
    \label{fig:PT_NJL_CBH}
\end{figure}

\begin{figure}[h]
    \centering
    \subfloat[]{\includegraphics[width=0.45\linewidth]{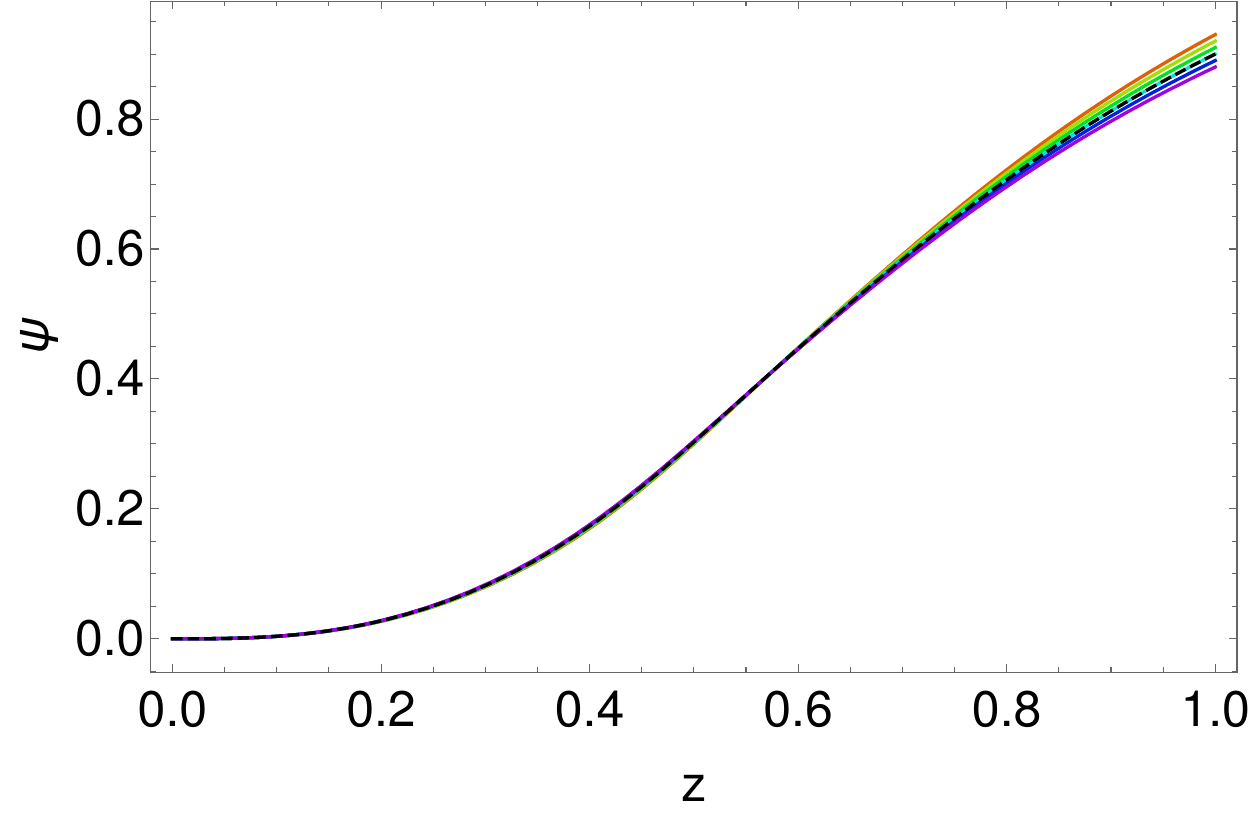}}\hfill
    \subfloat[]{\includegraphics[width=0.45\linewidth]{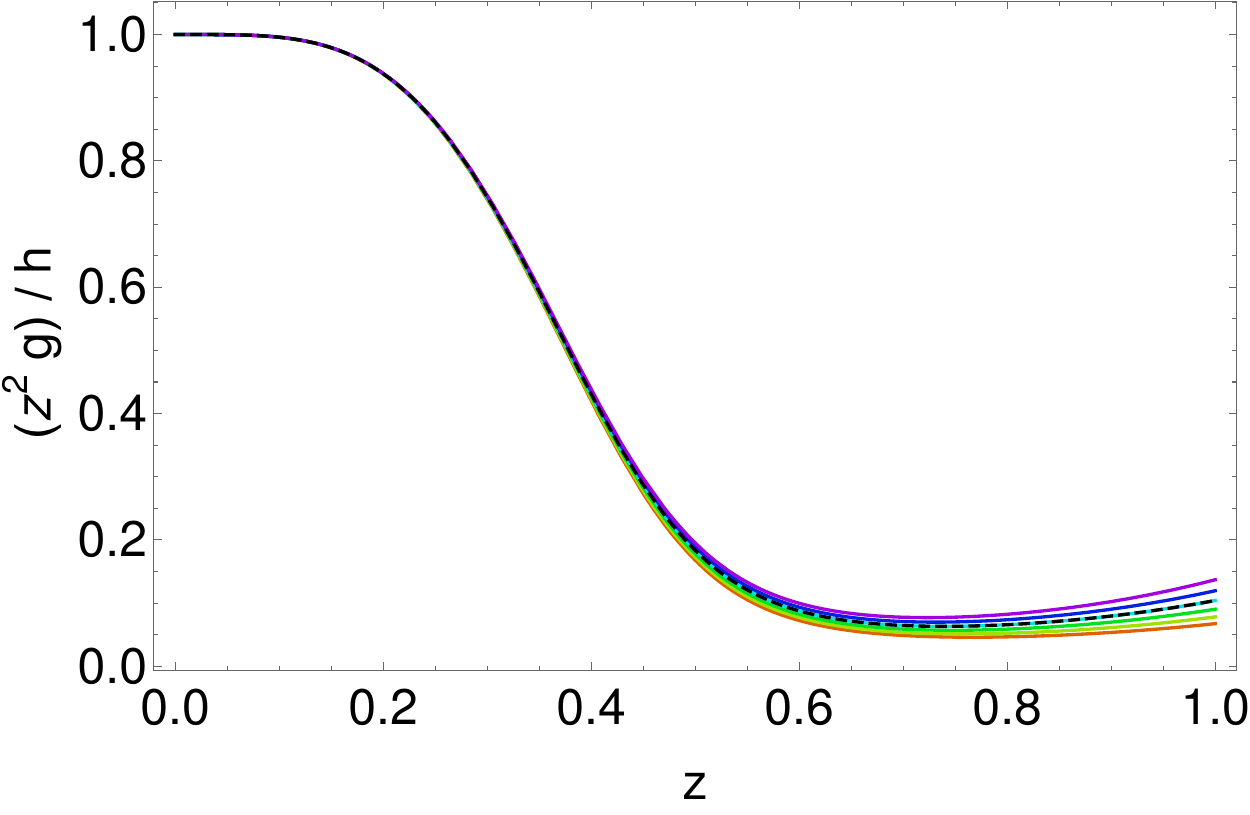}}
    \caption{Solutions for $\m_B=2685$ MeV. }
    \label{fig:sol2}
\end{figure}

The previous section demonstrated that the NJL condensate phase exhibits a higher pressure at increased chemical potentials. However, we have another possible phase described by a condensate in charged black hole geometry. Comparing the pressure of these phases, in figure \ref{fig:PT_NJL_CBH}
we see that the NJL model seems to evolve smoothly to the charged black hole at high densities. The solid green line marks the transition chemical potential $\m_B=2685$ MeV at $T=1$ MeV and $\z=0.77$. The solutions at this chemical potential are shown in \ref{fig:sol2}. It is evident from the figure that the maximum pressure solution (represented by the black dotted curve) is developing a singularity near the IR cutoff $z_0$ as $g_{tt}\to0.$

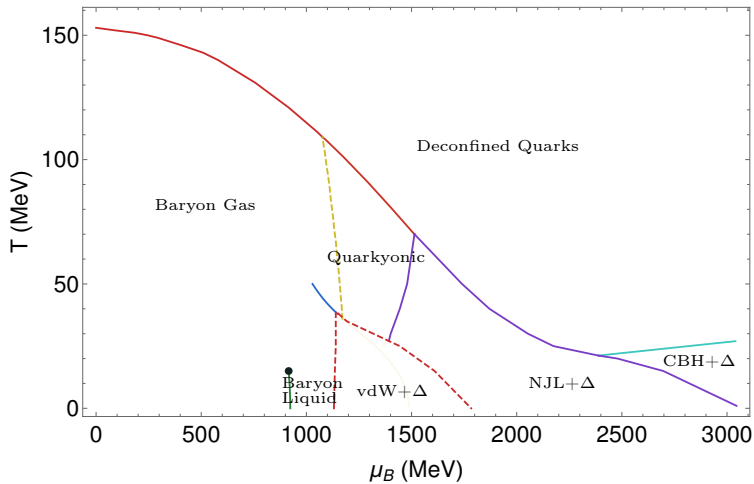
\begin{figure}[h]
\centering
\begin{overpic}[width=0.65\linewidth, unit=1mm]
{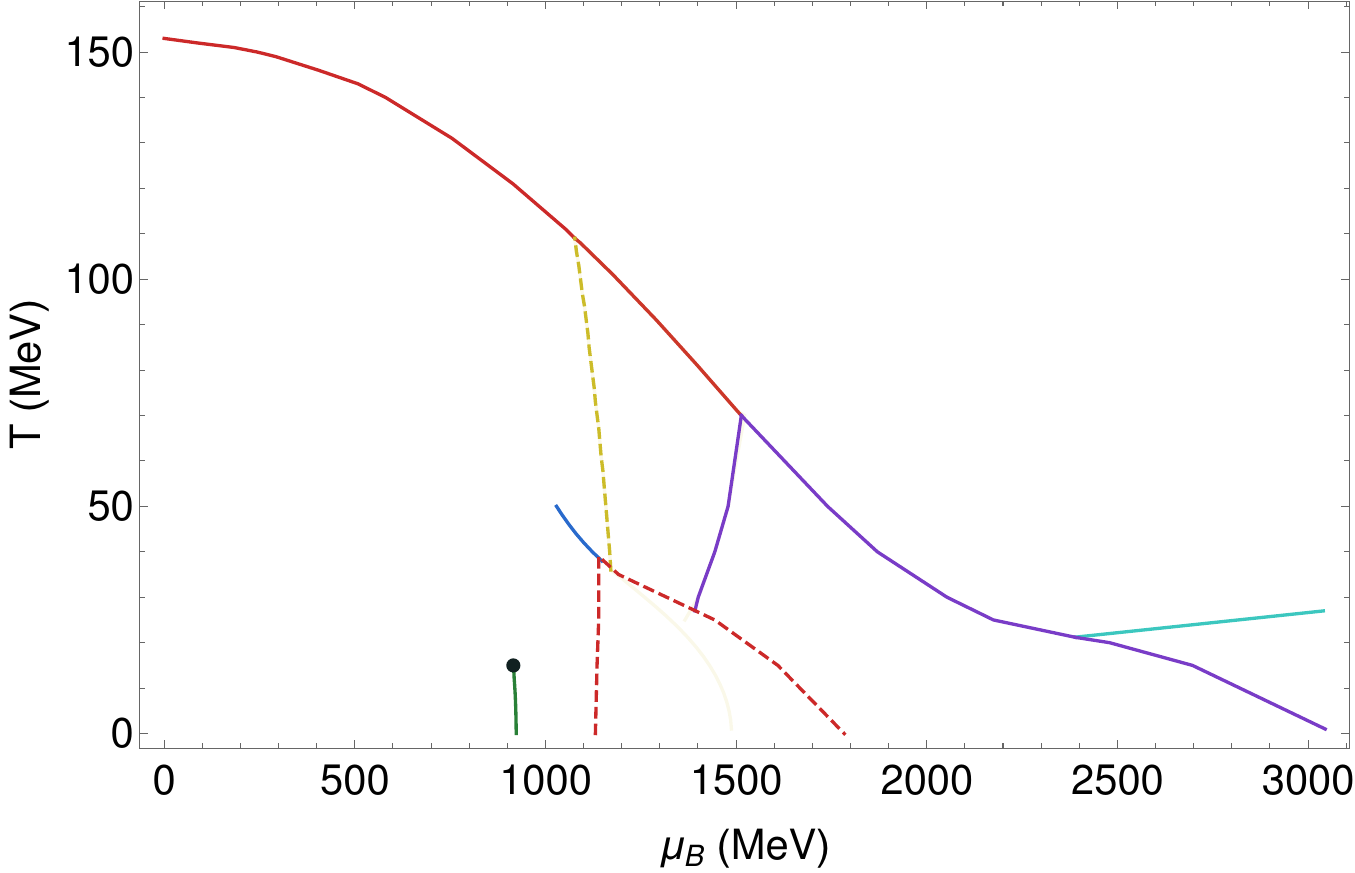}
\tiny
\put(20,37){Baryon Gas}
\put(37,13){Baryon}
\put(37,11){Liquid}
\put(43,30){Quarkyonic}
\put(47,12){vdW+$\D$}
\put(70,13){NJL+$\D$}
\put(55,45){Deconfined Quarks}
\put(88,16){CBH+$\D$}
\end{overpic}
    \caption{Phase Diagram}
    \label{fig:PD_Pheno}
\end{figure}

We now summarise these observations in the form of a phase diagram shown in figure \ref{fig:PD_Pheno} where the phases have been labeled. We will first describe these phases and then discuss the effects of various parameters and their validity in subsequent figures. Built on the phase diagram of \cite{Singh:2024amm} which has baryon gas, baryon liquid, Quarkyonic and deconfined quarks as phases, we observe that there are condensates at low temperatures. 

At intermediate chemical potential, we identify baryonic condensates in the van der Waals (vdW) phase with charge $q = 6$ and scaling dimension $5 \gtrsim \D \gtrsim 7$. These condensates are located in the region enclosed by the red dashed lines on the phase diagram. The onset of this phase is characterized by a second-order transition. As we increase the chemical potential, we find that the vdW condensate undergoes a first-order phase transition to the NJL condensate phase, marked by the region enclosed by the purple curves. The NJL condensate is associated with a charge $q = 2$ and a scaling dimension of $\D= 3$. Once again, the onset of this transition is a second-order transition. For sufficiently high chemical potential, we observe that the charged black hole condensate becomes favored over the NJL condensate. This represents another second-order phase transition.

\begin{figure}[h]
    \centering
    \subfloat[vdw $\D=6.3$]{\includegraphics[width=0.45\linewidth]{Paper3_FIG/PD_D6p3_zeta0p77.pdf}}\hfill
    \subfloat[vdw $\D=6.5$]{\includegraphics[width=0.45\linewidth]{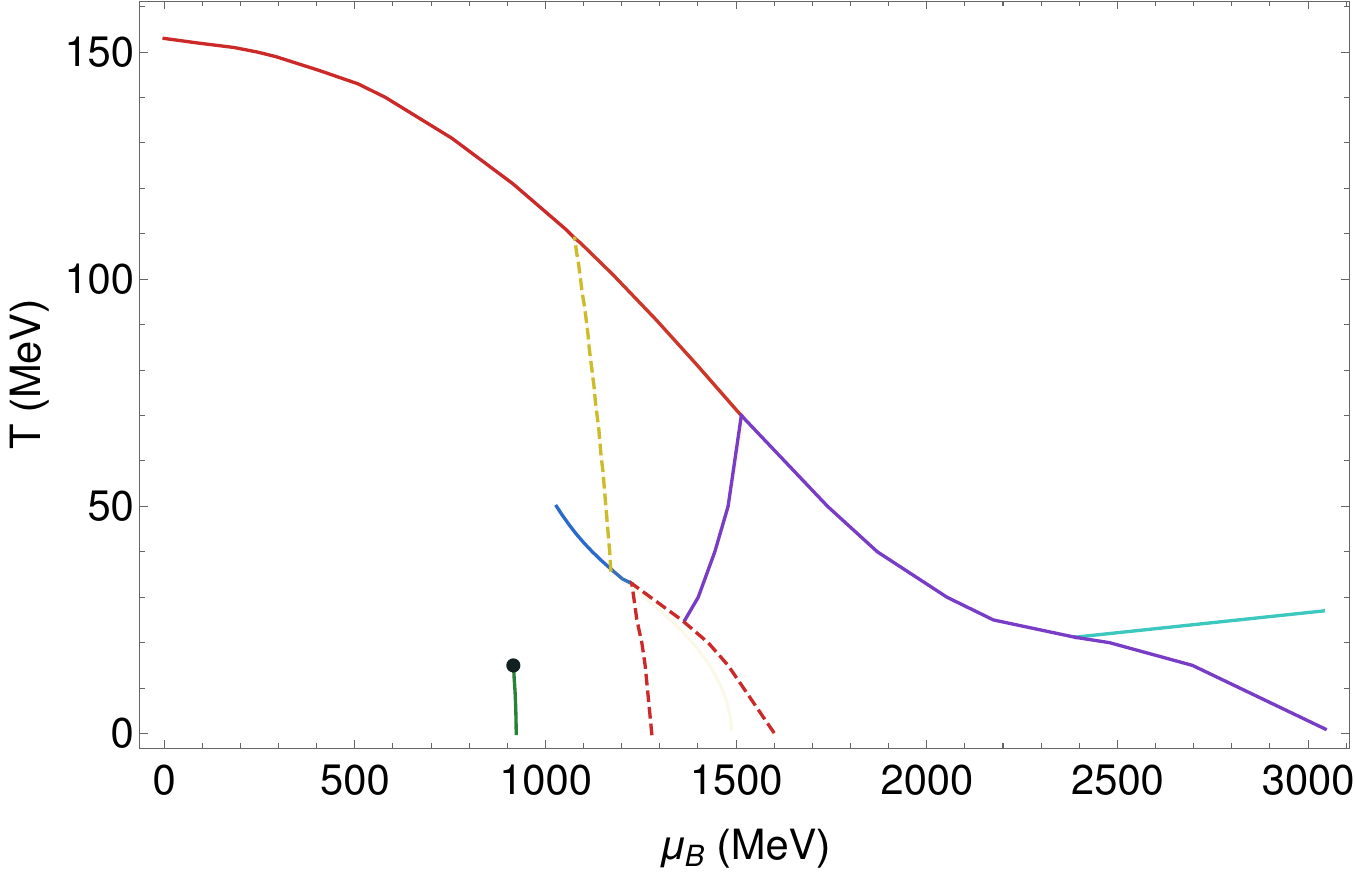}}
    \caption{Phase diagram with condensate, $\z=0.77$. }
    \label{fig:PD_zeta0p7}
\end{figure}

The phase diagram changes significantly with the parameters of the model. We plot these effects for two values of the scaling dimension in the vdW phase in figure \ref{fig:PD_zeta0p7}. It can be seen from the figure that there is an upper limit to $\D$ above which the vdW condensate disappears. 

\begin{figure}[h]
    \centering
    \subfloat[$\z=1$]
   { \includegraphics[width=0.45\linewidth]{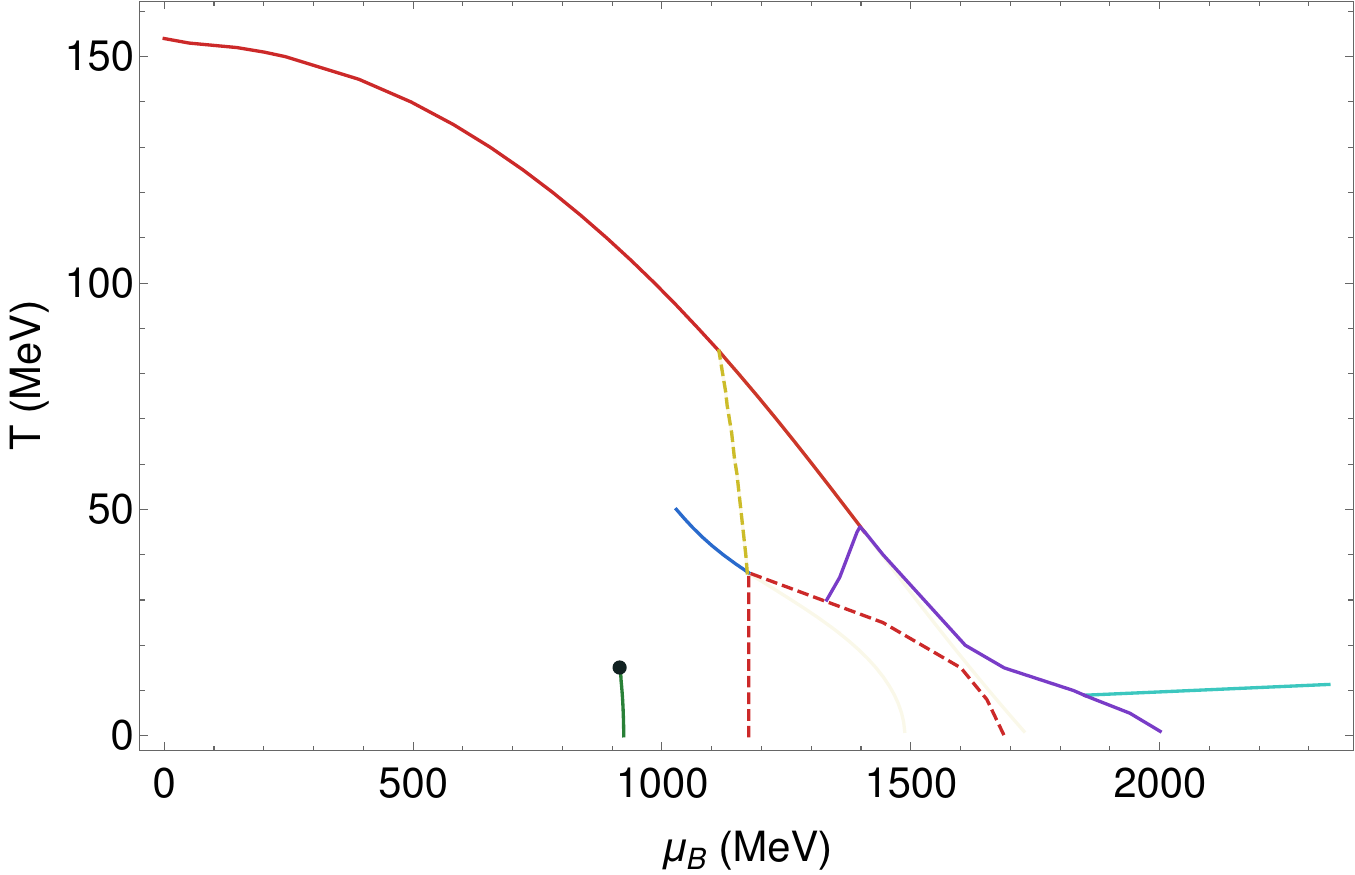}}\hfill
    \subfloat[$\z=0.77$]{
     \includegraphics[width=0.45\linewidth]{Paper3_FIG/PD_D6p5_zeta0p77.pdf}}
    \caption{Phase diagram with condensate, $\D=6.5$.}
    \label{fig:PD_zeta1}
\end{figure}

In figure \ref{fig:PD_zeta1}, we illustrate the effect of $\z$ on the phase diagram.  The condensates in both NJL and CBH are less favored as we increase $\z$. Whereas, the vdW condensates acquire a larger region on the phase diagram. 
We must remind the reader that for the smaller value of $\z$, we have seen that the transition to the black hole phase is not very reliable at very low temperatures (see figure \ref{fig:PT_NJL_CBH}). However, we do expect that the results are not qualitatively changed by an improved treatment by continuity with the slightly larger value of $\z$ and continuity of the phase boundary as a function of $\m_B, T$.

\section{Discussion}
We will present a summary of our results and an outlook towards the determination of the equation of state of nuclear matter at high densities. 

In this work, we have indicated the possibility that nuclear matter at high densities, either in the form of baryons or in the form of quarks, is unstable to condensation. We have analyzed the system using boundary conditions inspired by QCD phenomenology, as well as a simpler boundary condition where $\f_0=0$. The thermodynamically preferred description always involves condensates except possibly at very low densities. The maximum pressure condition for fixing the scalar field $\y$ at IR cutoff $z_0$ not only ensures that the onset of condensation is always a continuous second-order phase transition but also eliminates the possibility of offset of the condensate, which would be indicated by the appearance of a node in the scalar field profile. These conclusions seem to be robust over a range of the coupling $\z$, which controls the interaction between the matter and gravity in the bulk.

We have also explored the robustness of our conclusions by varying the scaling dimension of the operators that condense. In the nuclear phase forming a vdW liquid, an operator with $5\lesssim\D\lesssim7$ will condense.  Similarly, in the high-density (NJL or deconfined) phases, an operator with $\D=3$ condenses. A remarkable feature was that the NJL description seems to smoothly evolve to the deconfined quarks phase represented by the black hole.

We verify the thermodynamic consistency for the total baryon number density, given by \(\r_B = \frac{\partial p}{\partial \mu_B} = \Bar{\r} + \r_\y\). 
However, the entropy density, must be computed as \(s = \frac{\partial p}{\partial T}\)
. We may employ the Euler relation to compute the energy density, as the standard holographic method for obtaining the energy density from the boundary stress-energy tensor does not apply in the hard-wall framework.

We have obtained phase diagrams within the hardwall framework with both types of boundary conditions. The phase diagrams with $\f_0=0$ give useful insights for various parameters of the model, yet these are quite straightforward to obtain. Surprisingly, they resemble the phase diagrams already studied in the literature with other holographic models. The phenomenological boundary condition requires some effort, but it results in a satisfactory phase diagram.

In all these transitions, the density {\em increases} sharply after condensation. Similarly, the phase diagram has been shown to vary sharply with temperature. The phase diagram depends on the model parameter $\z$ as well as the scaling dimension of the condensing operator.

Even within the hardwall approach, our study does not explore all the aspects that are crucial for nuclear matter proper. For instance, we do not consider the effects of running coupling at all, and we did not include the field dual to the chiral condensate in the bulk. Additionally, we must study isospin chemical potentials, which will allow for yet another axis to the phase diagram. Some of these questions are clearly straightforward extensions that are likely to be of phenomenological interest.  

A key input to our study is the manner in which we incorporate the phenomenological equations of state in the hardwall model. This gives rise to two separate questions. Firstly, one can hope to study the fields sourced by a distribution of baryons in the AdS bulk. In earlier studies based on the Witten-Sakai-Sugimoto model, the baryons were represented by a dilute gas of instantons localized on D8-branes \cite{Kovensky:2021ddl}. It will be interesting to revisit these studies to determine an approximate truncated bulk solution which nevertheless treats the interactions among the instantons better. This could enable a better handling of the strong interactions among the baryons and avoid the questions arising from the choice of boundary conditions.

On the other hand, one can ask whether quantum gravity restricts the boundary conditions that can be applied to such truncated hardwall models. This might be natural by considering a path integral formulation of the bulk theory and focusing on gauge and diffeomorphism invariances and anomalies.

\appendix

\section{Holographic renormalization}\label{HRen}

To extract thermodynamic properties, we note that the bulk action in \eqref{action1} evaluates to
\begin{align}
    S_{\text{bulk}}&=\frac{1}{8\pi G_N}\int \frac{d}{dz} \left( \frac{1}{z} \sqrt{-g} g^{zz} \right) d^5x. \label{IFormS}
\end{align}
This form is particularly convenient for us because it expresses the on-shell action in terms of $g_0=g^{zz}(z_0).$

This action, however, yields divergences when evaluated on-shell, both from the gravity and matter components. To address this, we have implemented a holographic renormalization scheme \cite{deHaro:2000vlm}, which adds extra terms that we discuss below:

 \begin{align}
    S_{ren}=S_g+S_M+S_{GH}+S_{ct1}+S_{ct2}\label{S_ren}\\
    S_{GH}=-\frac{2}{2\k^2}\int d^{4}x\sqrt{\g}\theta\\
    S_{ct1}=\frac{2}{2\k^2}\int d^{4}x\sqrt{\g}\frac{3}{L}\\
    S_{ct2}=\frac{4-\Delta_+}{g_5^2}\int d^{4}x\sqrt{\g}\frac{\y ^2}{L}
\end{align}

\begin{itemize}
    \item The Gibbons-Hawking term is added for the well-definedness of the variational principle. We do not add a GH term at the IR boundary.  Here $\g$ is the induced metric on the UV surface $z=\e$ and $\theta$ is the trace of the extrinsic curvature associated with this surface.
    \item The counter term $S_{ct1}$ cancels divergences from the gravity part $S_g$ as well as from Gibbons-Hawking term $S_{GH}$. 
    \item In the matter part $S_M$, the field strength $F^2$ never diverges because of the asymptotic nature of the gauge field. However, the scalar field part diverges for $\Delta_n\leq2.5$. 
    \item We need another couterm $S_{ct2}$ which cancels the divergences from scalar field action. It is to be noted that the coefficient is always $\Delta_+$, independent of the choice of normalizable mode.
\end{itemize}



\bibliography{references}

\end{document}